\documentclass[11pt]{article}
\usepackage{amsmath,amssymb,amsthm}
\usepackage{graphics,epsfig}
\usepackage{hyperref}
\usepackage{natbib}
\usepackage{color}
\usepackage{graphicx}
\usepackage{caption}
\usepackage{subcaption}
\usepackage{float}
\usepackage{dsfont}
\usepackage{mathrsfs}
\usepackage{multirow}
\usepackage{comment}
\usepackage{setspace}
\usepackage{bbm}
\usepackage{bm}
\usepackage{amsfonts}
\usepackage{xcolor}
\usepackage{pslatex}
\usepackage{setspace}
\usepackage{geometry}
\usepackage{tikz}
\usetikzlibrary{arrows.meta, positioning}

\def \ra {\rightarrow}

\def \E {\mathbb{E}}

\def \a {\alpha}

\newtheorem{example}{\bf Example}

\newtheorem{definition}{\bf Definition}

\newtheorem{remark}{\bf Remark}

\newtheorem{theorem}{\bf Theorem}

\newtheorem{prop}{\bf Proposition}
\newtheorem{lem}[theorem]{\bf Lemma}
\newtheorem{as}{\bf Assumption}


\begin{document}
	
	\title{\bfseries The Quantum Structure of Markets: Linking   Hamiltonian-Jacobi-Bellman Dynamics to Schr\"odinger Equation through Feynman Action }
	
	\author{
		Paramahansa Pramanik \footnote{e-mail: {\small\texttt{ppramanik@southalabama.edu}} }\; \footnote{Department of Mathematics and Statistics, University of South Alabama, 411 North University Boulevard, Mobile, AL 36688, USA.}}
	
	\date{\today}
	\maketitle
		
	\subparagraph{Abstract.} We develop a Euclidean path-integral control to characterize optimal firm behavior in an economy governed by Walrasian equilibrium, Pareto efficiency, and non-cooperative Markovian feedback Nash equilibrium. The approach recasts the problem as a Lagrangian stochastic control system with forward-looking dynamics, thereby avoiding the explicit construction of a value function. Instead, optimal policies are obtained from a continuously differentiable It\^o process generated through integrating factors, which yields a tractable alternative to conventional solution methods for complex market environments. This construction is useful in settings with nonlinear stochastic differential equations where standard Hamilton-Jacobi-Bellman (HJB) formulations are difficult to implement. Consistent with Feynman-Kac-type representations, the resulting solutions need not be unique. In economies with a large number of firms, the analysis admits a natural comparison with mean-field game formulations. Our main contribution is to derive a noncooperative feedback Nash equilibrium within this path-integral setting and to contrast it with outcomes implied by mean-field interactions. Several examples illustrate the method's applicability and highlight differences relative to solutions based on the Pontryagin maximum principle generated by HJB.
	
	\subparagraph{Key words:} Dynamic profit; Euclidean path integral; Feynman Action; Stochastic differential games.
	
	\section{Introduction.} This paper studies the determination of optimal firm behavior in an environment characterized by Walrasian equilibrium, Pareto efficiency, and noncooperative feedback Nash interactions among a large population of firms. We formulate the firm’s problem as the maximization of a Lagrangian profit functional, interpreted as a dynamic objective without a terminal condition subject to nonlinear stochastic differential equations. To operationalize the problem, the stochastic Lagrangian is approximated over equally spaced finite time intervals, which allows us to exploit continuous-time methods while preserving tractability. Applying It\^o lemma to this construction yields a Wick-rotated Schr\"odinger-type equation that governs the evolution of the system. Optimal strategies are then recovered from the first-order conditions associated with this transformed representation. Because the method does not rely on restrictive linearity assumptions, it accommodates a broad class of nonlinear stochastic dynamics. Moreover, the resulting Schrödinger-type formulation is equivalent to a forward Fokker-Planck equation, which creates a natural bridge to mean-field game frameworks that combine forward Kolmogorov dynamics with backward Hamiltonian conditions \citep{lasry2007mean}.
	
	This analysis relates to a large body of work in dynamic optimization and recursive methods, which have become central tools across economics and related disciplines. Recursive formulations underpin a wide range of applications, including macroeconomic dynamics \citep{kydland1977,stokey1989,ljungqvist2012,marcet2019}, population processes \citep{yoshioka2018,yoshioka2019,pramanik2021effects}, repeated interactions in game theory \citep{abreu1990}, and stochastic differential games \citep{friedman1972,chow1996,yeung2006,theodorou2010,theodorou2011,ramachandran2012,pramanik2021dis,pramanik2023path}. In these settings, the standard approach typically characterizes optimal behavior as a policy rule derived from dynamic programming principles. Under conventional assumptions, such rules are time-invariant and emerge from the solution to a Hamilton-Jacobi-Bellman (HJB) equation, providing a compact recursive representation of the problem. Following \cite{marcet2019}, this class of problems is often described as the \emph{standard dynamic programming} case, in which the solution is time-consistent and well understood within the control literature \citep{bellman1952,bellman2013,bellman2015}. A key requirement for the application of HJB methods is that feasible actions are restricted by the current state variables and exogenous shocks \citep{intriligator1971,marcet2019}. This structure ensures that the optimization problem can be expressed recursively and that the value function satisfies a well-defined functional equation. While powerful, this framework can become difficult to implement in environments where such constraints are not easily expressed or where the underlying dynamics exhibit strong nonlinearities.
	
	These limitations are relevant in economic settings with forward-looking features. Examples include models in which a principal designs contracts subject to intertemporal participation constraints, as well as optimal policy problems with equilibrium restrictions, such as Ramsey-type frameworks \citep{aiyagari2002,kydland1977,marcet2019}. Many stochastic dynamic games share these characteristics, making standard recursive techniques less straightforward to apply. By contrast, the approach developed here sidesteps the need to explicitly construct a value function and instead relies on a path-integral representation that leads directly to equilibrium conditions. In doing so, it offers an alternative route to analyzing strategic interactions in stochastic environments while maintaining comparability with established mean-field and dynamic programming methods. An additional implication of this framework is that it provides a unifying perspective on equilibrium selection and aggregation in large economies with interacting agents. In particular, by exploiting the equivalence between the Wick-rotated Schr\"odinger representation and the forward Fokker-Planck equation, the model naturally captures the evolution of the cross-sectional distribution of firms while simultaneously characterizing individual optimal responses. This dual structure parallels the coupled systems studied in mean-field game theory, where a forward Kolmogorov (or Fokker-Planck) equation describes the distributional dynamics and a backward equation captures optimal decision-making at the agent level. In contrast to standard formulations that rely explicitly on a value function and the associated HJB equation, the present approach embeds optimality directly within a path-integral framework, thereby offering an alternative route to equilibrium characterization in high-dimensional settings. This feature is particularly advantageous in applications involving heterogeneous firms, strategic complementarities, or aggregate uncertainty, where solving the coupled HJB-Fokker-Planck system can be computationally burdensome. By recasting the problem in terms of stochastic flows and action functionals, the method highlights how equilibrium outcomes emerge from the interaction between individual stochastic trajectories and aggregate distributions, providing a transparent link between micro-level decision rules and macro-level dynamics.
	
	In many dynamic economic environments, current decisions are not isolated choices but instead embody implicit commitments regarding future behavior. Because an action taken today effectively constrains or signals future actions, agents must keep track of additional state variables that encode past promises and obligations. This feature introduces what are commonly referred to as \emph{forward-looking} dynamics, in which the evolution of the system depends not only on present states and shocks but also on previously made commitments. As emphasized by \cite{marcet2019}, such problems typically fall outside the scope of standard dynamic programming because they do not satisfy the conditions required for an HJB equation. In particular, the absence of time consistency and the dependence on commitment variables prevent the formulation of a conventional recursive structure. As a result, the solution cannot be summarized by a time-invariant policy function derived from a value function, complicating both theoretical analysis and computational implementation. The lack of a standard recursive form also poses significant challenges for stochastic modeling, since many of the tools developed for HJB-based systems rely critically on this structure.
	
	To address these difficulties, \cite{marcet2019} proposes an alternative formulation based on what is termed a saddle-point functional equation. Rather than casting the problem purely as an optimization task, this approach reframes it as a saddle-point problem in which optimality emerges from the interaction between maximizing and minimizing components of the functional. This distinction is important: while the HJB framework focuses on identifying a value function that satisfies a supremum condition, the saddle-point approach directly characterizes equilibrium conditions without requiring such a value function to exist or be well behaved. One advantage of this formulation is its ability to accommodate settings that are problematic for standard dynamic programming, including environments with non-concave objective functions, non-differentiable value functions, or multiple equilibria \citep{pramanik2024motivation}. By working with the integrand of the problem rather than relying on recursive decomposition, the method provides a more flexible analytical structure capable of handling a broader class of stochastic control problems. This flexibility becomes particularly valuable in economic models with complex incentive constraints or strategic interactions, where traditional assumptions ensuring concavity and smoothness are often violated.
	
	A different strand of the literature approaches forward-looking stochastic optimization through the so-called promised-utility framework, which has been widely applied in contract theory and repeated games \citep{green1987,thomas1988,abreu1990}. In this setting, the planner or contracting agent specifies current actions together with a schedule of promised continuation utilities contingent on future states of the world. These promised utilities effectively become additional state variables that evolve over time, thereby transforming the original problem into one that resembles a recursive formulation \citep{pramanik2025strategies}. In principle, this transformation allows the use of tools similar to those employed in HJB-based analyses, since the augmented state space captures the relevant intertemporal trade-offs. However, this apparent simplification comes at a cost \citep{pramanik2024stochastic,pramanik2025factors}. As noted by \cite{marcet2019,pramanik2024parametric}, ensuring that the continuation problem is well defined requires imposing restrictions on the set of admissible promised utilities. Without such restrictions, the resulting problem may fail to satisfy feasibility or incentive compatibility conditions, undermining the validity of the solution \citep{pramanik2025optimal}. The computational burden associated with the promised-utility approach is substantial, particularly when dealing with stochastic environments and high-dimensional state spaces. Because promised utility today must specify outcomes for every possible future contingency, it effectively defines a mapping from future states to continuation values, which then becomes part of tomorrow’s state variables. This structure necessitates the computation of a correspondence describing the set of feasible continuation utilities, a task that is often extremely demanding in practice \citep{kydland1977,marcet2019,pramanik2024measuring}. The difficulty is compounded by the need to verify that these correspondences satisfy incentive constraints and remain consistent over time. As a result, even though the promised-utility framework provides a conceptually appealing way to restore a recursive structure, its numerical implementation can be highly complex. These challenges highlight the need for alternative approaches such as the saddle-point formulation discussed above that can accommodate forward-looking dynamics without requiring the explicit construction of such intricate correspondences.
	
	The conceptual framework developed here extends naturally beyond economic systems and can be adapted to a wide range of problems in the life sciences where decision-making under uncertainty interacts with evolving biological states. Consider, for instance, the context of cancer treatment \citep{pramanik2024analysis,dasgupta2026frequent,kakkat2026angiotensin}, where the size or composition of a tumor can be interpreted as the state variable and the intensity or timing of chemotherapy serves as the control \citep{vikramdeo2024mitochondrial,valdez2025association,dunbar2026modeling}. In such settings, current treatment choices inherently influence not only immediate tumor dynamics but also future treatment possibilities \citep{pramanik2025dissecting,yusuf2025predictive}, toxicity accumulation, and resistance pathways \citep{vikramdeo2024abstract}, thereby introducing strong forward-looking features that are difficult to capture using standard recursive methods. By representing the treatment problem through a stochastic Lagrangian formulation, one can model the evolution of tumor cells under uncertainty while incorporating inter-temporal trade-offs between aggressive intervention and long-term patient outcomes \citep{pramanik2026bayesian}. The transformation of the problem into a Schr\"odinger-type equation provides a way to characterize the probabilistic evolution of tumor states \citep{powell2026role}, while optimal treatment strategies emerge from first-order conditions without requiring the explicit construction of a value function. This is particularly useful in oncology \citep{khan2024mp60,powell2025genomic,yusuf2025prognostic}, where biological responses are often nonlinear, heterogeneous, and subject to random fluctuations, making traditional dynamic programming approaches difficult to implement or calibrate. Moreover, the equivalence with a forward diffusion equation allows the framework to capture the distributional evolution of tumor cells \citep{valdez2025exploring}, including the emergence of resistant subpopulations, thereby offering a richer description of treatment dynamics than deterministic models.
	
	A similar line of reasoning applies in broader areas of mathematical biology, such as the study of animal movement and migration patterns. For example, in modeling fish migration, the spatial distribution of a population can be treated as the state variable \citep{bulls2025assessing,maki2025new}, while behavioral or environmental responses such as movement effort, navigation strategies, or responses to external stimuli like temperature gradients or predation risk can be interpreted as control variables. These systems are inherently stochastic due to environmental variability and incomplete information, and they often involve forward-looking behavior in the sense that current movement decisions affect future survival probabilities and reproductive success \citep{ellington2025metascorelens}. Within the proposed framework, one can construct a stochastic Lagrangian that captures both energetic costs and ecological benefits, discretize it over time, and derive a governing equation that describes the evolution of the population distribution \citep{ellington2025playmydata}. The resulting formulation enables the analysis of optimal migration strategies as emerging from underlying stochastic dynamics rather than being imposed exogenously. Importantly, because the method accommodates non-linearities and does not depend on restrictive assumptions about concavity or smoothness, it can incorporate realistic biological features such as crowding effects, spatial heterogeneity, and feedback between individuals and their environment. In both cancer modeling and ecological applications, this approach provides a unified way to link individual-level decision processes with aggregate outcomes, offering insights into how optimal strategies arise in complex, uncertain, and dynamically evolving systems.
	
	The intellectual foundations of game theory can be traced to the seminal contributions of \cite{morgenstern1944} and \cite{nash1950}, which together established the analytical framework for studying strategic interaction among rational agents. These early works formalized the idea that individual decision-makers must account for the anticipated responses of others, thereby introducing equilibrium as a central organizing concept. The notion of Nash equilibrium, in particular, provided a tractable and widely applicable solution concept that has since become a cornerstone of modern economic theory. Building on this foundation, a large body of research has examined the behavior of multi-player systems and the resulting equilibrium outcomes across a variety of settings \citep{ho1965,browne2000,yeung2006,ramachandran2012,ludkovski2015}. These developments expanded the scope of game theory beyond static environments to include dynamic and strategic considerations, paving the way for more sophisticated models in which agents interact over time and under uncertainty. As a result, game theory evolved into a versatile tool for analyzing competitive and cooperative behavior in economics, finance, engineering, and related disciplines. The formal study of dynamic strategic interactions now known as differential games, originated with the work of \cite{isaacs1954}, who provided the first systematic mathematical treatment of such problems. This framework was later refined and extended in \cite{isaacs1999}, further solidifying its role in the analysis of time-dependent strategic behavior and inspiring a wide range of subsequent contributions \citep{ramachandran2012}. Early on, it appeared natural to interpret differential games as extensions of optimal control problems, with the key distinction being that control variables are distributed across multiple decision-makers whose objectives may conflict. This perspective suggested that techniques from optimal control theory could be directly applied to strategic settings. Indeed, initial efforts to connect these fields can be found in \cite{isaacs1954}, and were later developed in works such as \cite{berkovitz1975} and \cite{krasovskii1988}. However, as the literature progressed, it became clear that differential games possess structural features most notably the presence of strategic interdependence that distinguish them fundamentally from single-agent control problems. Consequently, many of the analytical tools designed for optimal control do not readily extend to multi-agent environments, necessitating the development of new methods tailored specifically to game-theoretic contexts \citep{ramachandran2012}.
	
	The incorporation of uncertainty into dynamic strategic models marked a significant turning point in the evolution of the field. Beginning in the 1960s, researchers introduced stochastic elements into players’ observations and state dynamics, recognizing that real-world decision-making often takes place in environments characterized by incomplete information and random shocks \citep{ho1966,haurie1984}. This line of inquiry led to the formal definition of stochastic differential games in \citep{roxin1970,bafico1973}, which extended the deterministic framework to accommodate probabilistic dynamics. Subsequent work in the late 1970s focused on establishing rigorous theoretical foundations for these models, including conditions for existence and uniqueness of equilibrium solutions. These advances relied heavily on the use of martingale methods and variational inequality techniques, as demonstrated in contributions such as \citep{bensoussan1976,elliott1976,bensoussan1977,bensoussan1977a,elliott1977,elliott1981,ramachandran2012}. Together, these developments transformed stochastic differential games into a well-defined and analytically tractable field, capable of addressing complex strategic interactions under uncertainty and providing a bridge between probability theory, optimization, and economic modeling.
	
	Early contributions to the theory of differential games relied heavily on the machinery of dynamic programming, most notably through the Hamilton–Jacobi–Bellman–Isaacs (HJBI) equation, often referred to simply as the HJB framework in this context \citep{pramanik2025strategic}. Within this paradigm \citep{pramanik2025impact}, the central object of interest is the value of the game, which captures the equilibrium payoff associated with optimal strategies chosen by competing agents. A substantial portion of the early literature was devoted to formalizing this concept and deriving the associated HJB equation in a mathematically rigorous manner, particularly in settings where strategic interactions unfold over continuous time \citep{belal2007,ramachandran2012,pramanik2025optimal}. However, these efforts quickly revealed fundamental analytical difficulties. In many economically and practically relevant models, the HJB equation fails to admit classical (smooth) solutions, largely due to the presence of nonlinearities, discontinuities, and strategic conflicts embedded in the game structure. Foundational works such as \cite{berkovitz1975}, \cite{fleming1961}, and \cite{friedman1994} demonstrate that smooth solutions generally do not exist, and that even when solutions can be identified, they are often highly non-unique \citep{ramachandran2012}. This lack of regularity and uniqueness complicates both the interpretation of equilibrium outcomes and the implementation of computational methods, as standard techniques from optimal control theory rely critically on differentiability and well-behaved value functions. As a result, the early HJB-based approach, while conceptually appealing, proved insufficient for handling the full complexity of differential games.
	
	In response to these challenges, the development of viscosity solution theory in the 1980s marked a major breakthrough in the analysis of HJB-type equations. Rather than requiring classical differentiability, the viscosity framework provides a generalized notion of solution that is well suited to handling nonsmooth value functions and nonlinear partial differential equations. Pioneering contributions by \cite{crandall1983}, \cite{crandall1984}, \cite{souganidis1985}, \cite{lions1988}, and later \cite{buckdahn2008} established that, under appropriate boundary conditions, the value function of a differential game can be characterized as the unique viscosity solution of the corresponding HJBI equation. This result is particularly important because it restores a form of uniqueness that is otherwise absent in the classical setting, thereby providing a coherent interpretation of equilibrium outcomes. In addition, the viscosity solution framework offers powerful tools for analyzing numerical schemes derived from dynamic programming principles. In particular, it enables researchers to prove that discretization algorithms converge to the correct solution and to quantify the rate at which this convergence occurs. These advances significantly improved the tractability of differential games, making it possible to study more complex models with greater confidence in both analytical and computational results. Further extensions of this theory to infinite-dimensional settings, as developed in \cite{borkar1992} and \cite{swikech1996}, expanded its applicability to a broader class of problems, including those arising in stochastic control and functional differential equations, thereby solidifying viscosity methods as a central pillar in the modern theory of differential games.
	
	The structure and availability of information play a central role in shaping decision-making processes within stochastic differential games, as the strategies adopted by players depend critically on what they know about the underlying state variables, the actions of other agents, and the evolution of uncertainty over time \citep{pramanik2024measuring}. In much of the existing literature, including the works discussed above, it is typically assumed that agents operate under conditions of perfect and complete information, meaning that all relevant state variables and model parameters are fully observable and commonly known. While this assumption greatly simplifies the analysis and facilitates the application of established solution concepts such as Nash equilibrium and HJB-based methods, it represents an idealized benchmark that rarely holds in real-world environments. In practice, agents often face informational frictions, including noisy observations, delays in information transmission, and asymmetric access to relevant data, all of which can significantly alter strategic behavior and equilibrium outcomes. The interaction between information structures and dynamic strategic decision-making has been explored in contributions such as \cite{ho1974minimax}, \cite{friedman1975}, and \cite{ramachandran1995}, which highlight how different informational assumptions can fundamentally change the nature of optimal strategies and the resulting equilibria in differential games. Despite these important insights, the literature on stochastic differential games with incomplete or imperfect information remains relatively underdeveloped when compared to the corresponding body of work in stochastic control with partial observations. As noted by \cite{ramachandran2012}, while stochastic control theory has made substantial progress in addressing filtering, estimation, and decision-making under uncertainty with limited information, extending these advances to multi-agent strategic settings introduces additional layers of complexity. In particular, the presence of multiple decision-makers with potentially different information sets gives rise to issues such as signaling, belief formation, and higher-order expectations, which are not present in single-agent control problems. Consequently, developing a comprehensive framework for stochastic differential games under realistic informational constraints remains an open and challenging area of research, with significant implications for applications in economics, finance, engineering, and beyond.
	
	Among the key contributors to the development of stochastic differential games, Avner Friedman occupies a particularly important position, and his methodological contributions warrant separate discussion due to their distinct analytical perspective and lasting influence. In \cite{friedman1972}, Friedman introduces a novel way of formulating differential games by approximating them as a sequence of discrete-time games, thereby creating a bridge between continuous-time dynamics and tractable discrete representations. Within this framework, the control variables are assumed to enter separately into the kinetic equation governing the state evolution and into the integral component of the payoff functional, a structural feature that facilitates both analysis and computation \citep{ramachandran2012}. This decomposition allows the dynamic interaction between players to be studied in a stepwise manner while preserving the essential characteristics of the continuous-time game. One of the central results established in \cite{friedman1972} concerns pursuit–evasion games with general, potentially discontinuous payoff functions. Under suitable conditions, Friedman shows that such games admit a well-defined value and possess saddle-point equilibria, thereby ensuring that optimal strategies exist for both players despite the lack of smoothness in the payoff structure. Moreover, he demonstrates that the value function in this setting is Lipschitz continuous, a property that plays a crucial role in guaranteeing stability and robustness of the solution and can be extended to more general classes of survival differential games. In addition to these contributions, Friedman develops methods for determining the value of games with fixed time horizons and proposes a generalized procedure for computing saddle points, which are central to characterizing equilibrium behavior in zero-sum settings. These insights not only deepen the theoretical understanding of pursuit–evasion dynamics but also provide practical tools for analyzing a broader class of differential games, reinforcing the significance of Friedman’s approach within the literature \citep{ramachandran2012}.
	
	The Feynman path integral formulation represents an alternative quantization procedure that is grounded in the use of a quantum Lagrangian, in contrast to Schrödinger’s formulation, which is based on the Hamiltonian representation of dynamics \citep{fujiwara2017}. Rather than focusing on the evolution of a wave function through a differential equation, the path integral approach characterizes system behavior by aggregating contributions from all possible trajectories, thereby offering a fundamentally different conceptual lens on dynamical systems. This distinction has made the path integral framework a powerful and versatile tool extending well beyond its origins in quantum mechanics, with applications emerging in fields such as engineering, biophysics, economics, and finance, where stochastic processes and optimization under uncertainty play a central role \citep{kappen2005,anderson2011,yang2014path,fujiwara2017}. Although the Feynman and Schrödinger approaches are widely regarded as equivalent in terms of the physical predictions they generate, establishing this equivalence rigorously remains mathematically challenging. The primary difficulty stems from the fact that the Feynman path integral is not defined with respect to a countably additive measure, which complicates its formal interpretation within standard measure-theoretic frameworks \citep{linetsky1997,johnson2000,fujiwara2017,polansky2021motif}. Beyond these theoretical considerations, practical computational issues also arise when solving high-dimensional problems using traditional grid-based partial differential equation (PDE) methods associated with Hamiltonian formulations. In particular, the computational cost and memory requirements of such methods grow exponentially with the dimensionality of the system, rendering them infeasible for many real-world applications \citep{yang2014path,hua2019assessing}. As a result, alternative numerical strategies have been developed, among which Monte Carlo methods play a prominent role. This perspective underlies the development of path integral control, which leverages stochastic sampling techniques to approximate solutions to a class of stochastic control problems typically formulated via HJB equations \citep{kappen2005,theodorou2010,theodorou2011,morzfeld2015}. By avoiding the need to construct and evaluate a global grid over the state space, path integral control offers a scalable and computationally efficient approach, particularly well suited for high-dimensional systems where traditional PDE-based methods become impractical \citep{yang2014path}.
	
	If the objective function is quadratic and the differential equations are linear, then solution is given in terms of a number of Ricatti equations which can be solved efficiently \citep{kappen2007b,pramanik2021effects}. But the market dynamics is more complicated than just an ordinary linear differential equation and non-linear stochastic feature such as \emph{Merton-Garman-Hamiltonian dynamics} gives the optimal solution a weighted mixture of suboptimal solutions, unlike in the cases of deterministic or linear optimal control where a unique global optimal solution exists \citep{kappen2007b,pramanik2021effects}. In the presence of Wiener noise, \emph{Pontryagin Maximum Principle}, a variational principle, that leads to a coupled system of stochastic differential equations with initial and terminal conditions, gives a generalized solution \citep{kappen2007b,oksendal2019,pramanik2021effects}. Although incorporating randomness with its HJB equation is straight forward but difficulties come due to dimensionality when a numerical solution is calculated for both of deterministic or stochastic HJB \citep{kappen2007b}. General stochastic control problem is intractable to solve computationally as it requires an exponential amount of memory and computational time because, the state space needs to be discretized and hence, becomes exponentially large in the number of dimensions \citep{theodorou2010,theodorou2011,yang2014path}. Therefore, in order to calculate the expected values it is necessary to visit all states which leads to the summations of exponentially large sums \citep{kappen2007b,yang2014path,pramanik2021effects}. Following \cite{kappen2005a} and \cite{kappen2005} we know that, a class of continuous non-linear stochastic finite time horizon control problems can be solved more efficiently than Pontryagin's Maximum Principle. These control problems reduce to computation of path integrals interpreted as free energy because, of their various statistical mechanics forms such as Laplace approximations, Monte Carlo sampling, mean field approximations or belief propagation \citep{kappen2005a,kappen2005,kappen2007b,pramanik2021effects}.
	
	A class of non-linear HJB equations can be transformed into linear equations by imposing a logarithmic transformation. This transformation goes back to the early days of quantum mechanics and was first used by Schrödinger to relate HJB equation to the Schr\"odinger equation \citep{kappen2007b,pramanik2021effects}. Because of this linear feature, backward integration of HJB equation over time can be replaced by computing expectation values under a forward diffusion process which requires a stochastic integration over trajectories that can be described by a path integral \citep{kappen2007b,pramanik2021effects}. In this paper we are going to discuss this relationship rigorously. Furthermore, in more generalized case like Merton-Garman-Hamiltonian system, getting a solution through \emph{Pontryagin Maximum principle} is impossible and Feynman path integral method gives a solution \citep{baaquie1997}. Previous works using Feynman path integral method includes in motor control theory by \cite{kappen2005}, \cite{theodorou2010} and \cite{theodorou2011}. A rigorous discussion of this quantum approach in finance has been done in \cite{belal2007}. In \cite{pramanik2020} a Feynman-type path integral has been introduced to determine a feedback stochastic control. This method works in both linear and non-linear stochastic differential equations and a Fourier transformation has been used to find out solution of Wick-rotated Schr\"odinger type equation \citep{pramanik2020}. For these reasons we are going to use Feynman approach to find stochastic control solutions. Here we also discuss under which conditions Merton-Garman-Hamiltonian dynamics give solution and under which it does not. In that case we implement our Feynman-type path integral approach to get a solution.
	
	\begin{figure}[H]
		\centering
		\includegraphics[width=8.7cm]{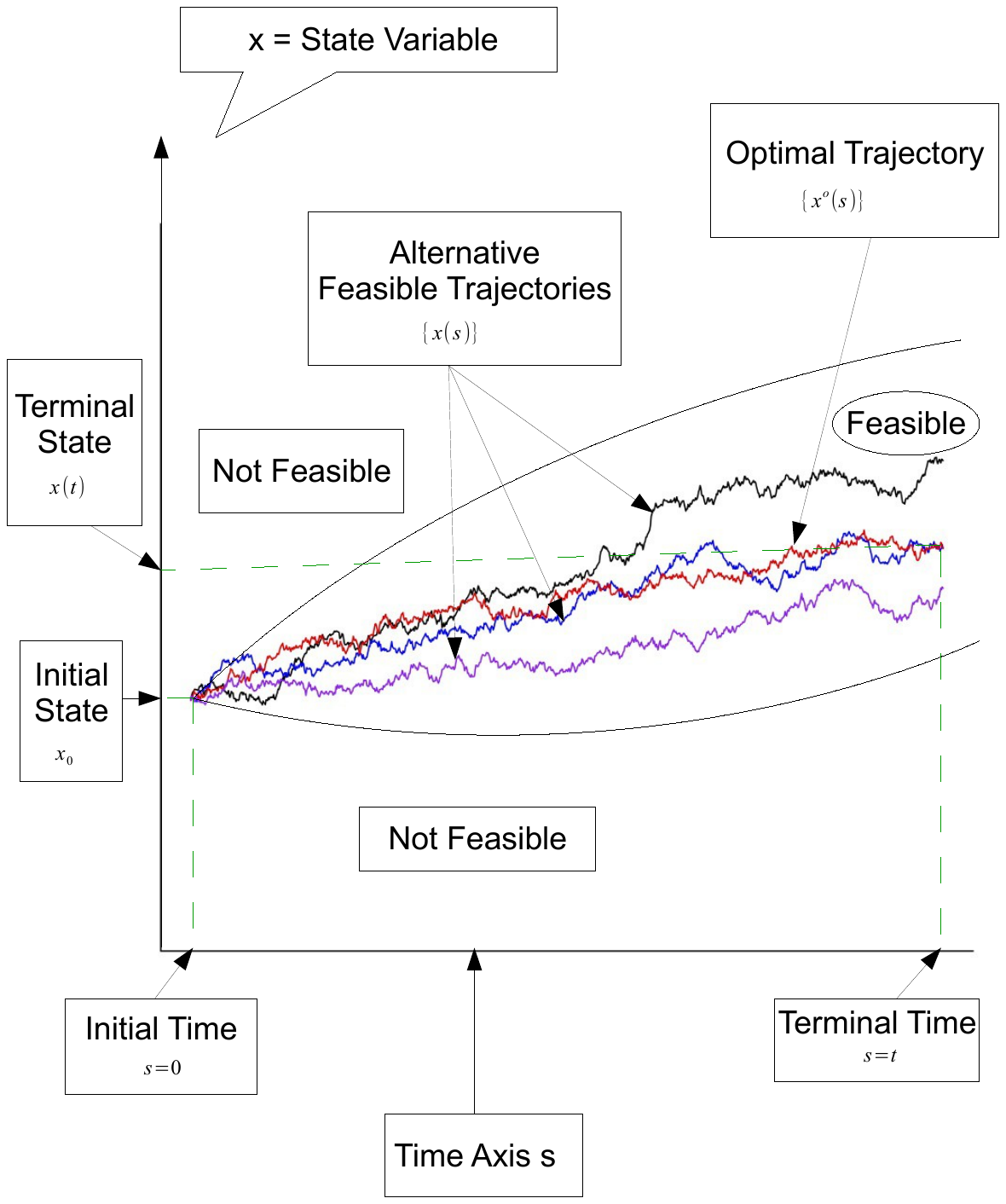}
		\caption{Forward-looking stochastic trajectories of the market share $X$ originating from the initial state $x_0$ at time $s=0$. The region bounded by the arcs represents the feasible set, within which trajectories may temporarily exit and re-enter over time. The red curve $\{x^o(s)\}$ denotes the optimal trajectory that reaches the terminal state $X(t)$ at time $t$, while the predominance of upward paths motivates the exponential action formulation. The Brownian motion component is treated by time-slicing $[0,t]$, applying It\^o's Lemma, and deriving the associated Schr\"odinger-type representation used to recover the optimal strategy.}
		\label{Traj}
	\end{figure}

Figure (\ref{Traj}) illustrates the aggregation of all admissible \emph{forward-looking} trajectories originating from the initial state $x_0$ at time $s=0$, where each path represents a feasible evolution of the state variable under stochastic dynamics. The shaded region enclosed by the two arcs characterizes the feasible set within which admissible trajectories must lie in order to satisfy the underlying constraints of the problem. It is important to note that individual trajectories are not necessarily confined to this region at all times; for instance, a given path may temporarily exit the feasible set and subsequently re-enter it, reflecting the stochastic nature of the system. This feature is exemplified by the blue trajectory, which departs from the feasible region shortly after originating at $x_0$, before eventually moving back toward admissible states. In contrast, the red curve represents the optimal trajectory, denoted by ${x^o(s)}$, corresponding to the evolution of the state variable $X$ interpreted here as a positive measure of market share which reaches the terminal state $X(t)$ at time $t$. Since the majority of trajectories exhibit an upward-sloping behavior, it is natural to represent the associated action functional as an exponential-type expression, which facilitates analytical tractability in the subsequent derivations. Given the forward-looking nature of the framework, no explicit terminal condition is imposed; instead, decision-making at each time $s$ depends only on the information available up to that point, consistent with a non-anticipative structure. The presence of Brownian motion in the dynamics of each trajectory introduces randomness that must be handled carefully. To this end, the time interval $[0,t]$ is partitioned into a collection of sufficiently small subintervals, allowing for a local approximation of the stochastic process. By applying It\^o Lemma to this discretized representation, one can derive a Schr\"odinger-type equation that governs the evolution of the system in a probabilistic sense. Finally, optimal strategies are obtained by differentiating this equation with respect to the control variable, thereby linking the stochastic dynamics with the underlying optimization problem in a manner consistent with the path-integral formulation.

In this paper we consider dynamic profit maximization over a time interval with finite horizon $t>0$. The objective is to find an optimal strategy for a firm in a system whose state dynamics are specified by a  stochastic differential equation. The instantaneous profit function we consider depends on the time $s$, a positive real-valued measure of the market share of the firm $X(s)$, and the real-valued dynamic strategy of the firm $u(s)$. Both $X(s)$ and $u(s)$ are functions of time $s$ and from the control theory point of view, $X$ is the state variable and $u$ is the control variable. The dynamic profit function is represented by a real valued function $\pi[s,X(s),u(s)]$ so that $\pi:[0,t]\times\mathbb R^2\ra\mathbb R$. Here $X\in \mathcal{X}$ and $ u\in \mathcal{U}$, where $\mathcal{X}$ is a functional space corresponding to the set of all  real-valued measure of market share trajectories taking the values from $\mathbb R^2$, and $\mathcal{U}$ is a functional space corresponding to the set of all possible strategies available to the firm. We assume the functional spaces $\mathcal{X}$ and $\mathcal{U}$ are bounded and complete so any choice outside of this set leads to a complete shut-down. The profit over the continuous time interval  $[0,t]$ is measured by the integral 
\[
\int_0^t\pi[s,X(s),u(s)]ds.
\]
The dynamics of the real-valued measure of market share are  given by
\begin{equation}\label{mkt}
dX(s)=\mu[s,X(s),u(s)]ds+\sigma[s,X(s),u(s)]dB(s),
\end{equation}
where $\mu[.]$ is the drift component, $\sigma[.]$ is the diffussion coefficient, and $B(s)$ is a Brownian motion process.  Hence, the real valued measure of market share is assumed to be an It\^o process. Let $$G[s,X(s),u(s)]ds=dX(s)-\mu[s,X(s),u(s)] ds-\sigma[s,X(s),u(s)] dB(s)$$
and
$$J_t(X,u):=\int_0^t\ G[s,X(s),u(s)] ds.$$

\begin{definition}\label{d1.1}
	\textbf{(Stochastic Isoperimetric Constraint \citep{edwards2012,haurie2006,torok2000})} 
	Let $(\Omega,\mathcal{F},\mathbb{P})$ be a probability space, and let $X=\{X(s)\}_{s\in[0,t]}$ denote a stochastic process taking values in a measurable state space $\mathcal{X}$, while $u=\{u(s)\}_{s\in[0,t]}$ denotes an admissible control process belonging to a control set $\mathcal{U}$. Consider a real-valued functional
	\[
	J_t(X,u) \;=\; \int_0^t \mathcal{L}(X(s),u(s),s)\,ds,
	\]
	where $\mathcal{L}$ is an integrable Lagrangian function ensuring that $J_t(X,u)$ is well-defined and finite almost surely. 
	
	An \emph{integrable functional constraint} of the form
	\[
	\mathbb{P}\big[J_t(X,u)=c\big]=1,
	\]
	for all admissible state trajectories $X\in\mathcal{X}$ and control processes $u\in\mathcal{U}$, where $c\in\mathbb{R}$ is a given positive constant, is called a \emph{stochastic isoperimetric constraint}.
\end{definition} 
	
	This definition requires that the accumulated value of the functional $J_t(X,u)$ remains fixed at the constant level $c$ almost surely, thereby imposing a pathwise restriction on the admissible pairs $(X,u)$ over the entire time horizon. In other words, despite the presence of stochasticity in the evolution of the state process, the constraint enforces that the total ``resource,'' ``cost,'' or ``measure'' represented by $J_t(X,u)$ is preserved with probability one. Such constraints generalize classical isoperimetric conditions from deterministic calculus of variations to stochastic environments, and they play a central role in problems where integral quantities must be maintained under uncertainty.
	
	\begin{definition}\label{dnonholo}
		(Stochastic Non-Holonomic Constraint) A constraint of the form
		\begin{equation}
		P\big(G[s,X(s),u(s)]=0\big)=1,\notag
		\end{equation}
		for all $s\in[0,t],\ X\in \mathcal X$ and $u\in \mathcal U$ is a stochastic non-holonomic constraint.	
	\end{definition}
	
	As constraint represented in Equation (\ref{mkt}) satisfies Definition \ref{d1.1} and Definition \ref{dnonholo}, and hence is a stochastic isoperimetric non-holonomic  constraint.
	
	In this paper we are interested in determining three scenario: Walrasian equilibrium, Pareto optimality and Nash equilibrium. The Walrasian system is a fundamental market structure in economics, and is the basis of many other market systems \citep{walras1900}. The main assumption under this system is that each firm is small when compared to the entire industry and have identical profit functions and market dynamics.  Under Walrasian structure each identical price taking firm maximizes its own profit subject to Equation (\ref{mkt}) and determines its optimal real-valued measure of market share. When every firm in the industry does so, the global policy maker finds the total optimal real-valued measure of market share and based on the market conditions determines the industry-price. Therefore, small firms do not influence the industry price. Furthermore, under this case a firm's real-valued measure of market share is so small compared to the industry, its inter-temporal movement is random which is analogous to a movement of a quantum particle in a vector field, and this leads us to use a stochastic market dynamics expressed in Equation (\ref{mkt}). The industry consists of all the firms, and its price is determined by the entire system. In this system, a single firm can earn at least zero profit in the long run where the industry allows free entry and exit of any firm. Therefore, if a firm  wants to survive (i.e., zero profit condition), it has to achieve its average cost \citep{walras1900}. We will discuss about these three definitions in the next section. In this paper, our main objective is to find an optimal strategy of a firm by solving a maximization problem with a filtration generated by $X$ for a non-negative time $s$ as $\mathcal{F}_0^X:=\{\mathcal F_s^X:s\geq 0\}$, and initial state variable $x_0\in\mathcal{X}$, is given by
	\begin{align}\label{n1}
	&\max_{u \in \mathcal U}\ \Pi(u,t)=\max_{u \in\mathcal U}\ \E_0\left\{ \int_{0}^{t}\ \pi[s,X(s),u(s)]ds\biggr|\mathcal{F}_0^X\right\},
	\end{align} 
	subject to the constraint given in Equation (\ref{mkt}), and initial condition $x(0)=x_0\in\mathcal X$. In Equation(\ref{n1}) $\E_0[.]=\E[.|x_0]$.
	
	\begin{as}\label{as_profit}
		Following set of assumptions of the profit function $\pi$ is considered:
		\begin{itemize}
			\item For continuous time interval $[s,t]$ with $s\geq0$, the filtration $\mathcal F_0^X$ takes values from a set $\mathfrak Z\subset \mathbb R$. $\{\mathcal F_s\}_{s=0}^t$ is an exogenous Markovian stochastic processes defined on the probability space $(\Omega,\mathfrak Z_\infty,\mathcal F,\mathcal F_0^X,\mathcal P)$, where $\Omega$ is sample space, $\mathfrak Z_\infty$ is the background smooth manifold generated by $X$, $\mathcal F$ is a Borel $\sigma$-algebra, and $\mathcal P$ is a canonical Wiener measure.
			\item For all $\{X(s),u(s)\}$, there exists an optimal strategy $\{ u^*(s)\}_{s=0}^t$, with initial condition $x_0$, which satisfies the stochastic dynamics represented by the Equation (\ref{mkt}) for continuous time $s\in[0,t]$.
			\item The profit function $ \pi[s,X(s),u(s)]$ is uniformly bounded, continuous on both the real-value measure of market share ($\mathcal X$) and strategy spaces ($\mathcal U$) and, for a given $\{X(s),u(s)\}\in\mathcal X\times\mathcal U$, it is $\mathcal F_0^X$-measurable.
			\item The profit function $\pi[s,X(s),u(s)]$ is strictly convex with respect to the real-valued measure of market shares and strategies.
			\item For all $\{X(s),u(s)\}\in\mathcal X\times\mathcal U$, there exists an interior optimal strategy $\{u^*(s)\}_{s=0}^t$, with initial condition $x_0\in\mathcal X$ satisfy Equation (\ref{mkt}), such that
			\[
			\E_0\left\{\pi[s,X(s),u(s)]\bigg|\mathcal F_0^X\right\}>0.
			\]
			\item In addition to the above argument, there exists an $\varepsilon>0$ such that for all $\{X(s),u(s)\}\in\mathcal X\times\mathcal U$,
			\[
			\E_0\left\{\pi[s,X(s),u(s)]\bigg|\mathcal F_0^X\right\}>\varepsilon.
			\]
		\end{itemize}
	\end{as}

\begin{remark}
	Assumption \ref{as_profit} guarantees the integrability conditions of the dynamic profit function $\pi[s,X(s),u(s)]$.	
\end{remark}

Following is the structure of the paper. Section 2 considers the definitions,assumptions related to three equililibria and Feynman-type path integrals. At the end of this section we have shown the existences of a Nash equilibrium under our case. Section 3 discusses about Pontryagin's maximum principle for three market equilibria. In section 4 we have shown with a simple example how HJB equation and path integral approach are related. Section 5 consists of the main results of our paper. Proofs of the main results are given in section 6. Finally, section 7 discusses our findings of this paper. Some important mathematical properties are included in the appendix section.

\section{Background markets and their dynamics.}

In this section, we provide a rigorous formulation of the core equilibrium concepts that underpin the analysis of stochastic economic systems with interacting agents. Specifically, we begin by defining stochastic Walrasian equilibrium, Pareto optimality, and noncooperative Nash equilibrium within a unified probabilistic framework that accommodates uncertainty and dynamic interactions. These definitions extend their classical deterministic counterparts by incorporating stochastic state variables and information structures, thereby allowing for a more realistic representation of economic environments subject to random shocks and evolving expectations. Following the formal definitions, we examine key properties of these equilibrium notions, including existence, efficiency, and their interrelationships, with particular attention to how stochastic dynamics and strategic behavior jointly influence outcomes. This analysis highlights both the similarities and the distinctions between competitive and strategic equilibria in uncertain settings, as well as the conditions under which efficiency and decentralization can be reconciled. In the latter part of the section, we introduce a Feynman-type path integral approach as an alternative analytical tool for studying these problems. This method offers a different perspective by focusing on the aggregation of possible stochastic trajectories and their associated action functionals, thereby providing a flexible framework that complements traditional approaches based on dynamic programming and partial differential equations.

Let $(\Omega,\mathcal F,\mathcal F_0^{\mathbf X},\mathcal P)$ be a probability space with sample space $\Omega$, Borel $\sigma$-algebra $\mathcal F$, filtration $\mathcal F_0^{\mathbf X}$ generated by a Brownian motion ${\bf B}(s)$ from $s=0$, and a canonical Wiener measure $\mathcal P$ on continuous time $[0,t]$. Since $\mathcal F_0^{\mathbf X}=\{\mathcal F_s^{\mathbf X}:s\geq 0\}$ the filtration generated by the measure of the market share $X$ alone, the problem of filtering is to construct and update expectations conditioned on $\mathcal F_s^{\mathbf X}$ for $s\in[0,t]$ \citep{hansen2022}. Let for $k$-number of firms, $\mathbf X$ be a Markov diffusion in $\mathbb R^k$ satisfies
\[
\mathbf X(t)=\mathbf{x}_0+\int_0^t \bm\mu[s,\mathbf X(s),\mathbf u(s)]ds+\int_0^t\bm\sigma [s,\mathbf X(s),\mathbf u(s)] d\mathbf B(s),
\]
where $\mathbf B=\{\mathbf B(s):s\geq 0\}$ is a m-dimensional vector of a standard Brownian motion defined on $(\Omega,\mathcal F,\mathcal F_0^{\mathbf X},\mathcal P)$, and $\mathbf{x}_0$ is an initial condition. The $k\times m$-dimensional process $\{\bm\sigma[s,\mathbf X(s),\mathbf u(s)]:s\geq 0\}$ is adapted to the filtration $\mathcal F_0^{\mathbf X}$ and we assume that, the $k\times k$-dimensional diffusion matrix $\bm\sigma\bm\sigma'$ is non-singular in nature for all $s$ almost surely, and the symbol $'$ represents the transposition of a matrix. The exact dimensionality of each element of the above integral equation is discussed after Equation (\ref{vectordynamics}). 

To clarify the computational advantage of the proposed approach, we briefly contrast the complexity of classical grid-based dynamic programming methods with the path integral Monte Carlo framework. In classical HJB solvers, the state space $\mathbb{R}^k$ is discretized along each dimension. If $n$ grid points are used per dimension, the total number of grid nodes grows exponentially as $O(n^d)$, which rapidly becomes computationally infeasible even for moderate $k$. This phenomenon is commonly referred to as the curse of dimensionality. In contrast, the path integral formulation replaces deterministic gridding with stochastic path sampling. The computational burden is primarily driven by the number of simulated trajectories rather than by an explicit discretization of the $k$-dimensional state space. For a fixed time discretization, the cost typically scales approximately linearly with the number of sampled paths, i.e., $O(n)$, where $n$ denotes the number of Monte Carlo trajectories \citep{pramanik2024estimation}. While the variance of the estimator may increase with dimension and thus require more samples for high accuracy, the exponential dependence on $k$ inherent in grid-based methods is avoided. Consequently, the PI approach offers substantial computational advantages in moderately to highly dimensional control problems, which motivates its use in the present market-share diffusion environment \citep{pramanik2024dependence}.

\begin{remark}
	Although the path integral Monte Carlo approach avoids the exponential grid growth associated with classical HJB solvers, the number of required samples generally depends on the desired statistical accuracy. In particular, for standard Monte Carlo estimators the root mean square error typically decays at the dimension-independent rate $O(n^{-1/2})$, where $n$ denotes the number of simulated trajectories. However, the variance of the underlying path-weighted estimator may increase with the state dimension $k$, especially in problems with strong coupling or highly concentrated importance weights. In such cases, achieving a fixed accuracy level may require the sample size $n$ to grow moderately with $k$. This reflects a trade-off between dimensionality and sampling effort while the path integral control removes the exponential $O(n^k)$ grid burden, high-dimensional problems may still necessitate larger Monte Carlo ensembles to maintain estimator stability. In the market-share diffusion setting considered here, the absence of explicit state-space discretization ensures that the computational burden grows primarily through sampling requirements rather than through exponential memory expansion, thereby preserving the practical scalability of the proposed framework.	
\end{remark}

\subsection{Equilibrium conditions.}	

Under Walrasian Equilibrium every firm maximizes its dynamic profit subject to the market dynamics defined in Equation (\ref{mkt}) and determines its optimal strategy such that the market clears. In other words, a \emph{Walrasian market clears} means total positive real-valued of measure of market shares of all firms equals the total endowment of the economy (i.e., the total available resource to all firms from the economy). Since all firms have identical profit functions and market dynamics, we choose a single firm's behavior to represent all other firms' behavior.

\begin{definition}\label{DWalrus}
	The continuous path of an optimal real-valued measure of market share  $x^*(s)$ and a continuous set of optimal strategies  $u^*(s)$ constitute a Walrasian Equilibrium $\{x^*(s),u^*(s)\}\in\mathcal X\times\mathcal U$  if each identical firm maximizes its own profit, and for every time point $s\in[0,t]$,
	\begin{equation}\label{w}
	\E_0 \left\{\int_{0}^t \pi[s,x^*(s),u^*(s)] ds\biggr|\mathcal F_0^{X}\right\}\geq\E_0 \left\{\int_0^t  \pi[s,X(s),u(s)] ds\biggr|\mathcal F_0^X\right\},
	\end{equation}
	with market dynamics defined in Equation (\ref{mkt}) such that total value of real-valued measure of market share equals to the total endowment of the industry.
\end{definition}

\begin{remark}\label{WalrasR}
	From Definition \ref{DWalrus} we understand that each identical firm maximizes its own profit subject to its own market dynamics and determines its own optimal strategy $\{u^*(s)\}_{s=0}^t$ as well as optimal real-valued measure of market share $\{x^*(s)\}_{s=0}^t$. Then global policy-maker (i.e., a Walrasian auctioneer) finds total optimal real-valued measure of market share and based on market conditions at time $s$ they decide industry price.	In other words, the search for a Walrasian equilibrium is similar to a maximization problem of a global planner.	
\end{remark}

Determining Pareto and Nash equilibria requires us to consider the other firms in the industry. Suppose, there are $k$-firms in an economy, where the strategy function of firm $\rho$ is given by $u_\rho(s)$ for $\rho=1,..., k$, $u_\rho\in \mathcal{U}_\rho\subset\mathcal{U}$, where $\mathcal{U}_\rho$ is the set of all available strategies of firm $\rho$, and $\mathcal{U}$ is the set of all possible strategies in the market. Let $X_\rho(s)\in\mathcal X_\rho\subset\mathcal X$ be the real-valued measure of market share for firm $\rho$, where $\mathcal X_\rho$ is firm $\rho$'s functional space of real-valued measure of market share. Let $\mathbf{X}(s)\in\mathcal X$ and $\mathbf{u}(s)\in\mathcal U$ be the vectors containing the elements $X_\rho(s)$ and $u_\rho(s)$ for $\rho=1,...,k$, respectively. Dynamic profit function of firm $\rho$ is expressed by $\pi_\rho[s,\mathbf{X}(s),\mathbf{u}(s)],$ 
with market dynamics specified by
\begin{equation}\label{vectordynamics}
d\mathbf{X}(s)=\bm{\mu}[s,\mathbf{X}(s),\mathbf{u}(s)]ds+\bm{\sigma}[s,\mathbf{X}(s),\mathbf{u}(s)]d\mathbf{B}(s),
\end{equation}
where $\bm{\mu}[s,\mathbf{X}(s),\mathbf{u}(s)]$ is an $k$-dimensional drift coefficient, $\bm{\sigma}[s,\mathbf{X}(s),\mathbf{u}(s)]$ is an $k\times m$-dimensional diffusion component, and $\mathbf{B}(s)$ is an $m$-dimensional Brownian motion. The initial condition is $\mathbf{x}(0)=\mathbf{x}_0\in\mathcal X$.

Under Pareto optimality each firm benefits at the expanse of others \citep{greenwald1986,mas1995}. Therefore, Pareto optimality insures mutual benefit for all of the firms simultaneously. Mathematically this is equivalent to maximizing the total dynamic profit,
\[
\overline{\Pi}_\text{P}(u,t)=
\E_0\left\{\int_0^t\sum_{\rho=1}^k\a_\rho\pi_\rho[s,\mathbf{X}(s),\mathbf{u}(s)]ds\biggr|\mathcal{F}_0^X\right\},
\]
where $\a_\rho$ is the profit weight corresponding to $\rho^{th}$ firm such that $\sum_{\rho=1}^k\a_\rho=1$. 

\begin{definition}\label{paretodef}
	For a strategy vector $\mathbf u(s)\in\mathcal U$ for $k$ firms, a set of optimal strategies $\mathbf{u}^*(s)$  is cooperative Pareto efficient $\{\mathbf{x}^*(s),\mathbf{u}^*(s)\}\in\mathcal X\times\mathcal U$ if there exists $(\a_\rho)_{1\leq\rho\leq k}$ with $\a_\rho\in[0,1]$ so that  
	\begin{equation}\label{p}
	\E_0\left\{ \int_{0}^t\ \sum_{\rho=1}^k\ \a_\rho\pi_\rho[s,\mathbf{x}^*(s),\mathbf{u}^*(s)]ds\biggr|\mathcal{F}_0^{X}\right\} \geq \E_0\left\{ \int_0^t \sum_{\rho=1}^k \a_\rho\pi_\rho[s,\mathbf{X}(s),\mathbf{u}(s)]ds\biggr|\mathcal{F}_0^X\right\},\ \text{for all $\mathbf u\in\mathcal U$} ,
	\end{equation}
	subject to the Equation (\ref{vectordynamics}) with initial condition $\mathbf{x}(0)=\mathbf{x}_0$, so that for at least one firm the inequality is strict and does not allow for any solution $\mathbf{u}(s)\in\mathcal U$, where $\mathcal{F}_0^{X}$ is the filtration corresponding to Pareto efficient strategies and $\a_\rho$ is the profit weight of $\rho^{th}$ firm such that $\sum_{\rho=1}^k \a_\rho=1.$
\end{definition}

\begin{remark}
	A cooperative Pareto solution is never dominated because there is always more than one Pareto solutions as dominance is a property which does not give total ordering in general sense \citep{engwerda2005lq}.		
\end{remark}

\begin{remark}
	Since a central planner chooses Pareto efficient strategies based on the welfare of the entire system, Definition \ref{paretodef} is weaker in the sense that we assume by cooperating, firms create some influences on central planner's decisions.
\end{remark}

\begin{remark}
	Under cooperation the profit of a firm is not uniquely determined. If all $k$ firms decide to use their strategies to increase the profit of firm $1$ as much as possible, a different maximum is attained for firm $1$ than in the case where all firms agree collectively to help a different firm to maximize its profit. Therefore, depending on how the firms choose to divide their strategies, a firm incurs different maxima which leads to a firm confront  with a whole set of possible outcomes from which somehow one is cooperatively selected. Let there be two strategies $u^1(s)\in\mathcal U^1\subset\mathcal U$ and $u^2(s)\in\mathcal U^2\subset\mathcal U$ such that every firm has higher profit if they choose $u^1(s)$. Then it is reasonable to assume that all the firms would choose strategy $u^1(s)$. Hence, the solution induced by strategy $u^1(s)$ dominates over the strategy $u^2(s)$ induced solution \citep{engwerda2005lq}. Furthermore, proceeding along this thought, it is reasonable to consider only those cooperative outcomes which have the property that if a different strategy than the one corresponding with this cooperative outcome is chosen, then at least one of  the firms has lower profit \citep{pramanik2025stubbornness,pramanik2025strategic}. In other words, only those solutions are concerned where no further Pareto improvement is possible. This leads us to Definition \ref{paretodef}. Since there are no central policy makers present in this environment, each firm has to consider other firms' profit to determine its Pareto optimal strategy.	
\end{remark}

Assuming $\pi_\rho[s,\mathbf{X}(s),\mathbf{u}(s)]$ is convex, non-negative and differentiable, the cooperative Pareto optimization problem of firm $\rho$ is
\begin{equation}\label{n6}
u_\rho^*=\max_{u_\rho\in \mathcal{U}^\rho\subset\mathcal U}\overline{\Pi}(\mathbf{u},t)=
\max_{u_\rho\in \mathcal U^\rho\subset\mathcal U}\E_0 \left\{\int_0^t\sum_{\rho=1}^k\a_\rho
\pi_\rho[s,\mathbf{X}(s),\mathbf{u}(s)]ds\biggr|\mathcal{F}_0^X\right\},
\end{equation}
subject to Equation (\ref{vectordynamics}), with initial condition $\mathbf{x}(0)=\mathbf{x}_0$. In other words, Equation (\ref{n6}) implies that $\rho^{th}$ firm performs its profit maximization in light of the optimal strategies of the others.

\begin{definition}
	For firm $\rho$, a set of optimal strategies $\mathbf{u}^*(s)=(u_\rho^*(s))_{\rho=1}^k\in\mathcal U$ constitutes a non-cooperative feedback Nash equilibrium $\{\mathbf{x}^*(s),\mathbf{u}^*(s)\}\in\mathcal X\times\mathcal U$, if there exists a set of optimal real-valued measure of market share $\mathbf{x}^*(s)=(x_\rho^*(s))_{\rho=1}^k\in \mathcal X$ such that
	\begin{align}
	\E_0\left\{\int_0^t\pi_\rho[s,\mathbf{x}^*(s),\mathbf{u}^*(s)]ds\biggr|\mathcal{F}_0^{X}\right\}\geq
	\E_0\left\{\int_0^t\pi_\rho[s,\hat{\mathbf{x}}_\rho^*(s),\hat{\mathbf{u}}_\rho^*(s)]ds\biggr|\mathcal{F}_0^x\right\},\notag
	\end{align}
	for all $\rho\in\{1,...,k\}$ where $t\in(0,\infty)$,
	subject to,
	\begin{align}\label{n11}
	d\hat{\mathbf{x}}_\rho^*(s)&=\mu[s,\hat{\mathbf{x}}_\rho^*(s),\hat{\mathbf{u}}_\rho^*(s)]ds+\sigma[s,\hat{\mathbf{x}}_\rho^*(s),\hat{\mathbf{u}}_\rho^*(s)]dB(s), 
	\end{align}
	with initial condition $\mathbf{x}(0)=\mathbf{x}_0$, where the strategy of $\rho^{th}$ firm $ u_\rho^*(s)$ is optimal if other firms choose their strategies according to  $\mathbf{u}^*(s)$ and the positive real-valued measure of market share of $\rho^{th}$ firm is optimal if other firms choose their positive real-valued measure of market shares according to $\mathbf{x}^*(s)$ such that
	\[\hat{\mathbf{x}}^*_\rho(s)=[x_1^*(s),...,x^*_{\rho-1}(s),x_\rho(s),x_{\rho+1}^*(s),\ldots,x_k^*(s)]',\] 
	and 
	\[\hat{\mathbf{u}}^*_\rho(s)=[u_1^*(s),...,u^*_{\rho-1}(s),u_\rho(s),u_{\rho+1}^*(s),\ldots,u_k^*(s)]',\]
	where $'$ represents the transposition of a matrix.
	\label{def3}
\end{definition}

In Definition \ref{def3}, $\hat{\mathbf{x}}^*_\rho(s)$ is the dynamics of positive real-valued measure of market share when other firms follow the Nash strategy and $\hat{\mathbf{u}}^*_\rho(s)$ represents all the other firms but firm $\rho$ has its Nash strategies. Hence, firm $\rho$ has the optimization problem 
\begin{equation}\label{n9}
u_\rho^*=\max_{u_\rho\in \mathcal U^\rho\subset\mathcal U}\widetilde{\Pi}(u_\rho,t)=
\max_{u_\rho\in \mathcal U^\rho\subset\mathcal U}\E_0\left\{\int_0^t\pi_\rho[s,\mathbf{X}(s),\hat{\mathbf{u}}_\rho(s)]ds\biggr|\mathcal{F}_0^X\right\},
\end{equation}
subject to the dynamics expressed in Equation (\ref{n11}) and initial conditions $\mathbf{x}(0)=\mathbf{x}_0$, where \[\hat{\mathbf{u}}_\rho(s)=[u_1(s),...,u_{\rho-1}(s),u_\rho(s),u_{\rho+1}(s),\ldots,u_k(s)]'.\]

Now we will show the existence of Nash Equilibrium under this construction. For $\rho=1,...,k$ and filtration $\mathcal{F}_0^X$ on interval $[0,t]$, let $\mathcal{U}_1,...,\mathcal{U}_k$ be compact convex strategy sets such that $\mathcal{U}=\prod_{\rho=1}^{k}\mathcal{U}_\rho$. Define $\mathcal{U}^\rho:=\prod_{\omega\neq\rho}\mathcal{U}_\omega$, such that the projections are $\phi_\rho:\mathcal{U}\ra\mathcal{U}_\rho$ and $\phi^\rho:\mathcal{U}\ra\mathcal{U}^\rho$. Suppose, $\hat{\mathbf{u}}_{-\rho}$ be the strategies of all the firms other than firm $\rho$. Then for all strategies $\{\mathbf{u}_1,\mathbf{u}_{2}\}\in\mathcal{U}^1\times\mathcal U^2\subset\mathcal U$, $\phi^\rho(u_1)=\hat{\mathbf{u}}_{-\rho}$, $\phi^\rho(\mathbf{u}_2)=u_\rho$, we can say
\[ \{u_\rho,\hat{\mathbf{u}}_{-\rho}\}=\{u_1,...,u_{\rho-1},u_\rho,u_{\rho+1},...,u_k\}\in\mathcal{U}=\mathcal{U}_\rho\times\mathcal{U}^\rho.\] 
It is important to note that, for any given map $M:\mathcal{U}^\rho\ra 2^{\mathcal{U}_\rho}$, its graph $\mathcal{G}$ is contained in $\mathcal{U}$. For each firm $\rho$, a set of real-valued measure of market share $\mathbf{X}\in\mathcal X$ and its strategy $u_\rho\in\mathcal{U}^\rho$ on $[0,t]$ the conditional expectation $\E_0[\pi_\rho|\mathcal{F}_0^X]:[0,t]\times\mathcal X\times\mathcal{U}\ra\mathbb R$. An optimal strategy set $\{u_\rho^*,\hat{\mathbf{u}}_{-\rho}^*\}$ is a feed back Nash equilibrium for the system $\{\E_0\int_0^t[\pi_1ds|\mathcal{F}_0^X],...,\E_0\int_0^t\left[\pi_kds|\mathcal{F}_0^X\right]\}$ provided for each $\rho=1,..,k$ we have
\[
\E_0\left\{\int_0^t\pi_\rho\left[s,\mathbf{x}^*,{u}_{\rho^*},\hat{\mathbf{u}}_{-\rho}^*\right]ds\biggr|\mathcal{F}_0^X\right\}=\max_{u_\rho\in \mathcal U^\rho\subset\mathcal U}\E_0\left\{\int_0^t\pi_\rho\left[s,\mathbf{X}(s),\hat{\mathbf{u}}_\rho(s)\right]ds\biggr|\mathcal{F}_0^X\right\}.
\]
Define $J\left(s,\mathbf{x}^*,u_\rho^*,\hat{\mathbf{u}}_{-\rho}^*\right):=\E_0\left\{\int_0^t\pi_\rho\left[s,\mathbf{x}^*,u_\rho^*,\hat{\mathbf{u}}_{-\rho}^*\right]ds\biggr|\mathcal{F}_0^X\right\}$ for all $\mathbf{x}^*\in\mathcal{X}\in\mathbb{R}^k$.

\begin{lem}\label{l0}
	Let $\mathcal{U}_1,...,\mathcal{U}_k$ be compact convex strategy sets such that $\mathcal{U}=\prod_{\rho=1}^k\mathcal{U}_\rho$, and for vector of optimal measure of market shares $\mathbf{x}^*\in\mathcal{X}\in\mathbb{R}^k$. For the filtration $\mathcal{F}_0^{X}$ on $[0,t]$ consider there exists a set valued map $M_\rho:\mathcal{U}^\rho\ra2^{\mathcal{U}_\rho}$	for all $\rho=1,...,k$ (equivalently $M_\rho^{-1}:\mathcal{U}_\rho\ra2^{\mathcal{U}^\rho}$). Then following conditions are equivalent:
	\begin{itemize}
		\item $\bigcap_{\rho=1}^k\mathcal{G}_\rho$ is non-empty.
		\item There exists an optimal strategy set $\left\{u_\rho^*,\hat{\mathbf{u}}_{-\rho}^*\right\}\in\mathcal{U}$ such that for all possible strategies $\left\{u_\rho,\mathbf{u}_{-\rho}\right\}\in M_\rho(\hat{\mathbf{u}}_\rho)$ for all $\rho=1,...,k$; $\mathbf{x}^*\in\mathcal{X}$ and filtration $\mathcal{F}_0^{X}$ on $[0,t]$, where $M_\rho(\hat{\mathbf{u}}_\rho)=\bigcup_{\rho=1}^k \left\{M_{\left\{u_\rho,\mathbf{u}_{-\rho}\right\}}\bigg|\left\{u_\rho,\mathbf{u}_{-\rho}\right\}\in\hat{\mathbf{u}}_\rho\right\}$.
	\end{itemize}
\end{lem}	

\begin{remark}
	Lemma~\ref{l0} provides an equilibrium–type consistency condition for the multi-firm control problem by linking the non-emptiness of the intersection $\bigcap_{\rho=1}^k\mathcal{G}_\rho$ to the existence of a joint optimal strategy profile in the product space $\mathcal U$. Economically, the result ensures that when each firm’s feasible-response correspondence $M_\rho$ is defined on compact and convex strategy sets, there exists a mutually compatible collection of strategies that supports the optimal market-share vector $\mathbf{x}^*$ under the available information filtration. Consequently, the lemma establishes the structural foundation required for the subsequent path-integral construction by guaranteeing that the stochastic control problem admits a well-defined and jointly admissible strategy configuration.
\end{remark}

\begin{lem}\label{l0.1}
	Let $\mathcal{U}_1,...,\mathcal{U}_k$ be compact convex strategy sets such that $\mathcal{U}=\prod_{\rho=1}^k\mathcal{U}_\rho$ and for all $\mathbf{x}^*\in\mathcal{X}\in\mathbb{R}^k$ on time interval $[0,t]$, let $l_1,...,l_k:[0,t]\times\mathcal{X}\times\mathcal{U}\ra\mathbb{R}$ satisfies following conditions:\\
	$(a)$ $\mathbf{u}_{-\rho}\mapsto l_\rho\left(s,\mathbf{x}^*,u_\rho,\hat{\mathbf{u}}_{-\rho}\right)$ is lower semi-continuous on $\mathcal{U}^\rho$ for all $u_\rho\in\mathcal{U}_\rho$ and $s\in[0,t]$.\\
	$(b)$ ${u}_\rho\mapsto l_\rho\left(s,\mathbf{x}^*,u_\rho,\hat{\mathbf{u}}_{-\rho}\right)$ is quasi-concave on $\mathcal{U}^\rho$ for all $\hat{\mathbf{u}}_{-\rho}\in\mathcal{U}^\rho$ and $s\in[0,t]$.\\
	$(c)$ For all $\rho=1,...,k$, the filtration process $\mathcal{F}_0^{X}$ on $[0,t]$, and $\hat{\mathbf{u}}_{-\rho}\in\mathcal{U}^\rho$ there $\exists\ u_\rho\in\mathcal{U}^\rho$ such that $l_\rho\left(s,\mathbf{x}^*,u_\rho,\hat{\mathbf{u}}_{-\rho}\right)>0$.
	Then there exists an optimal set of strategies $\left\{u_\rho^*,\hat{\mathbf{u}}_{-\rho}^*\right\}\in \mathcal{U}$ such that $l_\rho\left(s,\mathbf{x}^*,u_\rho^*,\hat{\mathbf{u}}_{-\rho}^*\right)>0,\ \forall s\in[0,t]$.
\end{lem}	

\begin{remark}
	Lemma~\ref{l0.1} establishes sufficient regularity conditions under which a jointly admissible and payoff-improving strategy profile exists in the multi-firm environment. The lower semicontinuity and quasi-concavity assumptions ensure well-behaved best-response correspondences on the compact convex strategy sets, thereby supporting existence of an optimal configuration. Condition (c) guarantees that each firm retains a strictly profitable deviation within the admissible filtration framework. Together, these properties provide the equilibrium foundation required for the subsequent stochastic and path-integral analysis.
\end{remark}

\begin{prop}[Existence of Nash Equilibrium]\label{existence}
	Let $\mathcal{U}_1,...,\mathcal{U}_k$ be compact convex strategy sets such that $\mathcal{U}=\prod_{\rho=1}^k\mathcal{U}_\rho$ and let $J_1,J_2,...,J_k:[0,t]\times\mathcal X\times\mathcal{U}\ra\mathbb{R}$ be continuous for a filtration $\mathcal{F}_0^X$ . If the mapping of $\rho^{th}$ firm's strategy $u_\rho\mapsto J_\rho(s,\mathbf{x}^*,u_\rho,\hat{\mathbf{u}}_{-\rho})$ is quasi-concave on $\mathcal{U}$ for all $\hat{\mathbf{u}}_\rho\in\mathcal{U}$ and $\forall\ \rho=1,...,k$, then the system $\{J_1,...,J_k\}$ has an admissible feedback Nash Equilibrium.	
\end{prop}	

\begin{remark}
	Proposition~\ref{existence} guarantees the existence of an admissible dynamic feedback Nash equilibrium under standard regularity conditions on the firms’ objective functionals. The compactness and convexity of the strategy sets, together with quasi-concavity of each firm’s payoff in its own control, ensure that best-response correspondences are nonempty and well behaved over the filtration $\mathcal{F}_0^X$. In the dynamic setting, this implies the existence of a time-consistent feedback profile in which no firm can profitably deviate given the rivals’ strategies and the current state $\mathbf{x}^*$. Consequently, the stochastic differential game admits a stable Markovian equilibrium foundation for the subsequent path-integral formulation.
\end{remark}

	\subsection{The Feynman-type path integral.}
	
	We now turn to the discussion of the Feynman-type path integral approach that forms the foundation of our analytical framework. At its core, a path is defined as a continuous mapping from a given time interval into a space of continuous functions, characterized by both an initial point and a terminal point, thereby representing a complete trajectory of the system under consideration. Within this general notion, one can distinguish between two types of path integrals. The first is the line path integral, which evaluates an integral along a specific trajectory between a fixed starting point and an ending point, typically within a finite-dimensional space. The second, and more general, is the functional path integral, where the domain of integration is itself a space of functions, meaning that the integration is carried out over an entire collection of possible paths rather than along a single trajectory. This distinction is crucial for applications involving stochastic dynamics, as it allows one to account for the full range of possible evolutions of the system rather than focusing on a single deterministic path. In this paper, we restrict attention exclusively to the functional path integral formulation originally introduced by Richard Feynman, commonly referred to as the Feynman path integral \citep{feynman1948}. This formulation provides a powerful framework for analyzing systems in which uncertainty plays a central role, as it aggregates contributions from all admissible trajectories and thereby captures the probabilistic structure of the underlying dynamics in a unified and tractable manner.
	
	\begin{definition}\label{d0.1}
		Let $\mathfrak L[s,p(s),\dot p(s)]=(1/2) m\dot p(s)^2-V(p)$ be a classical Lagrangian of a particle in generalized coordinate $p$ with mass $m$, where $(1/2) m\dot p^2$ and $V(p)$ are kinetic and potential energies respectively and, $\dot p(s)=\partial p/\partial s$. Then the classical action function is $\mathcal S=\int_0^T \mathfrak L(s,p(s),\dot p(s)) ds$ and the transition function of Feynman path integral is defined as $\Psi(p)=\int_{\mathbb R} \exp\{{\mathcal S}\} \mathcal{D}_P $, with $\mathcal{D}_P$ being an approximated Riemann measure and it represents the positions of a particle at different time points \citep{belal2007,pramanik2020}.
	\end{definition}
	
	Using Definition \ref{d0.1} and for a non-zero, Lagrangian multiplier $\lambda(s)$, we construct a stochastic Lagrangian action function from Equations (\ref{n1}) and (\ref{vectordynamics}) starting at time $s$  as 
	\begin{equation*}
	\mathcal L=\E_s\left\{\pi[s,X(s),u(s)]ds+\lambda(s) \left[\Delta \mathbf{X}(s)ds-\bm{\mu}[s,\mathbf{X}(s),\mathbf{u}(s)]ds-\bm{\sigma}[s,\mathbf{X}(s),\mathbf{u}(s)]d\mathbf{B}(s)\right]\right\},
	\end{equation*}
	with the transition function for small time interval $[s,s+\epsilon]$ for all $\epsilon\downarrow 0$
	\begin{equation}\label{M}
	\Psi_{s,s+\varepsilon}(\mathbf{X})=\frac{1}{L_\varepsilon} \int_{\Omega}\exp[-\varepsilon \mathcal{A}_{s,s+\varepsilon}(\mathbf{X})]\Psi_s(\mathbf{X})d\mathbf{X}(s),
	\end{equation}
	where $\Psi_s(\mathbf{X})$ is the value of the transition probability at time $s$ with the initial condition $\Psi_0(\mathbf{X})=\Psi_0$, $L_\varepsilon$ is a positive penalization constant, and the Feynman action function is defined as 
	\begin{equation*}
	\mathcal{A}_{s,s+\varepsilon}(\mathbf{X}):=
	\int_{s}^{s+\varepsilon}\E_\nu\left\{\pi[\nu,X(\nu),u(\nu)]d\nu+g[\nu+\Delta \nu,\mathbf{X}(\nu)+\Delta \mathbf{X}(\nu)]\right\},
	\end{equation*}
	where $g[\nu+\Delta \nu,\mathbf{X}(\nu)+\Delta \mathbf{X}(\nu)]\in C^{1,2}([0,t],\mathcal X)$ such that,	
	\begin{equation*}
	g[\nu+\Delta \nu,\mathbf{X}(\nu)+\Delta \mathbf{X}(\nu)]\approx \lambda(\nu) \left[\Delta \mathbf{X}(\nu)d\nu-\bm{\mu}[\nu,\mathbf{X}(\nu),\mathbf{u}(\nu)]d\nu-\bm{\sigma}[\nu,\mathbf{X}(\nu),\mathbf{u}(\nu)]d\mathbf{B}(\nu)\right].
	\end{equation*}

	\begin{remark}
		It is important to note that $g$ is not a function of $\mathbf{u}$.	This $g$ function is analogous to an integrating factor of partial differential equation. In other words, the integrating factor of the SDE (\ref{vectordynamics}) is chosen in such a way that the Lagrangian $\mathcal L$ exhibits a local solution \citep{pramanik2025construction}. Furthermore, we assume feedback system, then value of $\mathbf{X}$ is determined by $\mathbf{u}$.	
	\end{remark}

	\begin{as}\label{as0}
		For $t>0$, let ${\bm{\mu}}(s,\mathbf{u},\mathbf{X}):[0,t]\times \mathcal{U}\times \mathcal X \ra\mathbb{R}^k$ and $\bm{\sigma}(s,\mathbf{u},\mathbf{X}):[0,t]\times \mathcal{U}\times \mathcal X \ra\mathbb{R}^{k\times m}$ be some measurable function and, for a finite constant $K_1>0$, $\mathbf{u}\in\mathcal U\subset\mathbb{R}^k$, and $\mathbf{X}\in\mathcal X\subset\mathbb{R}^k$ we have linear growth as
		\[
		|{\bm{\mu}}(s,\mathbf{u},\mathbf{X})|+
		|\bm{\sigma}(s,\mathbf{u},\mathbf{X})|\leq 
		K_1(1+|\mathbf{X}|),
		\]
		such that, there exists another positive, finite, constant $K_2$ and for a different real-valued measure of market share 
		$\tilde{\mathbf{X}}_{k\times 1}$ such that the Lipschitz condition,
		\[
		|{\bm{\mu}}(s,\mathbf{u},\mathbf{X})-
		{\bm{\mu}}(s,\mathbf{u},\widetilde{\mathbf{X}})|+|\bm{\sigma}(s,
		\mathbf{u},\mathbf{X})-\bm{\sigma}(s,\mathbf{u},\widetilde{\mathbf{X}})|
		\leq K_2\ |\mathbf{X}-\widetilde{\mathbf{X}}|,\notag
		\]
		$ \widetilde{\mathbf{X}}\in\mathbb{R}^k$ is satisfied and
		\[
		|{\bm{\mu}}(s,\mathbf{u},\mathbf{X})|^2+
		\|\bm{\sigma}(s,\mathbf{u},\mathbf{X})\|^2\leq (K_2)^2
		(1+|\widetilde{\mathbf{X}}|^2),
		\]
		where 
		$\|\bm{\sigma}(s,\mathbf{u},\mathbf{X})\|^2=
		\sum_{\rho=1}^k \sum_{j=1}^m\left|{\sigma^{ij}}(s,\mathbf{u},\mathbf{X})\right|^2$.
	\end{as}

\begin{as}\label{as0.1}
	There exists a filtration probability space $(\Omega,\mathcal F,\mathcal{F}_s^{\mathbf X},\mathcal{P})$ with sample space $\Omega$, Borel $\sigma$-algebra $\mathcal F$, filtration at time $s$ of the positive real-valued measure of market share ${\mathbf{X}}$ as $\{\mathcal{F}_s^{\mathbf{X}}\}\subset\mathcal{F}_s$, a canonical Wiener measure $\mathcal{P}$ and a $m$-dimensional $\{\mathcal{F}_s\}$-adapted Brownian motion $\mathbf{B}$ where all available strategies $\mathbf{u}$ are $\{\mathcal{F}_s^{\mathbf{X}}\}$ adapted process such that Assumption \ref{as0} holds, for the feedback control measure of players there exists a measurable function $h$ such that $h:[0,t]\times C([0,t]):\mathbb{R}^k\ra\mathcal{U}$ for which $\mathbf{u}(s)=h[\mathbf{X}(s,u)]$ such that Equation (\ref{vectordynamics})	has a strong unique solution \citep{ross2008}.
\end{as}

\begin{definition}
	Player $\rho$'s strategy $u_\rho$ on time interval $[s,\tau]$ is an admissible process taking values in $\mathcal U$ and is progressively measurable with respect to the filtration $(\mathcal{F}_s^X,s_0\geq s)$, where
	\[
	\mathcal{F}_s^X=\sigma\left\{\mathbf{B}(s_1)-\mathbf{B}(s),s_1\in[s,s_0]\right\},s_0\in[s,\tau],
	\]
	augmented by all null sets in the canonical Wiener measure $\mathcal{P}$, where $\tau=s+\epsilon$ with $\epsilon\downarrow 0$ \citep{buckdahn2004}.
\end{definition}

\begin{definition}\label{de0}
	Suppose $\mathbf{X}(s,\mathbf u)$ is a non-homogeneous Fellerian semigroup in $\mathbb{R}^k$. The infinitesimal generator $A$ of $\mathbf{X}(s,\mathbf u)$ is defined by,
	\[
	Ah(X)=\lim_{\epsilon\ra 0}\frac{\E_s[h(\mathbf{X}(s,\mathbf u))]-h(X(\mathbf u))}{\epsilon},
	\]
	for $x\in\mathcal X$	where $h:\mathbb{R}^k\ra\mathbb{R}$ is a $C_0^2(\mathbb{R}^k)$ function, $\mathbf{X}$ has a compact support, and at $X(\mathbf u)$ the limit exists where $\E_s$ represents  conditional expectation of real-valued measure of all possible market shares $\mathbf{X}$ at time $s$. Furthermore, if the above Feller semigroup is homogeneous on $[0,t]$, then $Ah$ is exactly equal to the Laplace operator.
\end{definition}

The proposed path integral algorithm involves several structural and numerical parameters, including the time discretization step, diffusion scale, penalization constant, and the number of Monte Carlo trajectories. For the simulation-based implementation to remain well-posed and numerically stable, these quantities cannot be chosen arbitrarily. First, the drift $\mu(s,X,u)$ and diffusion $\sigma(s,X,u)$ must satisfy the linear growth and Lipschitz conditions stated in Assumption~\ref{as0} to guarantee the existence of a unique strong solution and to ensure stability of the time-discretized trajectories. Second, the time step used in the discretization of $[0,t]$ must be sufficiently small to control the \emph{Euler-Maruyama} approximation error and to preserve the non-negativity of the market-share process when required by the economic interpretation. Third, the penalization constant $L_\varepsilon$ and the parameter $\omega$ must be selected so that the exponential path weights remain numerically well-conditioned; excessively small values may lead to weight degeneracy, while excessively large values may reduce the effective selectivity of the importance sampling scheme. Finally, the Monte Carlo sample size $n$ must be chosen to balance statistical accuracy and computational cost, noting that the root mean square error decreases at rate $O(n^{-1/2})$.

Traditionally, these optimization problems are solved by using the Pontryagin principle \cite{pontryagin1987} after solving the Hamilton-Jacobi-Bellman equation. See \cite{bellman1952,bellman2013,bellman2015,ljungqvist2012,pontryagin1966,stokey1989} and \cite{yeung2006}. The main problem with this method is that finding a solution often requires obtaining a complicated value function.
An alternative method for solving optimal control problems is based on principles from quantum mechanics and path integrals. These methods have previously been used in motor control theory  \citep{kappen2005,theodorou2010,theodorou2011,pramanik2020m,pramanik2020,pramanik2019}, and finance \citep{belal2007}. There are three mathematical representations of this approach based on partial differential equations, path integrals, and stochastic differential equations  \citep{theodorou2011}. Partial differential equations give a macroscopic view of an underlying physical process, while path integrals and stochastic differential equations give a more microscopic view \citep{pramanik2023scoring}. Furthermore, the Feynman-Kac formula yields a special set of HJB equations which are backward parabolic partial differential equations \citep{kac1949}. Only a few problems in finance are directly tractable by Pontryagin maximum principle and solving the HJB equation usually involves solving a  system of differential equations which is often a difficult task \citep{pramanik2023optimization,pramanik2023omal}.  The potential advantage of the quantum approach is that a general non-linear system, such as Merton-Garman Hamiltonian, can be impossible to solve analytically (see page 17 for details). The quantum method allows a different approach to attack these problems and sometimes can give simplified solutions \citep{belal2007}. Path integrals are widely used in physics  as a method of studying stochastic systems. In finance, path integrals have been used to study the theory of options and interest rates \citep{linetsky1997,lyasoff2004}. A rigorous discussion of the application of different types of quantum path integrals in finance is given in \cite{belal2007}. The idea is that, in quantum mechanics a particle's evolution is random. This is analogous to the evolution of a stock price having non-zero volatility.

Motivated by \cite{belal2007} we consider a firm's real-valued measure of market share  as a stochastic process and use the principles of path integral as the basis for our mathematical model. The assumption is that since a firm is a very small part of an industry and  an economy, and is subject to many small stochastic perturbations, the movement of its share  can be approximated like a quantum particle in physics. 
Although these methods have been used in quantum approaches to financial problems we are not  aware of their use in stochastic optimization problems for the economic systems studied here.

\medskip

\section{The Pontryagin maximum principle through HJB.}

Traditionally, the optimization problems are solved by using the Pontryagin principle \cite{pontryagin1987} after solving the HJB equation (we will call this the P-method). See \cite{bellman1952,bellman2013,bellman2015,ljungqvist2012,pontryagin1966,stokey1989} and \cite{yeung2006}.  Following \cite{yeung2006}, the optimal strategies under Walrasian equilibrium, Pareto optimality and Nash equilibrium are given in Lemmas \ref{prop1}-\ref{prop3}.

\begin{lem}\label{prop1}
	(Walrasian Optimal Strategy \citep{yeung2006}) The strategy $u^*(s)=\phi^*(s,X)$ constitutes a firm's optimal solution to the problem (\ref{n1}) for some $C^{1,2}$ function $V(s,X):[0,t]\times\mathcal X\ra\mathbb{R}$ satisfying the following partial differential equation
	\begin{align}\label{n12}
	&- V_s(s,X)=\arg\max_{u\in \mathcal U}\ \biggr\{\pi(s,X,u)+V_X(s,X)\ \mu(s,X,u)+\mbox{$\frac{1}{2}$}\ \sigma^2(s,X,u)\ V_{XX}(s,X)\biggr\}, \forall (s,X)\in[0,t]\times\mathcal X,
	\end{align}
	where $V_s(s,X)=\partial V(s,X)/\partial s$, $V_X(s,X)=\partial V(s,X)/\partial X$ and, $V_{XX}(s,X)=\partial^2 V(s,X)/\partial X^2$.
\end{lem}

Clearly, the value function $V(s,X)$ takes an important role in the P-method. Furthermore, we need $V(s,X)$ to be $C^{1,2}$. According to \cite{ljungqvist2012} there are three main ways to obtain value function in P-method: value function iteration, guess the value function and verify, and Howard's improvement algorithm. Among these three, the guess the value function and verify method is the easiest, depending on the uniqueness of the system and the luck of making good guess. Therefore, difficulties arise in finding out the value function and  its derivatives with respect to state.

Before giving the optimal strategies under Pareto and Nash equilibria we need further assumptions. In our system the diffusion matrix $\sigma$ maps $\mathbb{R}^{2k\times m}$ to $\mathbb{R}^{k\times m}$. Let us assume the local covariance matrix  of the vector Markov Process \citep{fleming1969} is $\theta=\sigma[s,X(s),u(s)]\ \sigma'[s,X(s),u(s)]$ where $\theta$ is a matrix that has $k$ rows and $k$ columns with its $(\rho,j)^{th}$ element is represented as $\theta^{\rho j}$. Furthermore, $\E[dB(s)]=0,\ \E[dB(s)ds]=0$, and $\E[dB(s)'\ dB(s)]=ds$ \citep{oksendal2003}.

\begin{lem}\label{prop2}
	(Cooperative Pareto Optimal Strategy \citep{yeung2006,leitmann2013}) A set of strategies of $k$ firms $\big\{u_1^*(s)=\phi_1^*(s,x),...,u_k^*(s)=\phi_k^*(s,X),\ \text{for all},\ s\in[0,t]\big\}$ provides an optimal solution to the problem 
	\[
	\max_{u_\rho\in\mathcal U^\rho\subset\mathcal U}	\E_0\left\{\int_0^t\sum_{\rho=1}^k\a_\rho\pi_\rho[s,\mathbf{X}(s),\mathbf{u}(s)]ds\biggr|\mathcal{F}_0^X\right\},
	\]
	if there exists continuously differentiable function $V(s,X):[0,t]\times \mathcal X\ra\mathbb{R}$ satisfying the following partial differential equation:
	\begin{align}\label{n13}
	-V_s(s,X)-\mbox{$\frac{1}{2}$}\ \sum_{\rho=1}^k\sum_{j=1}^m\ \theta^{\rho\tilde\rho}\  V_{X_\rho X_{j}}(s,\mathbf X)=\arg\max_{u_\rho\in \mathcal U^\rho\subset\mathcal U}\ \left\{\sum_{\rho=1}^k\ \a_\rho\pi_\rho(s,\mathbf x,\mathbf u) +V_X\ \mu(s,\mathbf X,\mathbf u) \right\},
	\end{align}
	where, $V_s(s,\mathbf X)=\partial V(s,\mathbf X)/\partial s$, $V_{X_\rho X_{j}}(s,\mathbf X)=\partial^2V(s,\mathbf X)/(\partial X_\rho\partial X_{j})$ and, $V_X(s,\mathbf X)=\partial V(s,\mathbf X)/\partial x_\rho$.
\end{lem}

\begin{lem}\label{prop3}
	(Non-Cooperative Nash Equilibrium \citep{yeung2006}) A $k$-firm of optimal strategies 
	\[
	u_\rho^*(s)=\phi_\rho^*(s,X)\in \mathcal U^\rho\in \mathcal U,\ \text{for all}\  \rho\in\{1,...,k\}
	\]
	provides a non-cooperative Nash equilibrium solution to the game (\ref{n9}) if there exists suitably smooth functions of $\rho^{th}$ firm $V^\rho:[0,t]\times\mathcal X\ra\mathbb{R}$, for all $\rho\in\{1,...,k\}$, satisfying the semilinear parabolic partial differential equations:
	\begin{align}\label{n14}
	&-V_s^\rho(s,X)-\mbox{$\frac{1}{2}$} \sum_{\rho=1}^k\sum_{j=1}^m\ \theta^{\rho j}\  V_{X_\rho X_{\tilde\rho}}^\rho(s,X)\notag\\
	&\hspace{1cm}=\arg\max_{u_\rho\in \mathcal U^\rho\subset\mathcal U}\ \biggr\{\pi_\rho\left[s,X_\rho,\phi_1^*(s,X_1),...,\phi_{\rho-1}^*(s,X_{\rho-1}),u_\rho(s),\phi_{\rho+1}^*(s,X_{\rho+1}),...,\phi_k^*(s,X_k)\right]\notag\\
	&\hspace{1.5cm} + V_X^\rho(s,X)\ \mu\left[s,X,\phi_1^*(s,X_1),...,\phi_{\rho-1}^*(s,X_{\rho-1}),u_\rho(s),\phi_{\rho+1}^*(s,X_{\rho+1}),...,\phi_k^*(s,X_k)\right]\biggr\}\notag\\
	&\hspace{1cm}=\pi_\rho\left[s,X_\rho,\phi_1^*(s,X),...,\phi_n^*(s,X)\right]+ V_X^\rho(s,X)\ \mu\left[s,X,\phi_1^*(s,X_1),...,\phi_k^*(s,X_k)\right].
	\end{align}
\end{lem}

\begin{remark}
	Using the Pontryagin principle to find out a value function in order to get the appropriate strategy (control) fails to exploit the first order condition of the real valued measure of market share (state variable) at time $s$ which leads to more difficult problem to solve \citep{chow1996}.
\end{remark}

\section{A Comparison between HJB and Path integral approaches.}

This section develops an explicit connection between the HJB equation and the path integral formulation in the context of a Walrasian firm by means of a tractable illustrative example. The goal is to show how the two approaches traditionally viewed as distinct can be reconciled within a common analytical framework when the underlying dynamics are governed by linear stochastic differential equations. In this setting, the firm's objective is specified through a profit function with a quadratic structure, which plays a crucial role in simplifying the analysis. The use of a quadratic profit function is not merely a modeling convenience but is deliberately chosen to preserve analytical tractability within the path integral control framework. In particular, this functional form allows the associated HJB equation to be transformed into a linearly solvable representation, thereby avoiding the nonlinearities that typically complicate dynamic programming approaches. This transformation leads to a formulation in which the weights assigned to different trajectories follow a Gaussian structure, a key feature that enables the application of path integral techniques. As a result, the optimal control problem can be interpreted in terms of weighted averages over stochastic paths, rather than requiring the direct solution of a nonlinear partial differential equation. By working through this example, the section highlights how the path integral approach provides an alternative yet equivalent characterization of optimal behavior, while also offering computational advantages in settings where traditional HJB methods become difficult to implement.

From an economic modeling perspective, the quadratic specification can be viewed as a second-order local approximation of a more general nonlinear profit mechanism evaluated near an operating equilibrium. Such approximations are widely employed in dynamic industrial organization and stochastic control when firms operate in regimes where marginal responses are sufficiently smooth. Under this interpretation, the model captures the dominant curvature effects of profit with respect to market share and advertising effort. If the underlying market environment departs substantially from quadratic behavior- for instance, due to saturation phenomena, threshold effects in advertising response, or asymmetric adjustment frictions- the resulting optimal policy may be subject to approximation error. That said, the path integral control itself is not intrinsically confined to quadratic profit forms. More general nonlinear profit specifications can, in principle, be incorporated via importance sampling or alternative path-weight constructions, although doing so typically sacrifices the exact Gaussian structure and may increase Monte Carlo variability.

The extension of this connection to the case of a nonlinear SDE, specifically how the HJB equation is intricately linked to the Feynman path integral, remains an unresolved and open question \citep{fujiwara2017}. Furthermore, when the state variable has very high dimension, the constructing the HJB empirically is extremely difficult \citep{yang2014path}. Feynman approach bypasses these difficulties and gives an alternative set of solutions \citep{baaquie1997,theodorou2011,yang2014path}. Here we are assuming profit function of that firm is quadratic with respect to the strategy $u(s)$ and	the drift coefficient of its market dynamics is additively separable with respect to $u(s)$. Based on the above two assumptions the stochastic control problem reduces to a computational path integral and because of its statistical mechanical form it can be called as free energy \citep{kappen2007b}.

Let the objective of a Walrasian firm be 
\begin{equation}\label{r1}
\overline{\Pi}_w(\mathbf{X},t)=
\max_{u\in \mathcal U}\E_0 \left\{\int_0^t
\pi[s,{X}(s),{u}(s)]ds\biggr|\mathcal{F}_0^X\right\},
\end{equation}
with
\[
\pi[s,{X}(s),{u}(s)]=\exp(-\zeta s)R[s,{X}(s)]-\frac{c}{2}u^2(s),
\]
$R[s,X(s)]$ is the dynamic revenue function with discount rate $\zeta\in[0,1]$, $c$ is a positive, constant marginal cost and $u(s)$ represents some expenditure on advertisement at time $s\in[0,t]$. This firm also faces a stochastic market dynamics 
\begin{equation}\label{r2}
dX(s)=[\mu_1[s,X(s)]+u(s)]ds+\sigma dB(s),
\end{equation}
where without loss of generality we assume the diffusion coefficient $\sigma>0$ is a constant. Comparison of the drift coefficients between Equations (\ref{mkt}) and (\ref{r2}) yields $\mu[s,X(s),u(s)]=[\mu_1[s,X(s)]+u(s)]$, where $\mu_1[s,X(s)]$ is an arbitrary function representing the trend of real-valued measure of market share. In this example we restrict this problem as a simple-linear-quadratic form.	A Walrasian firm's objective is to maximize Equation (\ref{r1}) subject to its market dynamics represented by Equation (\ref{r2}). After setting $\tau=s+\varepsilon$ for all $\varepsilon>0$ such that $\varepsilon\downarrow 0$ a Taylor series expansion can be performed on $\overline{\Pi}_w(\mathbf{u},\tau)$ around $s$ with first order with respect to time and second order with respect to the firm's real-valued measure of market share $x$. The stochastic HJB equation of this system is
\begin{equation}\label{r3}
-\frac{\partial}{\partial s}\overline{\Pi}_w(\mathbf{X},s)=\max_{u\in\mathcal U}\biggr\{\left[\exp(-\zeta s)R[s,{X}(s)]-\frac{c}{2}u^2(s)\right]+\mu_1(s,X)\frac{\partial}{\partial X}\overline{\Pi}_w(\mathbf{X},s)+u(s)\frac{\partial}{\partial X}\overline{\Pi}_w(\mathbf{X},s)+\mbox{$\frac{1}{2}$}\sigma^2\frac{\partial^2}{\partial X^2}\overline{\Pi}_w(\mathbf{X},s)\biggr\}.
\end{equation}
After solving for the right hand side of the HJB Equation (\ref{r3}) yields the optimal strategy of the firm as 
\[
u^*(s)=\left(\frac{1}{c}\right)\frac{\partial}{\partial x}\overline{\Pi}_w(\mathbf{X},s).
\]
Plugging in $u^*$ in the Equation (\ref{r3}) yields,
\begin{equation}\label{r3.0}
-\frac{\partial}{\partial s}\overline{\Pi}_w(\mathbf{X},s)=\exp(-\zeta s)R[s,{X}(s)]-\frac{1}{2c}\left[\frac{\partial}{\partial X}\overline{\Pi}_w(\mathbf{X},s)\right]^2+\mu_1(s,X)\frac{\partial}{\partial X}\overline{\Pi}_w(\mathbf{X},s)+\mbox{$\frac{1}{2}$}\sigma^2\frac{\partial^2}{\partial X^2}\overline{\Pi}_w(\mathbf{X},s).
\end{equation}
It is important to note that above HJB equation is quadratic in $\overline{\Pi}_w(\mathbf{X},s)$.  For $c\ra\infty$ the above quadratic equation becomes a linear stochastic HJB equation:
\begin{equation}\label{r4}
-\frac{\partial}{\partial s}\overline{\Pi}_w(\mathbf{X},s)=\exp(-\zeta s)R[s,{X}(s)]+\mu_1(s,X)\frac{\partial}{\partial X}\overline{\Pi}_w(\mathbf{X},s)+\mbox{$\frac{1}{2}$}\sigma^2\frac{\partial^2}{\partial X^2}\overline{\Pi}_w(\mathbf{X},s).
\end{equation}
This equation can be thinking as the HJB equation of the wave function
\[
\Theta(\mathbf{X},s)=\exp\left\{-\frac{1}{\omega}\bar\Pi_w(\mathbf{X},s)\right\}
\]
with the Schr\"odinger equation given by,
\begin{equation}\label{r6}
-\frac{\partial}{\partial s}\Theta(\mathbf{X},s)\approxeq-\frac{1}{\omega}\exp(-\zeta s)R[s,{X}(s)]\Theta(\mathbf{X},s)+\mu_1(s,X)\frac{\partial}{\partial X}\Theta(\mathbf{X},s)+\mbox{$\frac{1}{2}$}\sigma^2\frac{\partial^2}{\partial X^2}\Theta(\mathbf{X},s).
\end{equation}
For $\Theta(\mathbf X,s)=\exp\{-\varepsilon \mathcal A_{s,\tau}(\mathbf X)\}$, with $\varepsilon=1/\omega$ the Equation (\ref{r6}) implies $\bar\Pi_w(\mathbf{X},s)=\mathcal A_{s,\tau}(\mathbf X)$.The reason behind assuming $\varepsilon=1/\omega$ is that we can compare $\Theta(\mathbf x,s)$ with the Equation (\ref{action.0}) which facilitates the proof of Proposition \ref{p1}. Taking $c\ra\infty$ in HJB Equation (\ref{r3.0}) helps to reverse the direction of computation. Replace $\Theta(\mathbf X,s)$ by a diffusion process $\Psi_{s}(\tilde{\mathbf{X}})=\Psi_{s}(\tilde{\mathbf{X}},\tau|\mathbf X,s)$ for $\tau>s$ represented by a Wick-rotated Schr\"odinger or a Fokker-Plank Equation
\begin{equation}\label{r7}
-\frac{\partial}{\partial s}\Psi_{s}(\tilde{\mathbf{X}})=-\frac{1}{\omega}\exp(-\zeta s)R[s,{X}(s)]\Psi_{s}(\tilde{\mathbf{X}})+\mu_1(s,X)\frac{\partial}{\partial X}\Psi_{s}(\tilde{\mathbf{X}})+\mbox{$\frac{1}{2}$}\sigma^2\frac{\partial^2}{\partial X^2}\Psi_{s}(\tilde{\mathbf{X}}),
\end{equation}
with $\Psi_{s}(\tilde{\mathbf{X}},s|\mathbf X,s)=\delta(\tilde{\mathbf{X}}-\mathbf X)$ being the \emph{ Dirac delta} function.  Define 
\[
\Phi(s,\mathbf X):=\int_{\mathbb{R}}\Psi_s^i(\tilde{\mathbf{X}},s|\mathbf X,s)\Theta(\tau,\tilde{\mathbf{X}})d\tilde{\mathbf{X}}.
\]
Clearly, $\Phi(\mathbf X,s)$ is independent of $\tau$ in both of the stochastic HJB Equation (\ref{r6}) and the Fokker-Plank Equation (\ref{r7}). Evaluating $\Phi(\mathbf X,s)$ for $\tau=s$ gives $\Phi(\mathbf X,s)=\Theta(\mathbf X,s)$. Further evaluation $\Phi(\mathbf X,s)$ for $\tau=s_{f}$ with $s\in(s_i,s_f)$, where  $s_i$ and $s_f$ stand for initial and final time of time interval $(s_i,s_f)\subset[0,t]$. Thus,
\begin{equation*}
\Phi(\mathbf X,s)=\int_{\mathbb R}\Psi_s(\tilde{\mathbf{X}},s_{f}|\mathbf X,s)\Theta(\mathbf X,s)d\tilde{\mathbf{X}}.
\end{equation*}
Thus,
\begin{equation*}
\Phi(\mathbf X,s)=\int_{\mathbb R}\exp\{-\varepsilon \mathcal A_{s,\tau}(\tilde{\mathbf{X}})\}\Psi_s(\tilde{\mathbf{X}},s_{f}|\mathbf X,s)d\tilde{\mathbf{X}}.
\end{equation*}
Finally after introducing the penalizing constant $L_\varepsilon>0$ above equation becomes,
\begin{equation*}
\Psi_{s,\tau}(\mathbf X)=\frac{1}{L_\varepsilon}\int_{\mathbb R}\exp\{-\varepsilon \mathcal A_{s,\tau}(\tilde{\mathbf{X}})\}\Psi_s(\tilde{\mathbf{X}},s_{f}|\mathbf X,s)d\tilde{\mathbf{X}},
\end{equation*}
which is similar to the expression represented by the Equation (\ref{M}).

In this paper after defining the objective function and the stochastic dynamics we construct a quantum Lagrangian such that we are treating the Lagrangian multiplier as penalization parameter which is explicitly defined and does not depend on time. Then we define a $g(s,\mathbf X)\in C^{1,2}([0,t],\mathbb R)$ such that it is an It\^o process and the stochastic differential equation is replaced by this $g$ function. A good guess of this function is to choose an integrating factor of the SDE. We directly utilize It\^o's Lemma to find a Wick-rotated Schr\"odinger type equation. Therefore, this method gives us a separate sets of optimal strategies than traditional HJB equation. The central part of our approach is to guess the $g$ function properly. Again this may be the limitation of this method as wrong guess might give us a sub-optimal strategy. Throughout this paper we assume the information is imperfect and incomplete. Therefore, a unique global optimal strategy does not exist in our case. 

\begin{figure}
	\centering
	\begin{tikzpicture}[
	node distance=1.8cm,
	every node/.style={font=\small},
	box/.style={rectangle, rounded corners, draw=black, thick, align=center, minimum width=8cm, minimum height=1.2cm},
	arrow/.style={->, thick}
	]
	
	
	\node[box, fill=blue!10] (hjb) {
		\textbf{HJB Equation}\\[2pt]
		$-\partial_s \overline{\Pi}_w(X,s)
		= \max_{u\in\mathcal U} \left\{\pi(s,X,u)
		+ \mu_1(s,X)\partial_X \overline{\Pi}_w(X,s)
		+ \tfrac{\sigma^2}{2} \partial_{XX}\overline{\Pi}_w(X,s) \right\}$
	};
	
	\node[box, fill=orange!15, below of=hjb] (control) {
		\textbf{Optimal control from HJB}\\[2pt]
		$u^*(s)=\dfrac{1}{c}\,\partial_X \overline{\Pi}_w(X,s)$
	};
	
	\node[box, fill=yellow!20, below of=control] (transform) {
		\textbf{Exponential transformation}\\[2pt]
		$\Theta(X,s)=\exp\!\left\{-\dfrac{1}{\omega}\overline{\Pi}_w(X,s)\right\}$
	};
	
	\node[box, fill=green!15, below of=transform] (linear) {
		\textbf{Linearized (Wick–rotated Schr\"odinger)}\\[2pt]
		Forward diffusion representation
	};
	
	\node[box, fill=purple!15, below of=linear] (paths) {
		\textbf{Monte Carlo path sampling}\\[2pt]
		Simulate trajectories $X^{(i)}(\cdot)$ under baseline dynamics
	};
	
	\node[box, fill=red!15, below of=paths] (weights) {
		\textbf{Exponential reweighting}\\[2pt]
		$w_i \propto \exp\left\{-\varepsilon \mathcal A_{s,\tau}\left(X^{(i)}\right)\right\}$
	};
	
	\node[box, fill=cyan!15, below of=weights] (estimate) {
		\textbf{Path integral estimator}\\[2pt]
		$\Psi_{s,\tau}(X)
		= \dfrac{1}{L_\varepsilon}
		\sum_{i=1}^n w_i$
	};
	
	\node[box, fill=gray!20, below of=estimate] (policy) {
		\textbf{Optimal strategy (path integral form)}\\[2pt]
		$u^*(s)\approx \sum_{i=1}^n w_i\,u^{(i)}(s)$
	};
	
	
	\draw[arrow] (hjb) -- (control);
	\draw[arrow] (control) -- (transform);
	\draw[arrow] (transform) -- (linear);
	\draw[arrow] (linear) -- (paths);
	\draw[arrow] (paths) -- (weights);
	\draw[arrow] (weights) -- (estimate);
	\draw[arrow] (estimate) -- (policy);
	
	\end{tikzpicture}
	
	\caption{Intuitive flow of the path integral control method for a firm. The nonlinear HJB equation is transformed via exponential change of variables into a linear diffusion representation. Optimal policies are then recovered through importance-weighted Monte Carlo trajectories rather than explicit state-space discretization.}
	\label{fig:pi_flowchart}
\end{figure}
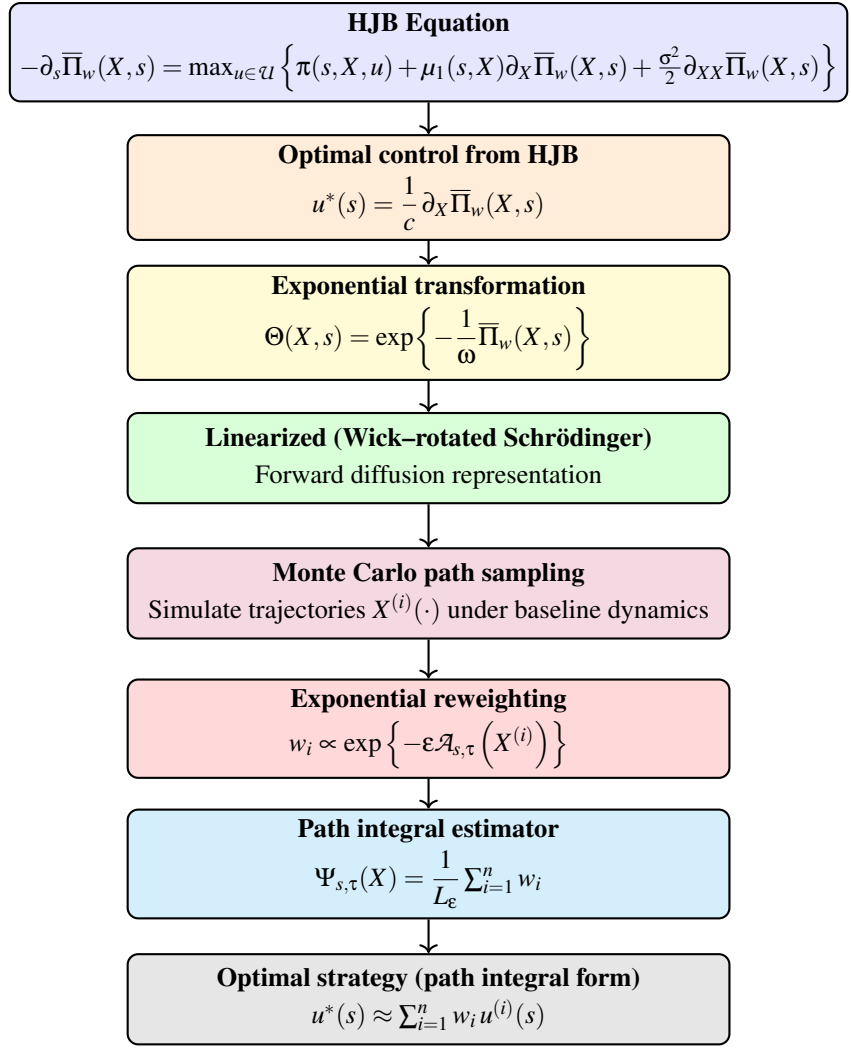

In order to build intuition for the proposed methodology, it is useful to interpret the path integral formulation as a probabilistic re-expression of the HJB equation rather than as a fundamentally different optimization paradigm (see Figure \ref{fig:pi_flowchart}). Starting from the nonlinear HJB equation associated with the firm’s dynamic profit maximization problem, the first-order condition yields the familiar feedback rule $u^*(s)=\frac{1}{c}\partial_X\overline{\Pi}_w(X,s)$, which, when substituted back into the HJB equation, produces a semilinear partial differential equation that is generally difficult to solve directly in high-dimensional environments. The key analytical step is the exponential transformation $\Theta(X,s)=\exp\{-\omega^{-1}\overline{\Pi}_w(X,s)\}$, which converts the nonlinear HJB equation into a linear backward evolution equation of Wick--rotated Schr\"odinger type. This linearization is crucial because it permits a forward probabilistic representation in terms of diffusion dynamics, thereby replacing deterministic value-function iteration with a statistical-mechanical expectation. Under this representation, the control problem can be viewed operationally as follows: one simulates an ensemble of stochastic market-share trajectories under the baseline dynamics, evaluates the cumulative action (or profit functional) along each path, and assigns exponential importance weights of the form $w_i \propto \exp\{-\varepsilon \mathcal A_{s,\tau}(X^{(i)})\}$. Paths that generate higher discounted revenues receive larger weights, while economically inferior trajectories are exponentially attenuated. The path integral estimator $\Psi_{s,\tau}(X)$ is then constructed as a normalized weighted average over these simulated trajectories, and the optimal feedback strategy is recovered from the corresponding weighted control ensemble $u^*(s)\approx \sum_{i=1}^n w_i u^{(i)}(s)$. From this perspective, the phrase ``running multiple paths'' acquires a precise economic meaning: rather than solving the HJB equation pointwise on a discretized state grid, the algorithm explores the space of feasible future market evolutions and performs the optimization implicitly through exponential reweighting. This viewpoint clarifies why the path integral approach can mitigate the curse of dimensionality associated with grid-based dynamic programming while remaining fully consistent with the underlying stochastic control structure of the Walrasian firm.

\section{Main results.}

Define a non-negative measurable discounted profit function for a single firm as 
\[
\pi[s,X(s),u(s)]=\exp(-\zeta s)\tilde{\pi}[s,X(s),u(s)].
\]
Assume that $\pi$ is a finite $C^\infty$ function with respect to $ X(s)$ and $u(s)$ where $\zeta\in[0,1]$ is a constant discount rate of profit over $s\in[0,t]$. The function $\tilde{\pi}[s,X(s),u(s)]$ is the actual profit at time $s$, and is assumed to be quadratic in terms of change in time, non-decreasing in output price, non-increasing in input price, homogeneous of degree one in output and input prices, convex in output and input prices, continuous in output and input prices, and is continuous with respect to $s$. We assume that $X(s)$ is a time dependent measure of a stochastic market dynamic and the strategy $u(s)$ is a deterministic function of $x$. Further technical assumptions are given in the Appendix.

To optimize the dynamic profit function $\Pi$ defined in Definition \ref{DWalrus}  
with respect to the strategy $u$ we need to specify a function $g:[0,t]\times\mathcal X\rightarrow\mathbb{R}$ to favor strategies that respect the dynamics specified by Equation (\ref{mkt}). In the standard Lagrangian framework this function is specified as $g(s,X)\approx\lambda(s)[\Delta X(s)ds-h(s,X)]$, where $h$ is a function that specifies the dynamics of the system and $\lambda(s)$ is the Lagrangian multiplier.

\begin{prop}[Walrasian Equilibrium]\label{p1} 
	An optimal strategy for maximizing the dynamic profit function $\Pi(u,t)$ with respect to the control $u$ and constraint  
	\[
	dX(s)=\mu[s,X(s),u(s)] ds+\sigma[s,X(s),u(s)] dB(s),
	\]
	with initial condition $x(0)=x_0$ is the solution of the equation 
	\begin{equation}\label{o}
	\left[\frac{\partial}{\partial u}f(s,X,u)\right]
	\left[\frac{\partial^2}{\partial X^2}f(s,X,u)\right]^2=
	2\left[\frac{\partial}{\partial X}f(s,X,u)\right]
	\left[\frac{\partial^2}{\partial X\partial u}f(s,X,u)\right],
	\end{equation}
	with respect to $u$ as a function of $x$ and $s$ evaluated at $X=X(s)$, where  
	\begin{equation}\label{w11}
	f(s,X,u)=
	\pi(s,X,u)+g(s,X)+\frac{\partial}{\partial s}g(s,X)+
	\mu(s,X,u)\ \frac{\partial}{\partial X}g(s,X)+
	\mbox{$\frac{1}{2}$}\sigma^2(s,X,u) \
	\frac{\partial^2}{\partial x^2} g(s,X),
	\end{equation}
	where $g(s,X)\in C^{1,2}\left([0,t]\times\mathbb R\right)$ is a chosen monotonic function on drift coefficient $\mu[s,X(s),u(s)]$ with $I(s)=g(s,X)$	being an It\^o process which is a positive, non-decreasing function vanishing at the infinity such that Assumptions \ref{as0}-\ref{as2}, Proposition \ref{existence} and Lemma \ref{l1} hold.
\end{prop}

\begin{remark}
	When a problem is formulated based on a linear SDE with additive white Gaussian noise and a quadratic profit function, it is well understood that the optimal feedback strategy can be obtained by solving a Riccati equation \citep{satoh2016}. Although the linear quadratic (LQ) dynamic optimization problem with Gaussian noise has brought many practical benefits, nonlinear optimal control theory has not yet led to sufficient practical use. According to Bellman’s principle, the optimal feedback strategy for a nonlinear stochastic control problem is given by solving a nonlinear PDE of second order, called the stochastic HJB equation, which is also very effective and rigorous in methodological perspective \citep{pham2009,satoh2016}. Due to the presence of nonlinearity and second order partial derivatives, the stochastic HJB equation is quite difficult to solve. Although some solution methods for the deterministic HJB equation (i.e. a first order nonlinear PDE) have been recently proposed, they are not directly applied to the stochastic HJB equation \citep{satoh2016}. The generalized HJB equation is solved by the Galerkin method. Proposition \ref{p1} gives an alternative approach to solve a very classical stochastic control problem. Our method is different than the classical path integral control approach in the sense that, former uses Feynman-Kac lemma directly to determine a Schr\"odinger equation, while we replace our stochastic market dynamics with a $C^{1,2}$-function $g:[0,t]\times\mathcal X\ra\mathbb R$, which is a monotonic transformation of the drift coefficient $\mu$, and utilize It\^o's lemma. After doing a bunch of Taylor series expansions we get a Wick-rotated Schr\"odinger-type (i.e., a Fokker-Plank type) equation. In our approach, the key concept revolves around the judicious selection of a suitable $g$ function. The policy maker plays a crucial role in determining this $g$ function by choosing an integrating factor, ensuring that it effectively addresses the SDE governing market dynamics. 
\end{remark}

\begin{example}\label{ex2}
	Let a Walrasian firm has the objective function 
	\[ 
	\E_0\left\{\int_0^t\exp(-\zeta s)[pX^2(s)-cX^2(s)u(s)]ds\biggr|\mathcal F_0^X\right\},
	\]
	where $\zeta\in(0,1]$ is a constant discount rate over continuous time interval $[0,t]$, $p>0$ is constant price, $X(s)$ is the positive real-valued market share, a twice differentiable function of $s$, $c$ is a positive constant marginal cost, and $u$ is the total expenditure on advertising. The quadratic profit function is adopted primarily for analytical tractability within the path integral control framework. In particular, the quadratic dependence ensures that the associated HJB equation can be transformed into a linearly solvable form with Gaussian path weights, which is a standard requirement in PI-based stochastic control. From an economic standpoint, the quadratic profit representation may be interpreted as a second-order local approximation to a more general nonlinear revenue and advertising cost structure around an operating equilibrium. Such approximations are commonly used in dynamic industrial organization and stochastic control models when marginal effects are smooth and firms operate near steady market shares. Under this interpretation, the present formulation captures the leading-order curvature of profit with respect to market share and advertising intensity.
		
		If the true market environment exhibits strongly non-quadratic features (for example, saturation effects, threshold advertising responses, or asymmetric adjustment costs), the optimal policy derived under the quadratic specification may experience approximation bias. Nevertheless, the path integral approach itself is not inherently limited to strictly quadratic profit functions. In principle, more general nonlinear profit structures can be accommodated through importance sampling or modified path-weight constructions, though at the expense of losing exact Gaussian structure and increasing Monte Carlo variance.
	
	Consider market dynamics given by 
	\begin{equation}\label{dynamic1}
	dX(s)=[aX(s)-u(s)]ds+\sqrt{\sigma u(s)}dB(s),
	\end{equation}
	where $a$ and $\sigma$ are two positive and finite constants. The negative terms in the drift part of Equation (\ref{dynamic1}) and the objective function reflect the firm's cost of advertising its product as its strategy. The diffusion component of Equation (\ref{dynamic1}) reflects the amount of variation in the system. To apply Proposition \ref{p1} we specify $g(s,x)$ to represent the market dynamics.To obtain the $g$ function we will go along the example $5.1.3$ of \cite{oksendal2003}. Multiplying $\exp(-as)$ on both sides of the Equation (\ref{dynamic1}) yields,
	\begin{equation}\label{dynamic1.0}
	\exp(-as)dX(s)-\exp(-as)aX(s)ds=\exp(-as)\left[-u(s)ds+\sqrt{\sigma u(s)}dB(s)\right].	
	\end{equation}
	The left hand side of the Equation (\ref{dynamic1.0}) can be related to $d(\exp(-as)X(s))$. If we choose $\hat g(s,X(s))=\exp(-as)X(s)$, then
	\[
	d(\exp(-as)X(s))=(-a)\exp(-as)X(s)ds+\exp(-as)dX(s).
	\]	
	Substituting the above result into Equation (\ref{dynamic1.0})	yields
	\[
	\exp(-at)X(t)-x_0=-\int_0^t\exp(-as)u(s)ds+\int_0^t\exp(-as)\sqrt{\sigma u(s)}dB(s).
	\]
	Therefore, by Theorem $4.1.5$ of \cite{oksendal2003} the solution of $X$ would be
	\[
	X(t)=\exp(at)\left[x_0+\exp(-at)\sqrt{\sigma u(s)}B(t)+\int_0^t\exp(-as)\left[-u(s)+a\sqrt{\sigma u(s)}dB(s)\right]ds\right].
	\]	
	Since $\hat g(s,X(s))=\exp(-as)X(s)$ gives a solution of $X(s)$ from the SDE (\ref{dynamic1}), for a fixed positive constant $\lambda^*$ let $g(s,X)=\lambda^* \exp(-as)X(s)$. Equation (\ref{w11}) yields 
	\[
	f(s,X,u)=x^2\exp(-\zeta s)(p-cu)+\lambda^*X\exp(-as)-a\lambda^*X\exp(-as)+\lambda^*(aX-u)\exp(-as).
	\]
	Therefore 
	\[
	\frac{\partial}{\partial x}f(s,X,u)= 2X\exp(-\zeta s)(p-cu)+\lambda^*\exp(-as),
	\]
	\[
	\frac{\partial}{\partial u}f(s,X,u)=-\left[cX\exp(\zeta s)+\lambda^*\exp(-as)\right],
	\]
	\[
	\frac{\partial^2}{\partial X^2}f(s,x,u)=2\exp(-\zeta s)(p-cu),
	\]
	and
	\[
	\frac{\partial^2}{\partial X\partial u}f(s,x,u)=
	-2cX\exp(-\zeta s). 
	\]
	Equation (\ref{o}) then implies that an optimal Walrasian strategy for this system is given by
	\begin{equation}\label{e3}
	\phi_w^*(s,x)=\frac{-1}{c\exp(-\zeta s)}\left[\exp(-\zeta s)\left(p-\frac{X}{A}\right)\pm\sqrt{\exp(-2\zeta s)\left(p-\frac{X}{A}\right)^2-\exp(-\zeta s)\left[\frac{2pX\exp(-\zeta s)}{A}+B\right]}\right],
	\end{equation}
	where $A=cX\exp(-\zeta s)+\lambda^*\exp(-as)$, and $B=\lambda^*\exp(-as)/A$.
\end{example}

In Example \ref{ex2} both the objective function and the market dynamics are linear continuous mappings from strategy space to the real line. According to the Generalized Weierstrass Theorem there exists an optimal strategy. One such strategy is given in Equation (\ref{e3}). To apply the Pontryagin maximum principle, it is necessary to bound the strategy $u$, because the optimal strategy is of bang-bang. 

\begin{remark}
	Even for a Walrasian firm, a profound understanding of its own profit function and market dynamics is crucial. Its decisions are intricately linked to the probabilities associated with various types of instantaneous information at its disposal. Recognizing this, the central policy-maker adeptly tailors the $g$ function, aligning it with the Walrasian system, to ascertain the optimal strategy.
	In contrast to situations where a bang-bang solution may arise under perfect and complete information, our scenario offers a closed-form solution for a specific value of $\lambda^*$. Moreover, we treat the constant $\lambda^*$ as a penalization parameter, explicitly determined based on the historical data of the economy. Once $\lambda^*$ is established, we leverage this value to derive the optimal Walrasian strategy. While the global policy-maker may have multiple values of $\lambda^*$ at their disposal, they judiciously select one based on the current market environment. This chosen $\lambda^*$ serves as the foundation for determining the optimal Walrasian strategy for each firm $\phi_w^*(s,X)$. It is crucial to emphasize that the key to attaining a solution lies in appropriately estimating $g$, a facet that may pose a limitation to this approach.	
\end{remark}	

\begin{example}\label{ex3}
	Suppose, a firm under the Walrasian system produces consumer goods with objective 
	function
	\[
	\E_0\left\{\int_0^t\exp(-\zeta s)\left[R(X)-cu^2\right]ds\biggr|\mathcal F_0^X\right\},
	\]
	where $\zeta\in(0,1]$ is a constant discount rate over $[0,t]$, $R(X)$ is the total revenue function such that it can be multiplicatively separable by $d^2/ds^2$ as discussed in the Appendix, $c$ is the constant marginal cost multiplied by squared strategy function $u(s)$. The main difference between this example with Example \ref{ex2} is that, the strategy $u(s)$ is a $C^{1,2}$ function and hence, we can calculate optimal strategy using Pontryagin's maximum principle. 
	
	Assume the market dynamics of a firm follow
	\begin{equation}\label{dynamic2}
	dX(s)=[bX(s)-u(s)]ds+\sqrt{2 b}dB(s),
	\end{equation}
	where $b$ is a positive constant. 
	
	We will undertake a comparative analysis between our approach and the conventional Pontryagin maximum principle to discern the optimal strategy for this Walrasian firm operating within the consumer goods industry.
	
	{\bf(i) Quantum Approach:}  In this section, we are going to determine an optimal strategy by using results of Proposition \ref{p1}. Since this Proposition utilizes a Wick-rotated Schr\"odinger type equation, we call this approach as \emph{quantum approach}. Since the consumption of consumer goods increases exponentially, a Walrasian firm under this sector should face the market dynamics which shows the behavior in Equation (\ref{dynamic2}). Since Equation (\ref{dynamic2}) is similar to the Equation (\ref{dynamic1}), for a fixed constant $\lambda^*$, $g(s,X)=\lambda^*\exp(-bs)X(s)$. 
	Equation (\ref{w11}) yields
	\[
	f(s,X,u)=\exp(-\zeta s)\left[R(X)-cu^2X\right]+\lambda^*X\exp(-bs)-b\lambda^*X\exp(-bs)+\lambda^*(bX-u)\exp(-bs).
	\]
	Therefore
	\[
	\frac{\partial}{\partial X}f(s,X,u)=\exp(-\zeta s)\left[\frac{\partial R(X)}{\partial X}-cu^2\right]+\lambda^*\exp(-bs),
	\]
	\[
	\frac{\partial}{\partial u}f(s,X,u)=-\left[2cXu\exp(-\zeta s)+\lambda^*\exp(-bs)\right],
	\]
	\[
	\frac{\partial^2}{\partial X^2}f(s,X,u)=\frac{\partial^2R(X)}{\partial X^2}\exp(-\zeta s),
	\]
	and 
	\[
	\frac{\partial^2}{\partial X\partial u}f(s,X,u)
	=2cu\exp(-\zeta s).
	\] 
	Equation (\ref{o}) yields a cubic strategy function $u$ such that,
	\[B_0(s,X) u^3+B_1(s,X) u^2+B_2(s,X) u+B_3(s,X)=0,\]
	with
	\[
	B_0(s,X)=4c^2\exp(-\zeta s),
	\]
	\[
	B_1(s,X)=
	4(cX)^2\frac{\partial^2 R(X)}{\partial x^2}\exp(-2\zeta s),
	\]
	\[
	B_2(s,X)=4c\left[\lambda^*X\exp(-(b+\zeta)s)-\frac{\partial R(X)}{\partial X}\exp(-\zeta s)-\lambda^*\exp(-bs)\right],
	\]
	\[
	B_3(s,X)=(\lambda^*)^2\frac{\partial^2 R(X)}{\partial X^2}\exp(-2bs),
	\]
	and the optimal Walrasian strategy becomes,
	\begin{multline*}
	\phi_w^*(s,X)=
	D_1(s,X)+\left\{D_2(s,X)+\left[D_2^2(s,X)+\left(D_3(s,X)-D_1^2(s,X)\right)^3\right]^{\frac{1}{2}}\right\}^{\frac{1}{3}}\\
	+\left\{D_2(s,X)-\left[D_2^2(s,X)+\left(D_3(s,X)-D_1^2(s,X)\right)^3\right]^{\frac{1}{2}}\right\}^{\frac{1}{3}},
	\end{multline*}
	such that 
	\[
	D_1(s,X)=-\frac{B_1(s,X)}{3B_0(s,X)},
	\]
	\[
	D_2(s,X)=D_1^3(s,X)+\frac{B_1(s,X)B_2(s,X)-3B_0(s,X)B_3(s,X)}{6B_0^2(s,X)},
	\] 
	\[
	D_3(s,X)=\frac{B_2(s,X)}{3B_0(s,X)},
	\]
	and $B_0(s,X)\neq 0$. The important part of this result is 
	that we start with a $g(s,X)$ function such that it is a $C^{1,2}$ 
	function within $[0,t]$ and we get the optimal strategy by 
	solving a cubic equation. 
	
	{\bf(ii) Pontryagin maximum principle:}	Following \cite{yeung2006} we know, an optimal strategy $u^*(s)=\phi^*(s,X)$ constitutes an optimal solution to the problem (\ref{n1}),  if there exists a continuously differentiable value function $V(s,X):[0,t]\times\mathcal X\ra\mathbb{R}$, satisfying the following HJB equation
	\begin{align*}
	-\mbox{$\frac{\partial}{\partial s}$}V(s,X)-\mbox{$\frac{1}{2}$} \sigma^2(s,X,u)\ \mbox{$\frac{\partial^2}{\partial x^2}$}V(s,X)=\arg\max_{u\in\mathcal U} \bigg\{\pi(s,X,u)+\mbox{$\frac{\partial}{\partial X}$}V(s,X)\ \mu(s,X,u)\bigg\}.
	\end{align*}
	After solving for 
	\[
	\arg\max_{u\in \mathcal U} \bigg\{\exp(-\zeta s)\left[R(X)-cu^2X\right]+(bX-u)\mbox{$\frac{\partial}{\partial X}$}V(s,X) \bigg\}=0,
	\]
	we get $$\phi_w^*(s,X)=\frac{V_X(s,X)}{2cX\exp(-\zeta s)},$$ where $V_X=\partial V/\partial X$. After using the value of $\phi^*(s,X)$, the right hand side of the HJB equation becomes,
	\begin{align}\label{a05}
	R(X)\exp(-\zeta s)+bX V_X(s,X)-\mbox{$\frac{3}{4cX}$}V_X^2(s,X)\exp(\zeta s).
	\end{align}
	For some functions $A(s)$ and $B(s)$ define a value function $V(s,X)=\exp(-\zeta s)\left[A(s)X+B(s)\right]$ so that
	\begin{align}\label{a06}
	-\mbox{$\frac{\partial}{\partial s}$}V(s,X)&=\exp(-\zeta s)\bigg\{\zeta[A(s)X+B(s)]-\big[X\mbox{$\frac{\partial}{\partial s}$}A(s)+\mbox{$\frac{\partial}{\partial s}$}B(s)\big]\bigg\},\notag\\\mbox{$\frac{\partial}{\partial x}$}V(s,X)&=\exp(-\zeta s)A(s)\ \text{and,}\ \mbox{$\frac{\partial^2}{\partial x^2}$}V(s,X)=0.
	\end{align} 
	After using Equations (\ref{a05}) and (\ref{a06}), the HJB Equation becomes,	
	\begin{equation}\label{a07}
	\exp(-\zeta s)\bigg\{\zeta[A(s)X+B(s)]-\big[X\mbox{$\frac{\partial}{\partial s}$}A(s)+\mbox{$\frac{\partial}{\partial s}$}B(s)\big]\bigg\}=R(X)\exp(-\zeta s)+bX V_X(s,X)-\mbox{$\frac{3}{4cX}$}V_X^2(s,X)\exp(\zeta s).
	\end{equation}
	After comparing terms in both sides of Equation (\ref{a07}) we get,
	\begin{equation*}
	R(X)=\zeta[A(s)X+B(s)]-\big[X\mbox{$\frac{\partial}{\partial s}$}A(s)+\mbox{$\frac{\partial}{\partial s}$}B(s)\big],
	\end{equation*}
	and,
	\begin{equation*}
	bx V_X(s,X)-\mbox{$\frac{3}{4cX}$}V_X^2(s,X)\exp(\zeta s)=0,
	\end{equation*}
	which yields $V_X(s,X)=0$ or 
	\[
	V_X(s,X)=\frac{4}{3}bcX^2\exp(-\zeta s).
	\]
	Therefore, Walrasian optimal strategy under Pontryagin  maximum principle is either $\phi_w^*(s,X)=0$ or $\phi_w^*(s,X)=2bX/3$.
\end{example}

\begin{remark}
	Our method yields a distinct solution compared to the traditional Pontryagin Maximum Principle. In our approach, the optimal Walrasian strategy emerges as the solution to a cubic polynomial, diverging from the Pontryagin maximum principle that derives optimal strategy from a value function characteristic equation denoted as $A(s)X+B(s)$. The crux of our methodology revolves around the pivotal task of correctly guessing an appropriate $g$ function (by taking the integration factor of an SDE).
	
	Unlike firms, a policy maker possesses superior insights into the drift coefficient of market dynamics derived from historical market data. Leveraging this knowledge, the policy maker selects a suitable $g$ function. Prior to finalizing the $g$ function, the policy maker ensures that it results in a unique solution to Equation (\ref{dynamic2}). This validation step underscores that even policy makers operate without perfect and complete information about the market. Consequently, our methodology manifests two distinct $g$ functions in Examples 1 and 2, showcasing the nuanced nature of strategic decision-making in the absence of perfect market information.
\end{remark}

Another important example considers problems involving European call options, which have been well studied in finance, and provide the basis for the Black-Scholes formula and further generalizations by Merton-Garman. In the generalized approach the stock volatility is stochastic and is derived by a parabolic partial differential equation \citep{baaquie1997,merton1973}. As  determining a solution through HJB equation becomes impossible in this case, methods of theoretical physics have been applied to get an optimal solution \citep{bouchaud1994}. For example, the Feynman-Kac lemma has been used in \cite{baaquie1997} and \cite{belal2007} to find a solution of a Merton-Garman-Hamiltonian type equations using the Dirac bracket method \cite{bergmann}. In Proposition \ref{p2} we use a path integral approach to a situation where the firm's objective is to maximize its portfolio subject to a Merton-Garman-Hamiltonian type stochastic volatility in an European call option with controls. Using the function $g$ as defined for Proposition \ref{p1}, the result given below provides an optimal investment strategy for this framework.

For this type of problem suppose that the firm has the objective 
of maximizing 
\[
\Pi_{\text{MG}}(u,t)=
\mathbb{E}_0\left\{\int_0^t\pi[s,H(s,K,V),V(s),u(s)]ds\biggr|\mathcal F_0^{K,V}\right\},
\]
where $u(s)\in\mathcal U$ is the strategy, $\mathcal F_0^{K,V}$ be the filtration of $K$ and $V$ starting at time $0$, $H$ is the European call option price which is a function of the time $s$, the stock price of the 
security at time $s$ is represented by $K(s)\in\mathcal K$, 
and the volatility at time $s$ is represented by $V(s)\in\mathcal V$, where $\mathcal K$ and $\mathcal V$ are the functional spaces of security and volatility respectively. It is assumed that the stock price and the volatility follow Langevin dynamics of the form 
\[
dK(s)=\mu_1[s,u(s)]K(s)ds+\sigma_1[s,u(s)]K(s)dB_1(s),
\]
and 
\[
dV(s)=\mu_2[s,u(s)]V(s)ds+\sigma_2[s,u(s)]V(s)dB_2(s),
\]
where $\mu_1[s,u(s)]$ is the expected return of the security, $\mu_2[s,u(s)]$ is the expected rate of increase in $V(s)$, and $B_1(s)$ and $B_2(s)$ are standard Brownian motion processes such that the correlation between $dB_1(\tilde s)$ and $dB_2(s)$ is zero unless $\tilde s=s$ for which case it equals a value $\gamma\in[-1,1]$. 

Now we will construct a Merton-Garman-Hamiltonian system from the above market dynamics. Before starting derivation, let us redefine the components involved in above two dynamics. Define $\mu_1(s,u)=\mu_1$, $V=\sigma_1^2(s,u)=\sigma_1^2$, $\mu_2(s,u)=\mu_2$, $\sigma_2(s,u)=\sigma_2$ and finally replace $V(s)$ function in the diffusion component of the volatility dynamics with $V^\alpha$ \citep{baaquie2013,arraut2021}. In this system $\mu_1$, $\mu_2$ and $\sigma_2$ are constants. Therefore, our new set of stochastic differential equations become,
\begin{align*}
dK&=\mu_1Kds+K\sqrt{V}dB_1,\\
dV&=\mu_2Vds+\sigma_2V^\a dB_2,
\end{align*}
or,
\begin{align*}
\frac{dK}{ds}=\mu_1K+K\sqrt{V}W_1,\\
\frac{dV}{ds}=\mu_2V+\sigma_2V^\a W_2,
\end{align*}
where $W_1$ and $W_2$ are two Gaussian Processes with zero means and volatilities $\sigma_1$ and $\sigma_2$ respectively, such that, their correlation has the following form
\begin{equation*}
\langle W_1(\tilde s),W_1(s)\rangle=\langle W_2(\tilde s),W_2(s)\rangle=\delta(s-\tilde s)=\frac{1}{\gamma}\langle W_1(s), W_2(\tilde s)\rangle.
\end{equation*}
Here, $\gamma\in[-1,1]$, and the brackets $\langle AB\rangle$ correspond to the correlation between A and B. Suppose there exists a  $C^2$-function $h(s,K,V,W_1,W_2)$ such that It\^o lemma yields,
\begin{multline}\label{mg0}
\frac{d h}{d s}=\frac{\partial h}{\partial s}+\mu_1 K\frac{\partial h}{\partial K}+\mu_2V\frac{\partial h}{\partial V}+\frac{\sigma_2^2K^2}{2}\frac{\partial^2 h}{\partial K^2}+\gamma V^{1/2+\a}\sigma_2\frac{\partial^2 h}{\partial K\partial V}+\frac{\sigma_2^2V^{2\a}}{2}\frac{\partial^2 h}{\partial V^2}
+\sigma_1 K\frac{\partial h}{\partial K}W_1+\sigma_2V^\a\frac{\partial h}{\partial V}W_2.
\end{multline}
Above equation can be represented in a more compact form, which separates the stochastic terms from the non-stochastic ones as follows,
\[
\frac{d h}{ds}=\xi+\phi W_1+\varphi W_2,
\]
where $\xi=\frac{\partial h}{\partial s}+\mu_1 K\frac{\partial h}{\partial K}+\mu_2V\frac{\partial h}{\partial V}+\frac{\sigma_2^2K^2}{2}\frac{\partial^2 h}{\partial K^2}+\gamma V^{1/2+\a}\sigma_2\frac{\partial^2 h}{\partial K\partial V}+\frac{\sigma_2^2V^{2\a}}{2}\frac{\partial^2 h}{\partial V^2}$, $\phi=\sigma_1 K\frac{\partial h}{\partial K}$ and, $\varphi=\sigma_2V^\a\frac{\partial h}{\partial V}$.

Let us consider two European call options $C_1$ and $C_2$ on the same underlying
security with strike prices and maturities.	Therefore our portfolio function becomes
\[
\pi=C_1+\Xi_1C_2+\Xi_2K,
\]
where $\Xi_1$ and $\Xi_2$ are constraints such that the white noise is removed. The total derivative of $\pi$ with respect to time becomes,
\begin{equation*}
\frac{d\pi}{ds}=\zeta_1+\Xi_1\zeta_2+\Xi_2\mu_1K+(\phi_1+\Xi_1\phi_2+\Xi_2\sigma_1K)W_1+(\varphi_1+\Xi_1\varphi_2)W_2.
\end{equation*}
Note that this result is obtained after recognizing $h(s)=C_1$ or $h(s)=C_2$ in Equation (\ref{mg0}) when it corresponds. It has been understood that even in this case of stochastic volatility, it is still possible to create a hedged portfolio and come to the condition $d\pi/ds=r\pi$, where $r$ is the constant spot interest rate \citep{arraut2021}. The solution of $\pi$ is non-trivial in this case. Let a parameter $\beta$ be
\begin{align}\label{mg1}
\beta(K,V,s,r)&=\frac{1}{\frac{\partial C_1}{\partial V}}\left[\frac{\partial C_1}{\partial s}+\mu_2V\frac{\partial C_1}{\partial K}+\frac{VK^2}{2}\frac{\partial^2 C_1}{\partial K^2}+\gamma\sigma_2 V^{\frac{1}{2}+\a}\frac{\partial^2 C_1}{\partial K\partial V}+\frac{\sigma_2^2V^{2\a}}{2}\frac{\partial^2C_1}{\partial V^2}-rC_1\right]\notag\\
&=\frac{1}{\frac{\partial C_2}{\partial V}}\left[\frac{\partial C_2}{\partial s}+\mu_2V\frac{\partial C_2}{\partial K}+\frac{VK^2}{2}\frac{\partial^2 C_2}{\partial K^2}+\gamma\sigma_2 V^{\frac{1}{2}+\a}\frac{\partial^2 C_2}{\partial K\partial V}+\frac{\sigma_2^2V^{2\a}}{2}\frac{\partial^2C_2}{\partial V^2}-rC_2\right],
\end{align}
where parameter $\beta$ in the Merton-Garman equation is defined as the market price volatility risk because the higher its value is, the lower the intention is of the investors to risk \citep{arraut2021}. As the stochastic volatility is not traded, it is impossible to make a direct hedging process over this quantity. Therefore, under the presence of stochastic volatility, it is necessary to consider the expectations of the investors through parameter $\beta$ \citep{arraut2021}. As $\beta$ is non-vanishing, it is always assumed that the risk of the market (in price) has been included inside the Merton-Garman equation. Rewriting Equation (\ref{mg1}) yields
\begin{equation}\label{mg2}
\frac{\partial C}{\partial s}+rK\frac{\partial C}{\partial K}+(\mu_2V-\beta)\frac{\partial C}{\partial V}+\frac{VK^2}{2}\frac{\partial^2 C}{\partial K^2}+\gamma\sigma_2 V^{\frac{1}{2}+\a}K\frac{\partial^2 C}{\partial K\partial V}+\frac{\sigma_2^2V^{2\a}}{2}\frac{\partial^2C_1}{\partial V^2}=rC.
\end{equation}
Now we will express the Hamiltonian form of the above Merton-Garman Equation or simply Merton-Garman-Hamiltonian equation. Define $K:=\exp(a)$ with $a\in(-\infty,\infty)$ and $V:=\exp(b)$ with $b\in(-\infty,\infty)$. Equation (\ref{mg2}) becomes,
\begin{multline}\label{mg3}
\frac{\partial C}{\partial s}+\left[r-\frac{\exp(b)}{2}\right]\frac{\partial C}{\partial a}+\left[\mu_2-\beta\exp(-b)-\frac{\sigma_2^2}{2}\exp[2b(\a-1)]\right]\frac{\partial C}{\partial b}+\frac{\exp(b)}{2}\frac{\partial^2 C}{\partial a^2}\\+\gamma\sigma_2\exp\left[b\left(\a-\frac{1}{2}\right)\right]\frac{\partial^2 C}{\partial a\partial b}+\sigma_2^2\exp\left[2b(\a-1)\right]\frac{\partial^2 C}{\partial b^2}=rC.
\end{multline}
After expressing Equation (\ref{mg3}) as an eigen value problem \citep{arraut2021} yields
\begin{multline}\label{mg4}
H_{MG}=r-\left[r-\frac{\exp(b)}{2}\right]\frac{\partial}{\partial a}-\left[\mu_2-\beta\exp(-b)-\frac{\sigma_2^2}{2}\exp[2b(\a-1)]\right]\frac{\partial }{\partial b}-\frac{\exp(b)}{2}\frac{\partial^2}{\partial a^2}\\-\gamma\sigma_2\exp\left[b\left(\a-\frac{1}{2}\right)\right]\frac{\partial^2 }{\partial a\partial b}-\sigma_2^2\exp\left[2b(\a-1)\right]\frac{\partial^2 }{\partial b^2}.
\end{multline}
Equation  (\ref{mg4}) is non-linear in nature and has two degrees of freedom \citep{arraut2021} and the exact solution is done in \citep{baaquie1997} by using path-integral technique when $\a=1$. In general, non-linear PDE like expressed in Equations (\ref{mg3}) and (\ref{mg4}) cannot be solved by Pontryagin maximum principle \citep{belal2007,carmona2016}. Our method can handle this type of equation and find optimal strategy of a Walrasian firm. For more detailed treatment of Merton-Garman-Hamiltonian equation see \cite{baaquie1997},\cite{belal2007},\cite{baaquie2013} and \cite{arraut2021}.

\begin{prop}[Merton-Garman Hamiltonian Type Equation]\label{p2}
	Suppose that a firm's objective portfolio is given by
	maximizing $\Pi_{\text{MG}}(u,t)$ with respect to the strategy $u\in \mathcal{U}$ such that Assumptions \ref{as0}-\ref{as2} and Lemma \ref{l1} hold. Let
	\begin{eqnarray}\label{m13.0}
	f(s,K,V,u) & = & \pi[s,H(s,K,V),V,u]+g(s,K,V)+
	\frac{\partial}{\partial s}g(s,K,V)\notag\\ & &+ 
	K\mu_1(s,u)\frac{\partial}{\partial K}g(s,K,V)+
	V\mu_2(s,u)\frac{\partial}{\partial V}g(s,K,V)\notag\\ & &+
	\mbox{$\frac{1}{2}$}K^2\sigma_1^2(s,u)
	\frac{\partial^2}{\partial K^2}g(s,K,V)+
	K\rho\sigma_1^3(s,u)\notag \\
	& &
	\times\frac{\partial^2}{\partial K\partial V}g(s,K,V) 
	+\mbox{$\frac{1}{2}$}V^2\sigma_2^2(s,u)
	\frac{\partial^2}{\partial V^2}g(s,K,V). 
	\end{eqnarray}
	An optimal Walrasian strategy is the functional solution of 
	\[
	-\left[
	\frac{\partial}{\partial u}f(s,K,V,u)
	\right]
	\Psi_s(K,V)=0
	\]
	where $\Psi_s(K,V)=\exp\{-s f(s,K,V,u)\}I(K,V)$
	is the transition probability at time $s$ and states $K(s)$ and $V(s)$ with initial condition $\Psi_0(K,V)=I(K,V)$.
\end{prop}
Proposition \ref{p2} is the extension of the framework of \cite{baaquie1997} that accounts for the firm's portfolio and has drift and diffusion components that are functions of the feedback control system and considers an optimal Walrasian strategy. Proposition \ref{p3} considers the case of the cooperative 
environment outlined in Definition \ref{paretodef}.

\begin{prop}[Cooperative Pareto Optimality]\label{p3}
	A cooperative Pareto optimal solution for firm $\rho$ where all the firms maximize the total dynamic profit $\overline{\Pi}_\text{P}(u,t)$ subject to 
	\begin{equation*}
	d\mathbf{X}(s)=\bm{\mu}[s,\mathbf{X}(s),\mathbf{u}(s)]ds+\bm{\sigma}[s,\mathbf{X}(s),\mathbf{u}(s)]d\mathbf{B}(s),
	\end{equation*}
	with initial condition $\mathbf{x}(0)=\mathbf{x}_0$
	is obtained by solving 
	\begin{equation}\label{sch3}
	-\frac{\partial f[s,\mathbf{X}(s),u(s)]}{\partial u_\rho}\ \Psi_s(\mathbf X)=0,
	\end{equation}
	with respect to $\rho^{th}$ firm's strategy, where $\Psi_s$ is the transition wave function defined as 
	\[
	\Psi_s(\mathbf{X})=\exp[-f(s,\mathbf{X},u)]\Psi_0(\mathbf{X})
	\]
	with initial condition $\Psi_0(\mathbf{X})$ and $f$ is defined as 
	\begin{equation*}
	f(s,\mathbf{X},u)=
	\sum_{\rho=1}^k\alpha_\rho\pi_\rho(s,\mathbf{X},u)+g(s,\mathbf{X})+
	\frac{\partial}{\partial s}g(s,\mathbf{X})+
	\bm{\mu}'(s,\mathbf{X},\mathbf{u})\mathcal{D}_{ \mathbf{X}}g(s,\mathbf{X})+
	\mbox{$\frac{1}{2}$}\bm{\sigma}'(s,\mathbf{X})\mathcal{H}_{\mathbf{X}}g(s,\mathbf{X})\bm{\sigma}(s,\mathbf{X}),
	\end{equation*}
	where $\mathcal{D}_\mathbf{X}$ is the gradient vector and $\mathcal{H}_\mathbf{X}$ is the Hessian matrix. 
\end{prop}

\begin{example}\label{ex5}
	Suppose that a firm under a Cooperative Pareto system has the objective function 
	\[ 
	\E_0\left\{\int_0^t\exp(-\zeta s)\sum_{\rho=1}^k\a_\rho\left\{p\left[X_\rho+\omega_1\sum_{\tilde\rho\neq\rho}X_{\tilde\rho}\right]u_\rho-cX_\rho\left[ u_\rho^2+\omega_2\sum_{\tilde\rho\neq\rho}u_{\tilde\rho}^2\right]\right\}ds\bigg|\mathcal F_0^X\right\},
	\]
	where $\zeta\in(0,1]$ is a constant discount rate over time interval $[0,t]$, $p>0$ is constant price, $\a_\rho$ is the weight corresponding to $\rho^{th}$ firm such that $\sum_{\rho=1}^k\a_\rho=1$, $X_\rho$ is $\rho^{th}$ firm's total output, $X_{\tilde\rho}$ is $\tilde\rho^{th}$ firm's output, $\omega_1>0$ is a constant reward to firm $\rho$ obtained by the actions of other firms, $\omega_2$ is constant weight of corresponding strategies other than than firm $\rho$, $c$ is a positive constant marginal cost for each firm, and $u_\rho$ is the strategy of firm $\rho$.

	Let the market dynamics be
	\[
	d\mathbf{X}(s)=
	[\mathbf{X}'(s)\mathbf{a}\mathbf{X}(s)-\mathbf{u}(s)]ds+
	\mathbf{X}(s)\bm{\sigma}d\mathbf{B}(s),
	\]
	where $\mathbf{X}$ and $\mathbf{u}$ both are $k$-dimensional vectors such that $X_\rho\in\mathcal X^\rho\subset\mathcal X$ and $u_\rho\in\mathcal U^\rho\subset \mathcal U$, $\mathbf a$ is a $k\times k$-dimensional constant symmetric matrix, $\bm\sigma$ is a $k\times m$-dimensional constant matrix and $\mathbf B$ is an $m$-dimensional Brownian motion. Since, the market dynamics described above is an SDE with non-linear drift, following \cite{oksendal2003} there exists an integrating factor
	\[\hat f(t)=\exp\left\{\frac{1}{2}trace(\bm\sigma'\bm\sigma)\int_0^tds-\bm\sigma\int_0^td{\bf B}(s)\right\}=\exp\left\{\frac{1}{2}trace(\bm\sigma'\bm\sigma)t-\bm\sigma{\bf B}(t)\right\},
	\]
	such that $\hat g(t,{\bf X})={\bf X}\hat f(t)$ which solves the the above SDE. Therefore, for a constant $\lambda^*$ assume $$g(s,\mathbf X)=\lambda^*{\bf X}\exp\left\{\frac{1}{2}trace(\bm\sigma'\bm\sigma)s-\bm\sigma{\bf B}(s)\right\}.$$
	Therefore,
	\begin{multline*}
	f(s,\mathbf X,\mathbf u)=\exp(-\zeta s)\sum_{\rho=1}^k\a_\rho\left\{p\left[X_\rho+\omega_1\sum_{\tilde\rho\neq\rho}X_{\tilde\rho}\right]u_\rho-cX_\rho\left[ u_\rho^2+\omega_2\sum_{\tilde\rho\neq\rho}u_{\tilde\rho}^2\right]\right\}+\lambda^*{\bf X}\exp\left\{\frac{1}{2}trace(\bm\sigma'\bm\sigma)s-\bm\sigma{\bf B}(s)\right\}\\
	+\lambda^*\frac{1}{2}trace(\bm\sigma'\bm\sigma){\bf X}\exp\left\{\frac{1}{2}trace(\bm\sigma'\bm\sigma)s-\bm\sigma{\bf B}(s)\right\}+\lambda^*[\mathbf{X}'(s)\mathbf{a}\mathbf{X}(s)-\mathbf{u}(s)]\exp\left\{\frac{1}{2}trace(\bm\sigma'\bm\sigma)s-\bm\sigma{\bf B}(s)\right\}.
	\end{multline*}
	Equation (\ref{sch3}) implies,
	\[
	\phi_{p\rho}^*(s,\mathbf X)=\frac{\exp(-\zeta s)\sum_{\rho=1}^kp\a_\rho \left[X_\rho+\omega_1\sum_{\tilde\rho\neq\rho}X_{\tilde\rho}\right]-\lambda^*\exp\left\{\frac{1}{2}trace(\bm\sigma'\bm\sigma)s-\bm\sigma{\bf B}(s)\right\}}{2c\a_\rho X_\rho \exp(-\zeta s)},
	\]
	such that $c\a_\rho X_\rho \exp(-\zeta s)\neq 0$.
	
	For comparison, Pareto optimal strategy under Pontryagin maximum principle is found by \cite{yeung2006} as $\phi_{p\rho}^*(s,X)=0$ or 
	\[
	\phi_{p\rho}^*(s,X)=\frac{\left(p\a_\rho-1\right)\left[X_\rho+\omega_1\sum_{\tilde\rho\neq\rho}X_{\tilde\rho}\right]+2cX_\rho \mathbf X'\mathbf a\mathbf X }{2cX_\rho},
	\]
	so that, $cX_\rho\neq 0$.
\end{example}

\begin{remark}
	Consistent with our earlier examples, our methodology produces distinct results. This disparity arises from the fact that Pontryagin's maximum principle assumes the characteristic equation corresponding to the value function to be linear. In contrast, our approach challenges this assumption, leading to a different outcome.
\end{remark}

\begin{example}\label{ex5.0}
	Consider a resource extraction problem of two players as discussed in the Section $7.2.1$ of \cite{yeung2006}. 
	Suppose, there are two players with objective function
	\begin{equation*}
	\max_{u_1,u_2\in\{\mathcal U^1\cup\mathcal U^2\}\subset\mathcal U}\E_0\left[\int_0^t\exp(-\zeta s)
	\left\{\left[\left(k_1u_1(s)\right)^{1/2}-
	\frac{c_1u_1(s)}{\mathbf{X}^{1/2}(s)}\right]
	\a_1^0\left[\left(k_2u_2(s)\right)^{1/2}-
	\frac{c_2u_2(s)}{\mathbf X^{1/2}(s)}\right]\right\}ds\biggr|\mathcal{F}_0^X\right],
	\end{equation*}
	subject to 
	\[
	d\mathbf X(s)=\left[a\mathbf X^{\frac{1}{2}}(s)-b\mathbf X(s)-u_1(s)-u_2(s)\right]ds+\left[\mathbf X(s)\mathbf X'(s)\right]\bm\sigma d\mathbf B(s).
	\]
	In the above problem $u^\rho\in\mathcal U^\rho\subset\mathcal U$ is the control strategy vector of player $\rho$ for $\rho\in\{1,2\}$, $a$ and $b$ are positive constant scalar, $\bm\sigma$ is a $k\times m$-dimensional constant, $\a_1^0\in[0,\infty)$ is the optimal cooperative weight corresponding to player $2$  and $B(s)$ is am $m$-dimensional Brownian motion. Here $[k_\rho u_\rho(s)]^{\frac{1}{2}}$ is player $\rho$'s level of satisfaction from the consumption of the resource extracted at time $s$ and $c-\kappa u_\rho(s)\mathbf X^{-\frac{1}{2}}(s)$ is the dissatisfaction level brought about by the cost extraction. Finally, $k_1,k_2,c_1,c_2$ are positive constant scalars.
	
	{\bf (i) Quantum approach:} Since the above SDE has a nonlinear drift, for a constant $\lambda^*$ assume 
	\[
	g(s,\mathbf X)=\lambda^*{\bf X}\exp\left\{\frac{1}{2}trace(\bm\sigma'\bm\sigma)s-\bm\sigma{\bf B}(s)\right\},
	\]
	with $\frac{\partial}{\partial s}g(s,\mathbf X)=\frac{1}{2}\lambda^*{\bf X}\ trace(\bm\sigma'\bm\sigma)\exp\{\frac{1}{2}trace(\bm\sigma'\bm\sigma)s\\-\bm\sigma{\bf B}(s)\}$, $\mathcal D_{\mathbf X}g(s,\mathbf X)=\lambda^*\exp\{\frac{1}{2}trace(\bm\sigma'\bm\sigma)s-\bm\sigma{\bf B}(s)\}$ and $\mathcal{H}_{\mathbf X}g(s,\mathbf X)=0$. Therefore,
	\begin{multline*}
	f(s,\mathbf X, u_1,u_2)=\exp(-\zeta s)\left\{\left[\left(k_1u_1\right)^{\frac{1}{2}}-\frac{c_1u_1}{\mathbf X^{\frac{1}{2}}}\right]+\a_1^0\left[\left(k_2u_2\right)^{\frac{1}{2}}-\frac{c_2u_2}{\mathbf x^{\frac{1}{2}}(s)}\right]\right\}+\lambda^*{\bf X}\exp\left\{\frac{1}{2}trace(\bm\sigma'\bm\sigma)s-\bm\sigma{\bf B}(s)\right\}\\
	+\frac{1}{2}\lambda^*{\bf X}\  trace(\bm\sigma'\bm\sigma)\exp\left\{\frac{1}{2}trace(\bm\sigma'\bm\sigma)s-\bm\sigma{\bf B}(s)\right\}+\lambda^*\left[a\mathbf X^{\frac{1}{2}'}-b\mathbf X'-u_1-u_2\right]\exp\left\{\frac{1}{2}trace(\bm\sigma'\bm\sigma)s-\bm\sigma{\bf B}(s)\right\}.
	\end{multline*}
	Equation (\ref{sch3}) gives us the cooperative Pareto optimal strategy of two players as
	\begin{align}
	\phi_{p1}^*(s,\mathbf X)&=\left[\frac{k_1\exp(-\zeta s)}{2\left[\lambda^*\exp\left\{\frac{1}{2}trace(\bm\sigma'\bm\sigma)s-\bm\sigma{\bf B}(s)\right\}+\exp(-\zeta s)c_1{\bf X}^{\frac{1}{2}}\right]}\right]^{\frac{2}{3}},\notag\\ 
	\phi_{p2}^*(s,\mathbf X)&=\left[\frac{\a_1^0k_2\exp(-\zeta s)}{2\left[\lambda^*\exp\left\{\frac{1}{2}trace(\bm\sigma'\bm\sigma)s-\bm\sigma{\bf B}(s)\right\}+\exp(-\zeta s)c_2{\bf X}^{\frac{1}{2}}\right]}\right]^{\frac{2}{3}},\notag
	\end{align}
	where $\lambda^*\exp\left\{\frac{1}{2}trace(\bm\sigma'\bm\sigma)t-\bm\sigma{\bf B}(t)\right\}+\exp(-\zeta s)c_1{\bf X}^{\frac{1}{2}}\neq 0$ and $\lambda^*\exp\left\{\frac{1}{2}trace(\bm\sigma'\bm\sigma)t-\bm\sigma{\bf B}(t)\right\}+\exp(-\zeta s)c_2{\bf X}^{\frac{1}{2}}\neq 0$.
	
	{\bf(ii) Pontryagin maximum principle:} From Example $7.2.1$ in \cite{yeung2006} we get cooperative Pareto optimal strategies of two players as,
	\begin{align}
	\phi_{p1}^*(s,\mathbf X)&=\frac{k_1\mathbf X}{4\left[c_1+\exp(-\zeta s)\mathbf X^{\frac{1}{2}}\ \mathcal D_{\mathbf X}W^{\a_1^0}(s,\mathbf X)\right]^2},\notag\\
	\phi_{p2}^*(s,\mathbf X)&=\frac{k_2\mathbf X}{4\left[c_2+\frac{1}{\a_1^0}\exp(-\zeta s)\mathbf X^{\frac{1}{2}}\ \mathcal D_{\mathbf X}W^{\a_1^0}(s,\mathbf X)\right]^2},\notag
	\end{align}
	where for $s\in[0,t]$ the value function is 
	\[
	W^{\a_1^0}(s,\mathbf X)=\exp(-\zeta s)\left[A^{\a_1^0}(s) \mathbf X^{\frac{1}{2}} +B^{\a_1^0}(s)\right],
	\]
	such that, $A^{\a_1^0}(s)$ and $B^{\a_1^0}(s)$ satisfy:
	\[
	\mbox{$\frac{\partial}{\partial s}A^{\a_1^0}(s)$}=\left[r+\mbox{$\frac{1}{8}$}\bm\sigma'\bm\sigma+\mbox{$\frac{1}{2}$}b\right]A^{\a_1^0}(s)-\mbox{$\frac{k_1}{4\left[c_1+\frac{1}{2}A^{\a_1^0}(s)\right]}$}-\mbox{$\frac{\a_1^0 k_2}{4\left[c_2+\frac{1}{2\a_1^0}A^{\a_1^0}(s)\right]}$}
	\]
	and,
	\[
	\mbox{$\frac{\partial}{\partial s}B^{\a_1^0}(s)$}=\zeta B^{\a_1^0}(s)-\mbox{$\frac{1}{2}$}a A^{\a_1^0}(s).
	\]
\end{example}

\begin{remark}
	In Example \ref{ex5.0}, we employ the function $g$ defined as $\lambda^*{\bf X}\exp\left\{\frac{1}{2}\text{trace}(\bm\sigma'\bm\sigma)t-\bm\sigma{\bf B}(t)\right\}$. This function is the result of multiplying the integrating factor with $\lambda^*{\bf X}$. It is important to note that our approach relies on the It\^o lemma rather than the Feynman-Kac lemma, leading to the derivation of a distinct set of equilibria for the two players. 
\end{remark}

Finally, we find optimal strategy of the $\rho^{th}$ firm using a non-cooperative feedback Nash equilibrium. We assume that a firm is rational in decision making and earns more profit at the cost of the profit of the other firms in the market. Hence, Firm $\rho$ seeks to maximize 
\[
\Pi_{\text{N}}(u,t)=\E_0\left\{\int_0^t\pi_\rho[s,\mathbf{X}(s),u_\rho(s),\hat{\mathbf{u}}^*_{-\rho}(s)]ds\biggr|\mathcal F_0^{\bm X}\right\}
\]
with respect to the strategy $u_\rho$ where $\hat{\mathbf{u}}^*_{-\rho}(s)$ is the optimized strategies for firms other than the $\rho^{th}$ firm and $\mathcal F_0^{\bm X}$ is the filtration process of all real-valued measure of market shares starting at time $0$. Following Proposition \ref{fixed} guarantees the existence of a fixed point under the Nash environment. 

\begin{prop}\label{fixed}
	Suppose, the domain of firm $\rho$'s Lagrangian $\mathcal L_\rho$ has a non-empty, convex and compact denoted as $\widetilde\Xi$ such that $\widetilde\Xi\subset [0,t]\times\mathcal U^\rho\times\mathcal X^\rho\subset \mathbb R^{3}$. As $\mathcal L_k: \widetilde\Xi\ra\widetilde\Xi$ is continuous, then for any given positive constants $\zeta$ and $\lambda^*$, there exists a vector of state and control variables $\bar Z_\rho^*=[\bm u^*,\bm x^*]'$ in continouous time $s\in[0,t]$ such that $\mathcal L_\rho$ has a fixed-point in Brouwer sense, where $'$ denotes the transposition of a matrix.	
\end{prop}	

\begin{prop}\label{pr1}
	A non-cooperative Nash optimal solution for maximizing $\Pi_{\text{N}}(u,t)$
	subject to 
	\begin{equation*}
	d\mathbf{X}(s)=\bm{\mu}[s,\mathbf{X}(s),u_\rho(s),\hat{\mathbf{u}}^*_\rho(s)]ds+\bm{\sigma}[s,\mathbf{X}(s),u_\rho(s),\hat{\mathbf{u}}^*_{-\rho}(s)]d\mathbf{B}(s),
	\end{equation*} 
	with initial condition $\mathbf{x}(0)=\mathbf{x}_0$ is the solution of
	\begin{equation}\label{na2.3}
	-\frac{\partial f^\rho[s,\mathbf{X}(s),u_\rho(s),\hat{\mathbf{u}}^*_{-\rho}(s)]}{\partial u_\rho}\ \Psi_s(\mathbf{X})=0,
	\end{equation}
	where $\Psi_s$ is the transition probability defined as 
	\[
	\Psi_s(\mathbf{X})=\exp[-f(s,\mathbf{X},u_\rho(s),\hat{\mathbf{u}}^*_{-\rho}(s))]\Psi_0(\mathbf{X})
	\]
	with initial condition $\Psi_0(\mathbf{X})$ and
	\begin{eqnarray*}
		f^\rho[s,\mathbf{X},u_\rho(s),\hat{\mathbf{u}}^*_{-\rho}(s)] & = & \pi_\rho[s,\mathbf{X}(s),u_\rho(s),\hat{\mathbf{u}}^*_{-\rho}(s)]\\ & &+g^\rho[s,\mathbf{X}(s)]+\frac{\partial}{\partial s}g^\rho[s,\mathbf{X}(s)]\\ & &+\bm{\mu}'[s,\mathbf{X}(s),u_\rho(s),\hat{\mathbf{u}}^*_{-\rho}(s)]\mathcal{D}_{\mathbf{X}}g^\rho[s,\mathbf{X}(s)]\\
		& & +\mbox{$\frac{1}{2}$}\bm{\sigma}'[s,\mathbf{X}(s),u_\rho(s),\hat{\mathbf{u}}^*_{-\rho}(s)]\mathcal{H}_{\mathbf{X}}g^\rho[s,\mathbf{X}(s)]\\ & &\times\bm{\sigma}'[s,\mathbf{X}(s),u_\rho(s),\hat{\mathbf{u}}^*_{-\rho}(s)],
	\end{eqnarray*}
	such that Assumptions \ref{as0}-\ref{as2}, Proposition \ref{existence} and Lemma \ref{l1} hold.
\end{prop}

\begin{remark}
	Since the stochastic differential equation follows Assumptions \ref{as0} and \ref{as0.1}, the non-cooperative Nash equilibrium is locally stable. As our information set is incomplete and imperfect, we cannot get a global stable solution. This equilibrium is not globally unique. The important fact is that a firm chooses a $g$ function so that it solves for the market dynamics and continues to expect same market behavior in the future.  	
\end{remark}

\begin{example}\label{ex6}
	Consider an economy endowed with a renewable resource with $k\geq 2$ firms such as in section $2.6$  in \cite{yeung2006}. We can compare our Nash equilibrium strategy through quantum approach with traditional Pontryagin maximum principle in \cite{yeung2006}. Suppose, $\rho^{th}$ firm's resource extraction in time $s\in[0,t]$ is $u_\rho(s)$ for all $\rho=\{1,2,...,k\}$. Define $\hat{\mathbf{u}}^*_{-\rho}(s)=\sum_{q=1}^{k}u_{q}^*(s)$ where $\rho\neq q$ and $k$-dimensional vector $\mathbf{X}(s)$ is the size of the resource stock at time $s$ such that $\mathbf X(s)>\bm 0$. Under this construction $\rho^{th}$ firm's objective function is
	\[
	\E_0\left\{\int_0^t\exp(-\zeta s)\left[\left(\sum_{q=1}^{k}u_q^*(s)+u_\rho\right)^{-\frac{1}{2}}u_\rho(s)-\frac{c}{\mathbf X^{\frac{1}{2}}(s)}u_\rho(s)\right]ds\biggr|\mathcal F_0^{\bm X}\right\},
	\]
	subject to the resource dynamics
	\[
	d\mathbf X(s)=\left[a\mathbf x^{\frac{1}{2}}(s)-b\mathbf X(s)-\sum_{q=1}^{k}u_q^*(s)-u_\rho(s)\right]ds+\mathbf X(s)\mathbf X'(s)\bm\sigma d\mathbf B(s),
	\]
	where $cu_\rho(s)/[\mathbf X^{\frac{1}{2}}(s)]$ is $\rho^{th}$ firm's cost of resource extraction at time $s$, $\bm\sigma$ is a $k\times m$-dimensional constant diffusion vector component and, vector $\mathbf B(s)$ is an $m$-dimensional Brownian motion. In this model assume $a,b$ and $c$ are the scalars. Since the resource dynamics has non-linear drift and same diffusion coefficient like before, for a constant $\lambda^*$ assume $g^\rho(s,\mathbf X)=\lambda^*{\bf X}\exp\left\{\frac{1}{2}trace(\bm\sigma'\bm\sigma)s-\bm\sigma{\bf B}(s)\right\}$, with the integrating factor $\hat g^\rho=\exp\left\{\frac{1}{2}trace(\bm\sigma'\bm\sigma)s-\bm\sigma{\bf B}(s)\right\}$. Hence, $\frac{\partial}{\partial s}g^\rho(s,\mathbf X)=\frac{1}{2}\lambda^*\ trace(\bm\sigma'\bm\sigma){\bf X}\exp\left\{\frac{1}{2}trace(\bm\sigma'\bm\sigma)s-\bm\sigma{\bf B}(s)\right\}$, $\mathcal D_{\mathbf X}g^\rho(s,\mathbf X)=\lambda^*\exp\left\{\frac{1}{2}trace(\bm\sigma'\bm\sigma)s-\bm\sigma{\bf B}(s)\right\}$, and $\mathcal{H}_{\mathbf X}g^\rho(s,\mathbf X)=0$. Therefore,
	\begin{multline*}
	f^\rho(s,\mathbf X,u_\rho,u_{-\rho}^*)=\exp(-\zeta s)\left[\left(\sum_{q=1}^{k}u_q^*+u_\rho\right)^{-\frac{1}{2}}u_\rho-\frac{c}{\mathbf X^{\frac{1}{2}}}u_\rho\right]+\lambda^*{\bf X}\exp\left\{\frac{1}{2}trace(\bm\sigma'\bm\sigma)s-\bm\sigma{\bf B}(s)\right\}\\
	+\frac{1}{2}\lambda^*\ trace(\bm\sigma'\bm\sigma){\bf X} \exp\left\{\frac{1}{2}trace(\bm\sigma'\bm\sigma)s-\bm\sigma{\bf B}(s)\right\}+\lambda^*\left[a\mathbf X^{\frac{1}{2}}(s)-b\mathbf X(s)-\sum_{q=1}^{k}u_q^*(s)-u_\rho(s)\right]\\
	\times\exp\left\{\frac{1}{2}trace(\bm\sigma'\bm\sigma)s-\bm\sigma{\bf B}(s)\right\}.
	\end{multline*}
	Finally, Equation (\ref{na2.3}) implies the feedback Nash Equilibrium as
	\begin{align}
	\phi_{NQ}^{\rho*}(s,\mathbf X)=2\exp(\zeta s)\left(\sum_{q=1}^{k}u_q^*\right)^{\frac{3}{2}}\left[\exp(-\zeta s)\left[\left(\sum_{q=1}^{k}u_q^*\right)^{-\frac{1}{2}}-\frac{c}{\mathbf x^{\frac{1}{2}}}\right]-\lambda^*\exp\left\{\frac{1}{2}trace(\bm\sigma'\bm\sigma)s-\bm\sigma{\bf B}(s)\right\}\right].\notag
	\end{align}
	From section $2.6$ of \cite{yeung2006} we know, the feedback Nash equilibrium from Pontryagin maximum principle is
	\begin{align}
	\phi_{NP}^{\rho*}(s,\mathbf X)&=\frac{\mathbf X(2k-1)^2}{2\left[\sum_{q=1}^k\left(c+\exp(\zeta s)\mathcal D_{\mathbf X}V^q\mathbf X^{\frac{1}{2}}\right)\right]} \left\{\sum_{q=1}^k\left[c+\frac{\mathcal D_{\mathbf X}V^q\mathbf X^{\frac{1}{2}}}{\exp(-\zeta s)}\right]-\left(k-\frac{3}{2}\right)\left[c+\frac{\mathcal D_{\mathbf X}V^\rho\mathbf X^{\frac{1}{2}}}{\exp(-\zeta s)}\right]\right\},\notag
	\end{align}
	where $V^\rho$ and $V^q$ are the value function of firms $\rho$ and $q$ with their gradients $\mathcal D_{\mathbf X}V^\rho$ and $\mathcal D_{\mathbf X}V^q$ respectively. By Corollary $2.6.1$ in \cite{yeung2006} HJB equation has a solution
	\[
	V^\rho(s,\mathbf X)=\exp(-\zeta s)\left[A(s)\mathbf X^{\frac{1}{2}}+B(s)\right],
	\]
	where $A(s)$ and $B(s)$ satisfies,
	\begin{align}
	\mbox{$\frac{\partial}{\partial s}$}A(s)&=\left[\zeta+\mbox{$\frac{1}{8}$}\bm\sigma'\bm\sigma-\mbox{$\frac{b}{2}$}\right]A(s)-\frac{2k-1}{2k^2}\left[c+\mbox{$\frac{1}{2}$}A(s)\right]^{-1}+\frac{c(2k-1)^2}{4k^3}\left[c+\mbox{$\frac{1}{2}$}A(s)\right]^{-2}+\frac{(2k-1)^2A(s)}{8k^2\left[c+\mbox{$\frac{1}{2}$}A(s)\right]^2},\notag\\\mbox{$\frac{\partial}{\partial s}$}B(s)&=\zeta B(s)-\mbox{$\frac{1}{2}$}a A(s).\notag
	\end{align}
\end{example}

\section{Numerical results.}

This section provides a systematic numerical investigation of the proposed path integral control framework. While the preceding sections establish the theoretical structure and equilibrium properties of the model, the present analysis evaluates its computational performance and robustness under alternative specifications. In particular, we examine the behavior of the algorithm across varying state dimensions, perturbations of key structural parameters, and nonlinear market dynamics. The objective of these experiments is twofold. First, we assess the numerical stability and scalability of the method relative to the dimensionality of the state space, thereby addressing the computational considerations discussed earlier. Second, we evaluate the sensitivity of the resulting equilibrium strategies and market-share trajectories to variations in diffusion intensity, discounting, and penalization parameters. Together, these experiments provide quantitative evidence that the path integral approach remains well-behaved beyond the baseline examples and supports its applicability in more general stochastic differential game environments.

\begin{figure}
	\centering
	\includegraphics[width=0.75\textwidth]{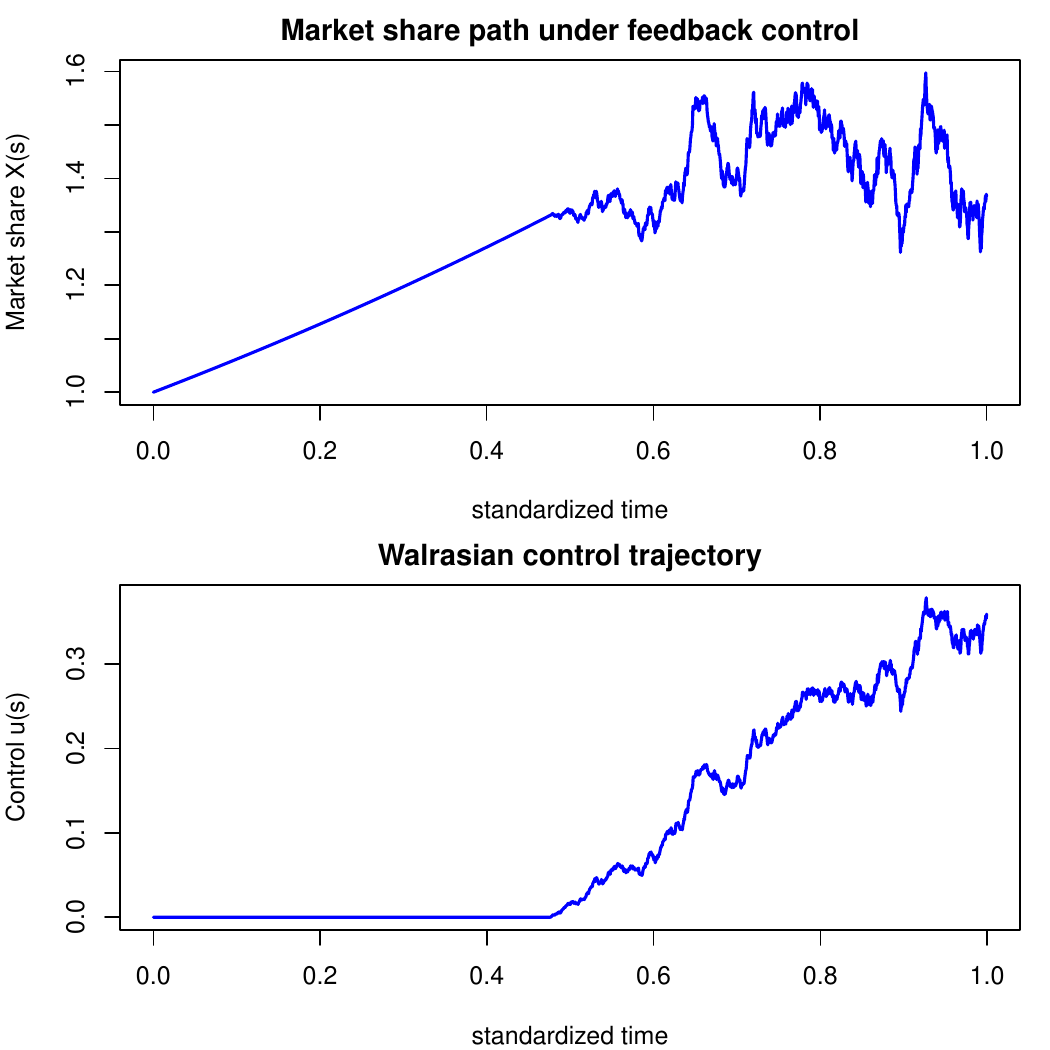}
	\caption{Simulated market-share trajectory and corresponding Walrasian feedback control under the parameter configuration of Example~\ref{ex2}. The upper panel reports the evolution of $X(s)$ over standardized time, while the lower panel displays the optimal advertising intensity $u(s)=\phi_w^*(s,X(s))$.}
	\label{fig:walrasian_simulation}
\end{figure}

\begin{figure}
	\centering
	\includegraphics[width=0.85\textwidth]{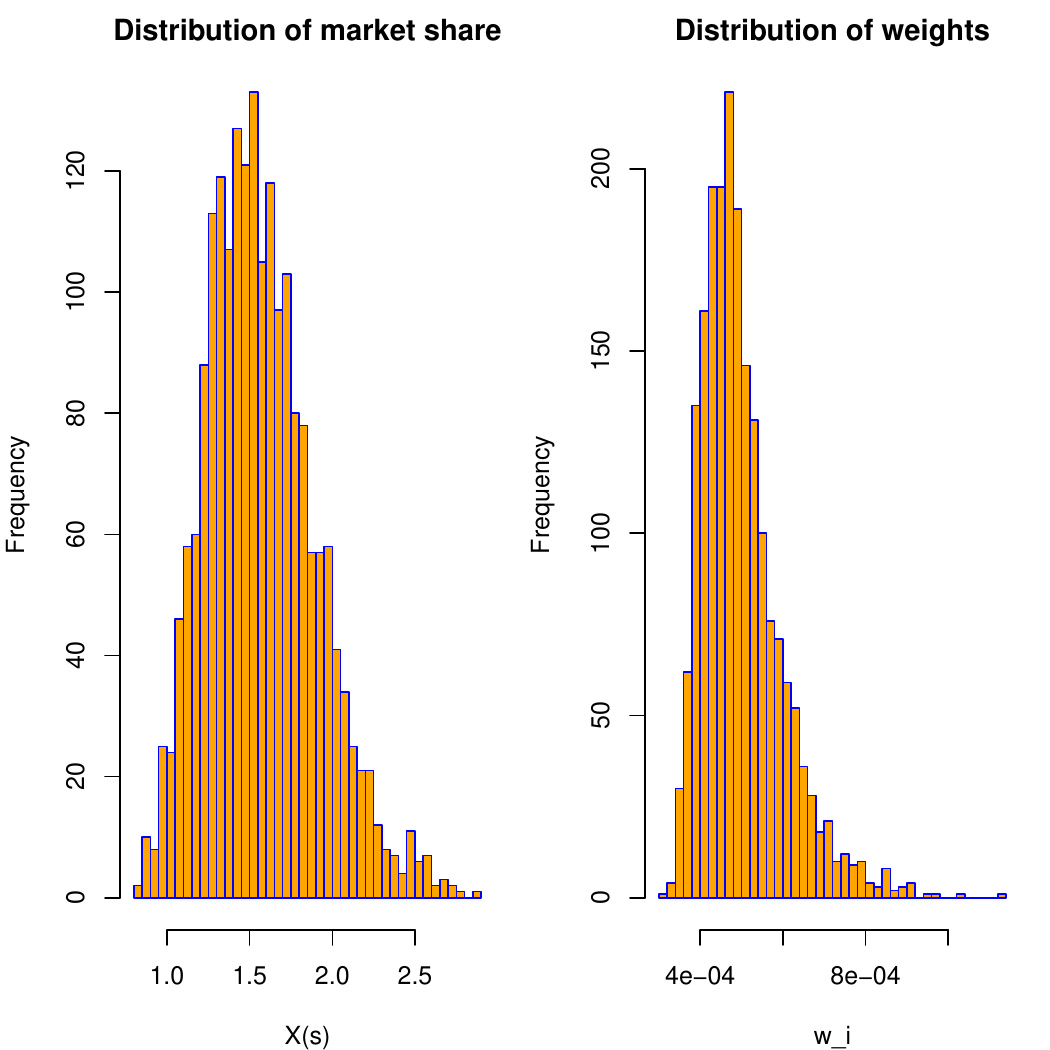}
	\caption{Monte Carlo simulation for Example~\ref{ex2}. The left panel shows the empirical distribution of the terminal market share $X(s)$ under the Walrasian feedback strategy, while the right panel reports the distribution of exponential importance weights used in the path integral reweighting scheme.}
	\label{fig:montecarlo_diagnostics}
\end{figure}

\begin{figure}
	\centering
	\includegraphics[width=0.85\textwidth]{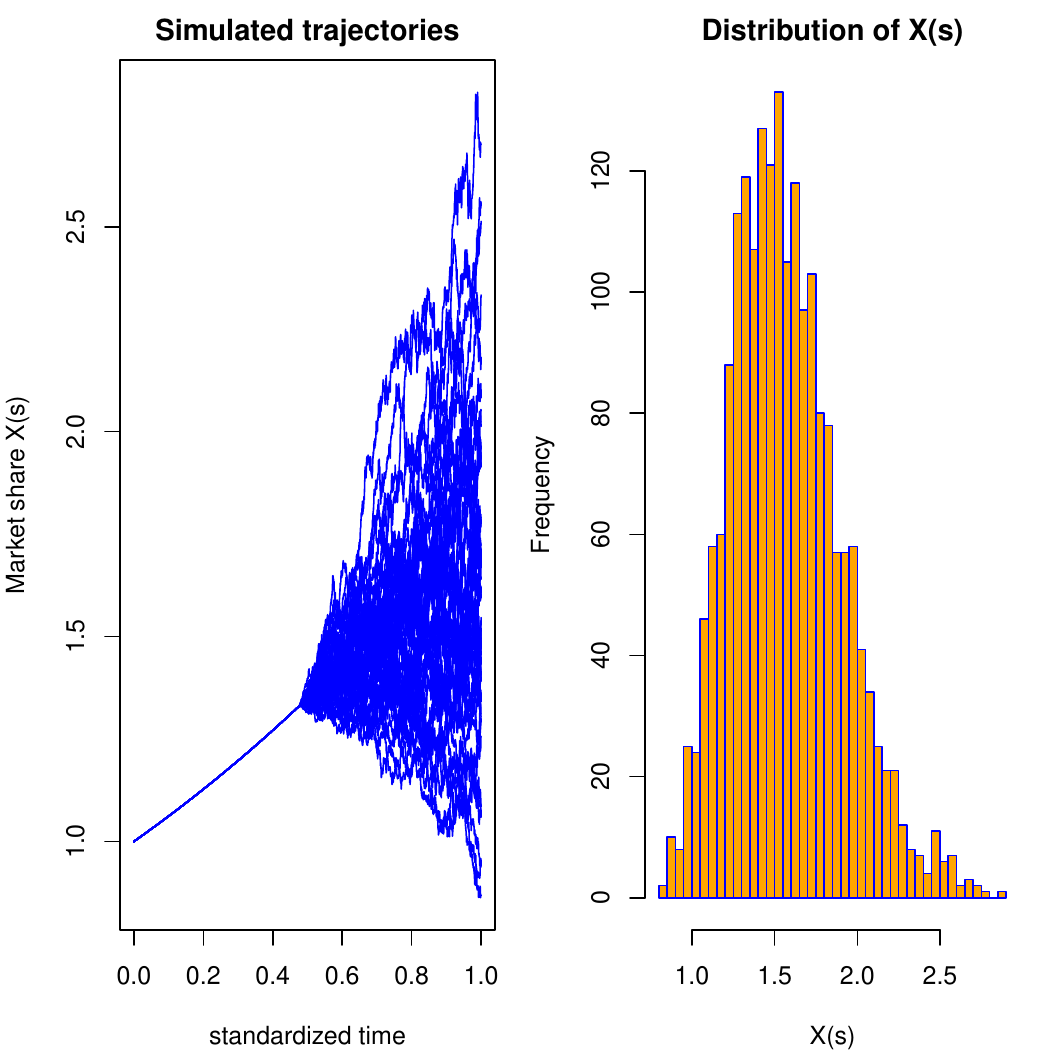}
	\caption{Monte Carlo simulation for Example~\ref{ex2}. The left panel displays a subset of simulated market-share trajectories under the Walrasian feedback control, plotted over standardized time. The right panel reports the empirical distribution of terminal market shares, illustrating the dispersion generated by stochastic diffusion and endogenous advertising dynamics.}
	\label{fig:montecarlo_trajectories}
\end{figure}

Figure \ref{fig:walrasian_simulation} illustrates a representative simulated trajectory of market share under the Walrasian feedback advertising policy derived in Example~\ref{ex2}. The upper panel reports the evolution of the state variable $X(s)$ over standardized time $s/t \in [0,1]$, while the lower panel displays the corresponding optimal advertising expenditure $u(s)=\phi_w^*(s,X(s))$. The trajectory reflects the interaction between deterministic growth, endogenous advertising control, and stochastic diffusion driven by $\sqrt{\sigma u(s)}\,dB(s)$. The standardized time axis facilitates comparison across alternative horizon lengths and parameter configurations considered in subsequent robustness experiments. Figures \ref{fig:walrasian_simulation}-\ref{fig:montecarlo_trajectories} based on the values from Table~\ref{tab:walrasian_params}.

\begin{table}[H]
	\centering
	\caption{Simulation Parameters for Example~\ref{ex2}}
	\begin{tabular}{lll}
		\hline
		Parameter & Description & Value \\ 
		\hline
		$a$ & Intrinsic growth rate of market share & $0.30$ \\
		$\sigma$ & Diffusion intensity parameter & $0.50$ \\
		$p$ & Constant price coefficient & $1.00$ \\
		$c$ & Marginal advertising cost parameter & $0.80$ \\
		$\zeta$ & Continuous-time discount rate & $0.20$ \\
		$\lambda^*$ & Scaling parameter in $g(s,X)$ transformation & $0.60$ \\
		$x_0$ & Initial market share & $1.00$ \\
		$t$ & Time horizon & $1.00$ \\
		$\Delta s$ & Time discretization step & $0.001$ \\
		$N$ & Number of discretization steps & $2000$ \\
		\hline
	\end{tabular}
	\label{tab:walrasian_params}
\end{table}

Figure \ref{fig:montecarlo_diagnostics} reports Monte Carlo diagnostics for Example~\ref{ex2} based on $n=2000$ simulated paths. The left panel displays the empirical distribution of the terminal market share $X(s)$ under the Walrasian feedback control, while the right panel shows the distribution of exponential path weights used in the path integral reweighting scheme. The terminal distribution reflects the combined impact of endogenous advertising effort and stochastic diffusion, whereas the weight dispersion provides information about effective sample size and numerical stability of the Monte Carlo approximation. Together, these diagnostics illustrate the robustness and computational behavior of the proposed path integral framework.

The Monte Carlo experiment (see Figure \ref{fig:montecarlo_trajectories}) for Example~\ref{ex2} is conducted to evaluate the statistical behavior and robustness of the Walrasian feedback control under stochastic market dynamics. Using $n=2000$ simulated paths over a time horizon $t=1$ with discretization step $\Delta s = 0.001$, we generate sample trajectories of the controlled diffusion process described in Equation \eqref{dynamic1}, where the structural parameters are set to $a=0.30$, $\sigma=0.50$, $p=1.00$, $c=0.80$, $\zeta=0.20$, $\lambda^{*}=0.60$, and initial market share $x_0=1.00$. The optimal advertising intensity $u(s)=\phi_w^*(s,X(s))$ is computed at each time step using the derived feedback expression, and discounted profits are accumulated along each trajectory. The left panel of the resulting figure displays a subset of simulated market-share paths (standardized to the unit interval for comparability), illustrating the dispersion induced by the stochastic component and the stabilizing role of the endogenous control. The right panel presents the empirical distribution of terminal market shares $X(s)$, summarizing the cross-sectional variability generated by the diffusion intensity $\sigma$ and the growth parameter $a$. In addition to reporting the mean and selected quantiles of $X(s)$ and discounted profits, we compute exponential importance weights $w_i \propto \exp(\varepsilon_w \pi_i)$ with $\varepsilon_w=1.0$ to assess effective sample size (ESS) and numerical stability within the path integral framework. The resulting ESS ratio provides a diagnostic measure of weight concentration, confirming that under the chosen parameter configuration the Monte Carlo approximation remains well-conditioned and does not suffer from severe degeneracy. Together, these simulations demonstrate that the feedback strategy yields economically meaningful trajectories and stable numerical performance under realistic stochastic perturbations.

\begin{figure}
	\centering
	\includegraphics[width=0.90\textwidth]{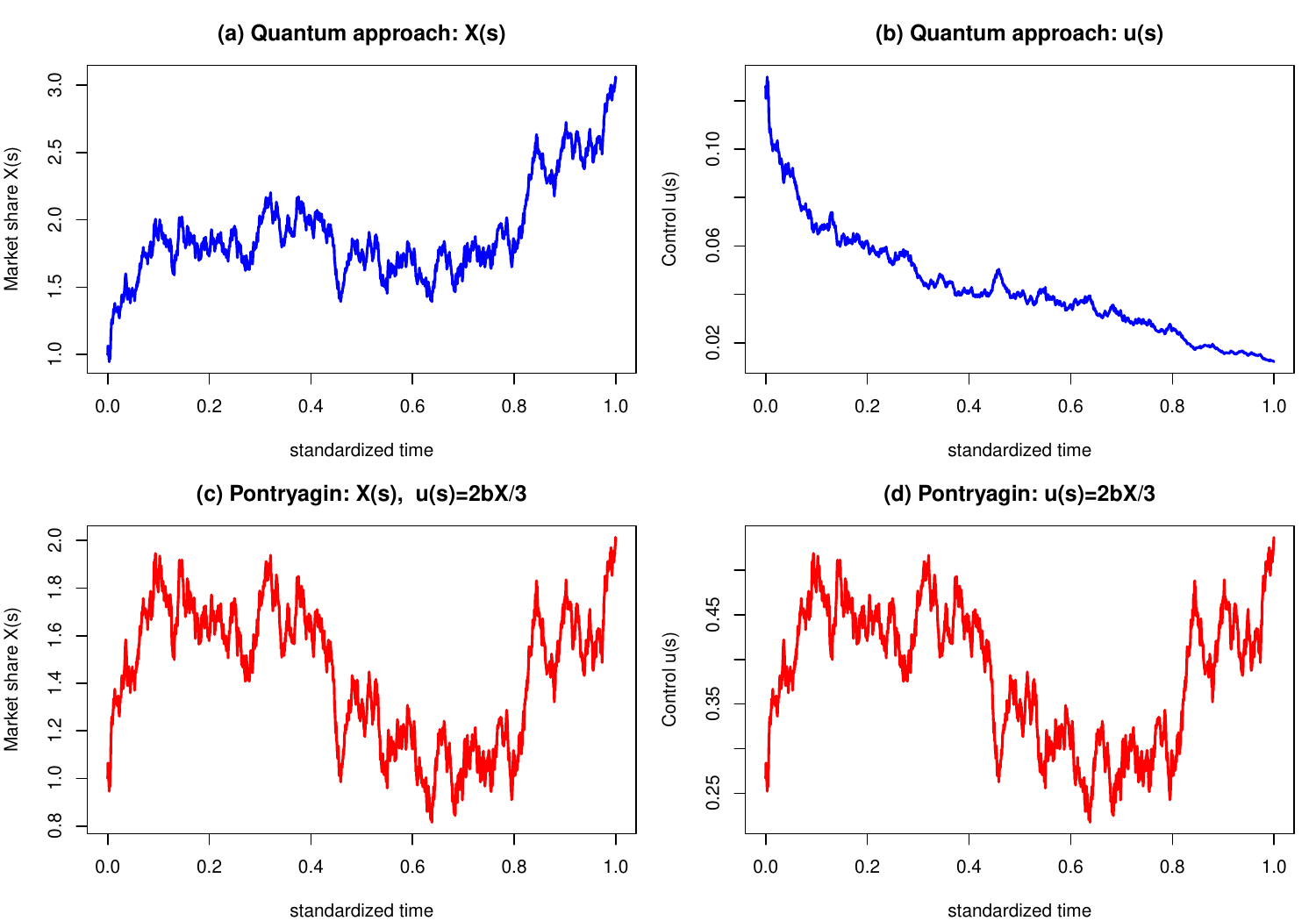}
	\caption{Comparison of optimal strategies in Example~\ref{ex3}. Panels (a) and (b) report the simulated market-share trajectory $X(s)$ and control path $u(s)$ under the quantum (path integral) approach obtained from the cubic feedback equation. Panels (c) and (d) display the corresponding dynamics under the Pontryagin maximum principle with linear feedback rule $u(s)=2bX(s)/3$. All simulations use identical stochastic shocks, highlighting structural differences between nonlinear cubic feedback and linear Pontryagin approach.}
	\label{fig:ex3_compare}
\end{figure}

Figure~\ref{fig:ex3_compare} presents a four-panel comparison between the quantum (path integral/Wick-rotated) approach and the classical Pontryagin maximum principle for the Walrasian firm described in Example~\ref{ex3}. The upper panels display the simulated market-share trajectory $X(s)$ and corresponding control $u(s)$ under the quantum strategy obtained by solving the cubic equation derived from Proposition~\ref{p1}, while the lower panels report the dynamics under the Pontryagin feedback rule $u(s)=2bX(s)/3$. All simulations are conducted under the structural and numerical configuration summarized in Table~\ref{tab:ex3_params}, including growth parameter $b$, marginal cost $c$, discount factor $\zeta$, auxiliary scaling parameter $\lambda^{*}$, and quadratic revenue specification $R(X)=pX^2$. The comparison highlights how the cubic quantum control produces a nonlinear feedback adjustment relative to the linear Pontryagin rule, leading to distinct transient dynamics despite identical stochastic shocks and diffusion intensity $\sqrt{2b}$. Together, the panels illustrate both the qualitative differences in feedback structure and the quantitative implications for stochastic market-share evolution.

\begin{table}
	\centering
	\caption{Simulation Parameters for Example~\ref{ex3}}
	\label{tab:ex3_params}
	\begin{tabular}{lll}
		\hline
		Parameter & Description & Value \\
		\hline
		$b$ & Deterministic growth parameter & $0.40$ \\
		$c$ & Marginal cost coefficient & $0.80$ \\
		$\zeta$ & Continuous-time discount rate & $0.20$ \\
		$\lambda^{*}$ & Auxiliary scaling parameter in $g(s,X)$ & $0.60$ \\
		$p$ & Revenue curvature parameter ($R(X)=pX^2$) & $1.00$ \\
		$x_0$ & Initial market share & $1.00$ \\
		$t$ & Time horizon & $1.00$ \\
		$\Delta s$ & Time discretization step & $0.001$ \\
		Diffusion & Noise coefficient & $\sqrt{2b}$ \\
		\hline
	\end{tabular}
\end{table}

\begin{figure}
	\centering
	\includegraphics[width=0.95\textwidth]{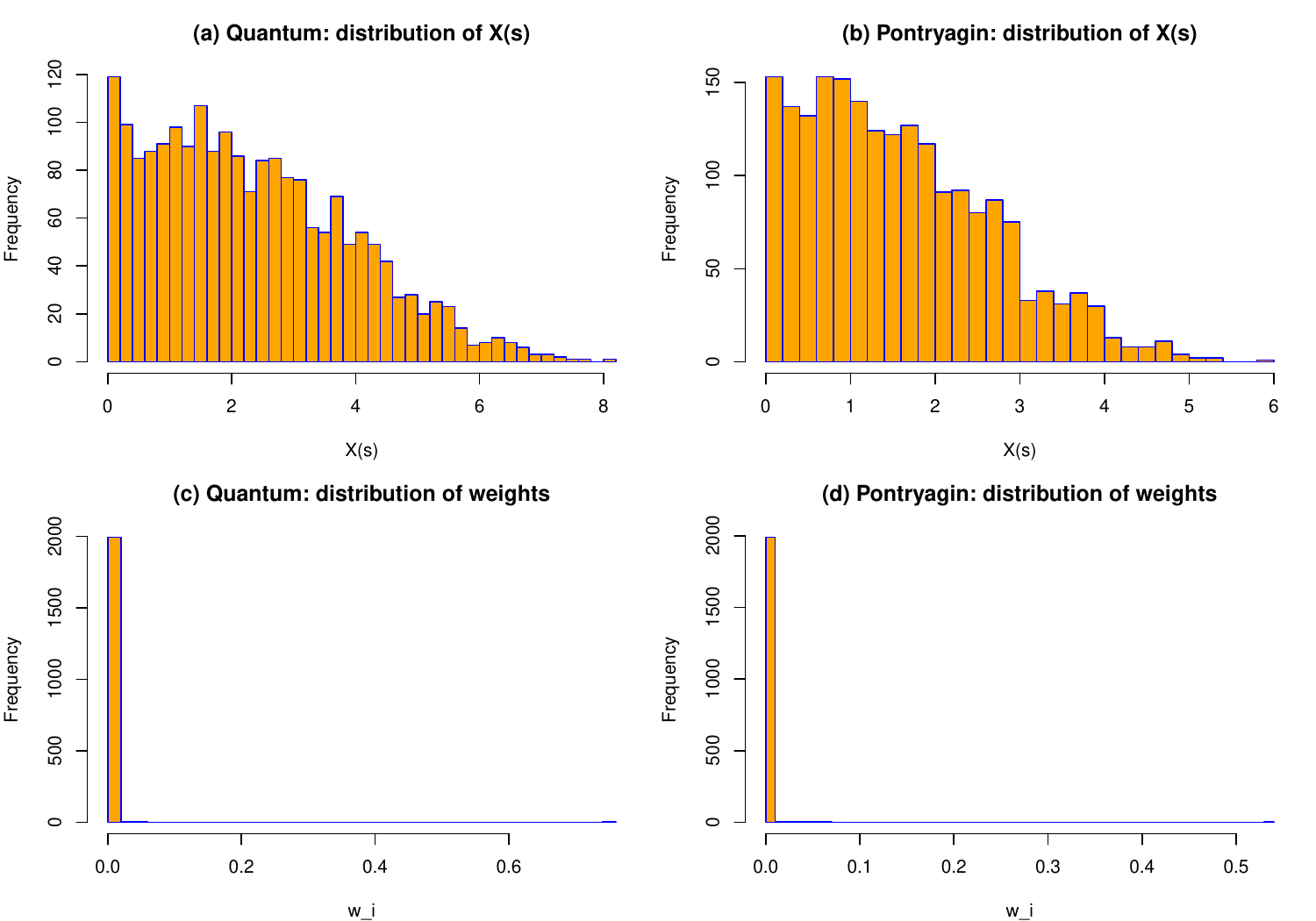}
	\caption{Monte Carlo comparison of the quantum and Pontryagin strategies in Example~\ref{ex3}. Panels (a) and (b) display the empirical distributions of terminal market share $X(t)$ under the cubic quantum feedback and the linear Pontryagin rule, respectively. Panels (c) and (d) show the corresponding distributions of exponential path-integral weights, illustrating differences in weight concentration and effective sample behavior.}
	\label{fig:ex3_mc_compare}
\end{figure}

Figure~\ref{fig:ex3_mc_compare} describes the Monte Carlo comparison between the quantum (path integral/Wick-rotated) control and the Pontryagin maximum principle for Example~\ref{ex3}. The simulations are conducted under the structural configuration $b=0.40$, $c=0.80$, $\zeta=0.20$, $\lambda^{*}=0.60$, quadratic revenue $R(X)=pX^2$ with $p=1$, initial market share $x_0=1$, time horizon $t=1$, and discretization step $\Delta s=0.001$, yielding $N=t/\Delta s$ time steps per trajectory. For each approach, $n_{\text{paths}}=2000$ stochastic paths are generated using the Euler–Maruyama scheme applied to the diffusion $dX(s)=[bX(s)-u(s)]ds+\sqrt{2b}\,dB(s)$, where identical Brownian increments are used path-by-path to ensure a fair structural comparison. The upper-left and upper-right panels display the empirical distributions of terminal market share $X(t)$ under the cubic quantum feedback rule and the linear Pontryagin rule $u(s)=2bX(s)/3$, respectively, illustrating how nonlinear feedback curvature alters dispersion and tail behavior. The lower panels report the corresponding path-integral importance weights $w_i\propto \exp(\varepsilon_w \Pi_i)$, where $\Pi_i$ denotes total discounted profit and $\varepsilon_w=1$ governs weight concentration; these histograms visualize sampling efficiency and concentration through ESS. Collectively, Figure~\ref{fig:ex3_mc_compare} demonstrates how differences in feedback structure translate into distinct terminal distributions and weight concentration patterns, thereby providing a quantitative robustness assessment of the proposed quantum approach relative to the classical Pontryagin benchmark.

\begin{figure}
	\centering
	\includegraphics[width=0.95\textwidth]{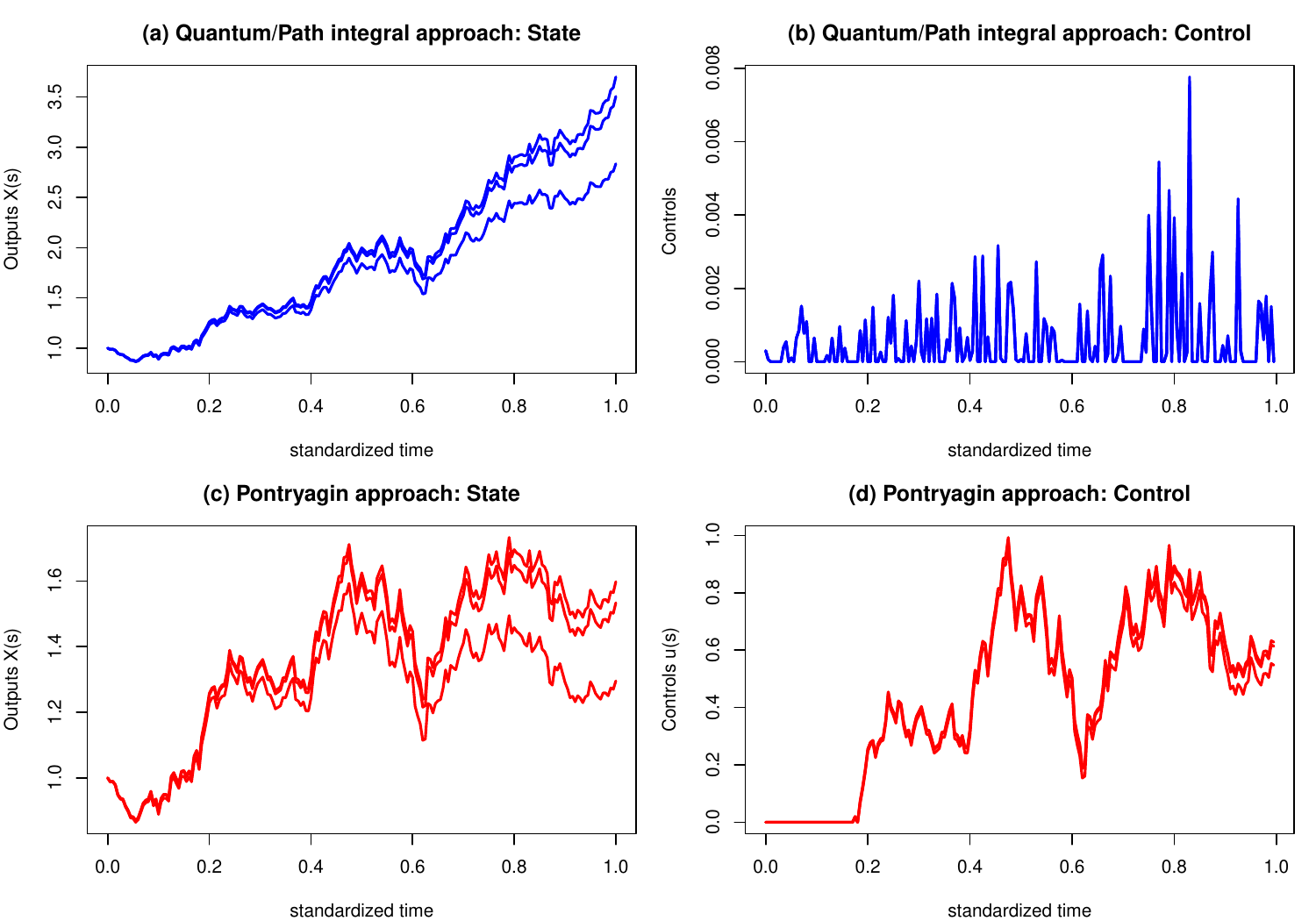}
	\caption{Stability comparison between the path integral control and the Pontryagin maximization principle in Example~\ref{ex5}. Panels (a) and (b) display the state trajectories and control paths obtained from the Monte Carlo path integral controller, while panels (c) and (d) show the corresponding results generated using the Pontryagin feedback strategy. }
	\label{fig:ex5_compare}
\end{figure}

Figure~\ref{fig:ex5_compare} provides a four-panel stability comparison between the proposed quantum/path integral control implementation and the classical Pontryagin Pareto rule for the cooperative $k$-firm environment in Example~\ref{ex5}. Table~\ref{tab:ex5_params} is used for the parameters. In panels (a)–(b), the blue curves report the closed-loop evolution of the vector state $\mathbf{X}(s)=(X_1(s),\dots,X_k(s))$ and the corresponding control vector $\mathbf{u}(s)=(u_1(s),\dots,u_k(s))$ generated by a robust Monte Carlo PI-style controller, while panels (c)-(d) display the analogous trajectories under the red Pontryagin feedback rule derived in the example. For both approaches, the underlying market dynamics are simulated in discrete time via an \emph{Euler–Maruyama} approximation to the multiplicative diffusion system $d\mathbf{X}(s)=[\bm{\mu}(\mathbf{X}(s))-\mathbf{u}(s)]ds+\Sigma(\mathbf{X}(s))\,dB(s)$, where the nonlinear drift is constructed from the symmetric interaction matrix $\mathbf{A}$ as $\bm{\mu}(\mathbf{X})=(\mathbf{A}\mathbf{X})\odot \mathbf{X}$ (componentwise product) and the diffusion channel is $\Sigma(\mathbf{X})=\sigma_0\,\mathbf{X}$ under a one-factor Brownian driver.

\begin{table}
	\centering
	\caption{Parameter values used to generate Figure~\ref{fig:ex5_compare} (Example~\ref{ex5}).}
	\label{tab:ex5_params}
	\begin{tabular}{lll}
		\hline
		Parameter & Description & Value \\
		\hline
		$k$ & Number of firms & $3$ \\
		$\zeta$ & Discount rate & $0.20$ \\
		$p$ & Price parameter & $1.00$ \\
		$c$ & Marginal cost parameter & $0.80$ \\
		$\omega_1$ & Cross-firm reward weight in revenue & $0.30$ \\
		$\omega_2$ & Cross-firm interaction weight in costs & $0.20$ \\
		$\alpha_\rho$ & Cooperative Pareto weights & $\alpha_\rho=1/3,\ \rho=1,2,3$ \\
		$\sigma_0$ & Diffusion intensity in $\Sigma(\mathbf{X})=\sigma_0\mathbf{X}$ & $0.25$ \\
		$\gamma$ & Robustness parameter (temperature selection) & $0.50$ \\
		$M$ & Number of Monte Carlo rollouts per update & $800$ \\
		$H$ & Receding-horizon lookahead (steps) & $60$ \\
		$\kappa_u$ & Control update scaling & $1.0$ \\
		$u_{\min}$ & Lower bound on control & $0$ \\
		$u_{\max}$ & Upper bound on control & $5$ \\
		$x_0$ & Initial state (each firm) & $(1,1,1)'$ \\
		$t$ & Time horizon & $1.0$ \\
		$\Delta s$ & Time step size & $0.01$ \\
		$\mathbf{A}$ & Symmetric interaction matrix & $\begin{pmatrix}
		0.15&0.05&0.02\\
		0.05&0.12&0.04\\
		0.02&0.04&0.10
		\end{pmatrix}$ \\
		\hline
	\end{tabular}
\end{table}

The algorithm for the path integral controller is implemented in a receding-horizon fashion: at each decision time $s_n=n\Delta s$ with current state $\mathbf{X}(s_n)$, the controller samples $M$ independent disturbance sequences $\left\{\varepsilon^{(i)}_{n:h}\right\}_{i=1}^M$ over a lookahead window of length $H$ and rolls out $M$ hypothetical future trajectories under a baseline control (set to zero in the code) using the same discrete-time state update, each rollout accumulates a discounted running cost $$J^{(i)}=\sum_{h=1}^H q\left[s_{n+h},\mathbf{X}^{(i)}(s_{n+h}),\mathbf{u}_{\text{base}}\right]\Delta s,$$ where $q=-\exp(-\zeta s)\sum_{\rho=1}^k \alpha_\rho\left\{p[X_\rho+\omega_1\sum_{\tilde\rho\neq\rho}X_{\tilde\rho}]u_\rho-cX_\rho[u_\rho^2+\omega_2\sum_{\tilde\rho\neq\rho}u_{\tilde\rho}^2]\right\}$ is the negative of the cooperative Pareto profit integrand. A robust parameter $\hat\theta$ is then computed by minimizing the entropic-robust objective given as 
\[
\gamma\theta+\theta\ln\left[\frac{1}{M}\sum_{i=1}^M\exp\left(J^{(i)}/\theta\right)\right],
\] over a grid of $\theta>0$,
producing normalized importance weights $r^{(i)}\propto \exp(J^{(i)}/\hat\theta)$. Finally, the PI control update at time $s_n$ is obtained by pushing the weighted disturbance through the diffusion channel, $\mathbf{u}(s_n)=\kappa_u\,\Sigma(\mathbf{X}(s_n))\big(\sum_{i=1}^M r^{(i)}\varepsilon^{(i)}_{n}\big)\sqrt{\Delta s}$, followed by admissibility enforcement through componentwise truncation $u_\rho(s_n)\in[u_{\min},u_{\max}]$ and a nonnegativity constraint. In contrast, the Pontryagin benchmark computes $\mathbf{u}(s_n)$ deterministically from the closed-form Pareto feedback reported in Example~\ref{ex5}, namely
\[
u_\rho(s_n)=\max\!\left\{0,\frac{(p\alpha_\rho-1)\left[X_\rho+\omega_1\sum_{\tilde\rho\neq\rho}X_{\tilde\rho}\right]+2cX_\rho\,\mathbf{X}'\mathbf{A}\mathbf{X}}{2cX_\rho}\right\},
\]
with the same upper bound imposed for numerical comparability). To ensure that the visual comparison isolates differences attributable to feedback structure rather than stochastic realization, both closed-loop systems are driven by the same Brownian increment $\Delta B_n\overset{iid}{\sim}\mathcal{N}(0,\Delta s)$ at each time step. Under the parameter configuration summarized in Table~\ref{tab:ex5_params}, the four panels jointly demonstrate stability in the sense that neither approach induces explosive behavior in $\mathbf{X}(s)$ or $\mathbf{u}(s)$ over the horizon $[0,t]$; however, the path integral controller exhibits a distinctly sample-driven, adaptive adjustment pattern because the control is computed from the weighted ensemble of simulated disturbances, whereas the Pontryagin rule produces a smoother deterministic feedback response governed by the curvature term $\mathbf{X}'\mathbf{A}\mathbf{X}$ and the cross-firm reward coupling through $\omega_1$. In the similar fashion, the above images can be created for the rest of the examples of this paper.

\section{Proofs.}

\subsection{Proof of Proposition \ref{existence}}

For $\epsilon>0$ and for firm $\rho$ define $\hat{l}_\rho,l_\rho:[0,t]\times\mathcal X\times\mathcal{U}\ra\mathbb{R}$ by
\[
\hat{\mathbf{u}}_\rho\mapsto\hat{l}_\rho(s,\mathbf{x}^*,\hat{\mathbf{u}}_\rho)=\hat{l}_\rho(s,\mathbf{x}^*,u_\rho,\hat{\mathbf{u}}_{-\rho})=\max_{u_\rho\in \mathcal U^\rho\subset\mathcal U}\E_0\left\{\int_0^t\pi_\rho\left[s,\mathbf{X}(s),\hat{\mathbf{u}}_\rho(s)\right]ds\biggr|\mathcal{F}_0^X\right\},
\]
\[
\hat{\mathbf{u}}_\rho\mapsto l_\rho(s,\mathbf{x}^*,\hat{\mathbf{u}}_\rho)=\E_0\left\{\int_0^t\pi_\rho\left[s,\mathbf{X}(s),\hat{\mathbf{u}}_\rho(s)\right]ds\biggr|\mathcal{F}_0^X\right\}-\hat{l}_\rho(s,\mathbf{x}^*,\hat{\mathbf{u}}_\rho)+\epsilon,
\]
and let $O_\epsilon=\{\mathbf{u}^1\in\mathcal{U}\big|l_\rho(s,\mathbf{x}^*,\hat{\mathbf{u}}_\rho)\geq 0,\ \forall\rho=1,...,k\}$. As for each $\rho=1,...,k$, the uniform continuity of \[
\E_0\left\{\int_0^t\pi_\rho\left[s,\mathbf{X}(s),\hat{\mathbf{u}}_\rho(s)\right]ds\biggr|\mathcal{F}_0^X\right\}
\]
implies continuity of $\hat{l}_\rho$, therefore, we conclude that the set $O_\epsilon$ is compact.

Now we will discuss the conditions of Lemma \ref{l0.1}. For continuous functions $l_1,...,l_k$, the filtration $\mathcal{F}_0^X$ and for all $\rho=1,..,k$ following conditions hold:\\
$(a)$ $u_\rho\mapsto l_\rho(s,\mathbf{x}^*,u_\rho,\hat{\mathbf{u}}_{-\rho})=\E_0\left\{\int_0^t\pi_\rho\left[s,\mathbf{X}(s),\hat{\mathbf{u}}_\rho(s)\right]ds\biggr|\mathcal{F}_0^X\right\}-\max_{u_\rho\in \mathcal U^\rho\subset\mathcal U}\E_0\left\{\int_0^t\pi_\rho\left[s,\mathbf{X}(s),\hat{\mathbf{u}}_\rho(s)\right]ds\biggr|\mathcal{F}_0^X\right\}+\epsilon$ is a quasi-concave on $\mathcal{U}_\rho$ for all $\hat{\mathbf{u}}_\rho\in\mathcal{U}^\rho$.\\
$(b)$ For any $\hat{\mathbf{u}}_{-\rho}\in\mathcal{U}^\rho$ there exists a strategy $u_\rho\in\mathcal{U}_\rho$ such that $l_\rho(s,\mathbf{x}^*,u_\rho,\hat{\mathbf{u}}_{-\rho})>0$. Now with above conditions Lemma \ref{l0.1} implies the existence of a set of optimal strategies $\left\{u_\rho^*,\hat{\mathbf{u}}_{-\rho}^*\right\}\in\mathcal{U}^\rho\subset\mathcal U$ such that for all $\rho=1,...,k$, $l_\rho\left(s,\mathbf{x}^*,u_\rho^*,\hat{\mathbf{u}}_{-\rho}^*\right)>0$. Therefore, $O_\epsilon$ is non-empty.

Now let us consider the decreasing family of $\{O_\epsilon|\epsilon>0\}$ compact nonempty sets and a point $\{u_\rho^*,\hat{\mathbf{u}}_{-\rho}^*\}=\hat{\mathbf{u}}_\rho$ in the intersection $O=\bigcap\{O_\epsilon|\epsilon>0\}$. By the definition of $O$, for all $\epsilon>0$,
\[
J\left(s,\mathbf{x}^*,u_\rho^*,\hat{\mathbf{u}}_{-\rho}^*\right)\geq \max_{u_\rho\in \mathcal U^\rho\subset\mathcal U}\E_0\left\{\int_0^t\pi_\rho\left[s,\mathbf{X}(s),\hat{\mathbf{u}}_\rho(s)\right]ds\biggr|\mathcal{F}_0^X\right\}-\epsilon,
\]
for all $\rho=1,...,k$. Thus,
\[
J\left(s,\mathbf{x}^*,u_\rho^*,\hat{\mathbf{u}}_{-\rho}^*\right)= \max_{u_\rho\in \mathcal U^\rho\subset\mathcal U}\E_0\left\{\int_0^t\pi_\rho\left[s,\mathbf{X}(s),\hat{\mathbf{u}}_\rho(s)\right]ds\biggr|\mathcal{F}_0^X\right\}.
\]
Therefore, the optimal strategy set $\{u_\rho^*,\hat{\mathbf{u}}_{-\rho}^*\}=\hat{\mathbf{u}}_\rho^*$ is a Nash equilibrium of the system $J_1,...,J_k$. $\square$

\subsection{Proof of Proposition \ref{p1}}

The arguments here are based on the use of the quantum Lagrangian action function. Further details are given in the Appendix. Equation (\ref{mkt}) implies
$\Delta X(s)=X(s+ds)-X(s)=\mu[s,X(s),u(s)]ds+\sigma[s,X(s),u(s)]dB(s)$. Following \cite{chow1996} from Equation (\ref{w6}), the Euclidean action function is, 
\begin{equation*}
\mathcal{A}_{0,t}(X)=\int_0^t\E_s\{\pi[s,X(s),u(s)]ds+\lambda(s) [\Delta X(s)-\mu[s,X(s),u(s)]ds-\sigma[s,X(s),u(s)]dB(s)]\}.
\end{equation*}
Let $\varepsilon>0$, and for a normalizing constant $L_\varepsilon>0$ from Lemma \ref{l1} in the Appendix, let
\begin{equation}\label{w16}
\Psi_{s,s+\varepsilon}(X)=\frac{1}{L_\varepsilon} \int_{\Omega}\exp[-\varepsilon \mathcal{A}_{s,s+\varepsilon}(X)]\Psi_s(X)dX(s),
\end{equation}
where $\Psi_s(X)$ is the value of the transition probability at time $s$ and state $X(s)$ with the initial condition $\Psi_0(X)=\Psi_0$.

Fubini's Theorem implies that the action function on time interval $[s,s+\varepsilon]$ is
\[
\mathcal{A}_{s,s+\varepsilon}(X)=
\E_s\int_{s}^{s+\varepsilon}\pi[\nu,X(\nu),u(\nu)]d\nu+g[\nu+\Delta \nu,X(\nu)+\Delta X(\nu)].
\]
where 
\begin{equation*}
g[\nu+\Delta\nu,X(\nu)+\Delta X(\nu)]\approx\lambda(\nu)[\Delta X(\nu)d\nu-\mu[\nu,X(\nu),u(\nu)]d\nu-\sigma[\nu,X(\nu),u(\nu)]dB(\nu)+o(1).
\end{equation*}
This conditional expectation is valid when the strategy $u(\nu)$ is determined at time $\nu$ and the measure of firm's share $X(\nu)$ is known \citep{chow1996}. The evolution of a process takes place as if the action function is stationary. Therefore, the conditional expectation with respect to time only depends on the expectation of initial time point of this time interval.

It\^o's Lemma implies,
\begin{eqnarray}\label{action.0}
\varepsilon\mathcal{A}_{s,s+\varepsilon}(X) & = & 
\E_s\biggr\{\varepsilon\pi[s,X(s),u(s)]+\varepsilon g[s,X(s)] + \varepsilon\frac{\partial}{\partial s}g[s,X(s)]+ \varepsilon\mu[s,X(s),u(s)]\frac{\partial}{\partial x}g[s,X(s)] \notag\\
& &  +\varepsilon\sigma[s,X(s),u(s)]\frac{\partial}{\partial X}g[s,X(s)]dB(s)+
\mbox{$\frac{1}{2}$}\varepsilon\sigma^2[s,X(s),u(s)]\frac{\partial^2}{\partial X^2}g[s,X(s)]+o(\varepsilon)\biggr\},
\end{eqnarray}
and $[\Delta X(s)]^2\sim\varepsilon$ as $\varepsilon\downarrow 0$. \cite{feynman1948} uses an interpolation method to find an approximation of the area under the path in $[s,s+\varepsilon]$. Using a similar approximation, 
\begin{eqnarray}
\mathcal{A}_{s,s+\varepsilon}(X)& = & \pi[s,X(s),u(s)]+g[s,X(s)]+
\frac{\partial}{\partial s}g[s,X(s)]+
\mu[s,X(s),u(s)]
\frac{\partial}{\partial X}g[s,X(s)] \notag \\
& & +\mbox{$\frac{1}{2}$}
\sigma^2[s,X(s),u(s)]
\frac{\partial^2}{\partial X^2}g[s,X(s)]+o(1), \label{action}
\end{eqnarray}
where $\E_s[dB(s)]=0$ and $\E_s[o(\varepsilon)]/\varepsilon\ra 0$ as $\varepsilon\downarrow 0$.  Combining Equations (\ref{w16}) and (\ref{action}) yield
\begin{eqnarray}\label{action9}
\Psi_{s,s+\varepsilon}(X) & = & \frac{1}{L_\varepsilon}\int_{\mathbb{R}}
\exp\left[-\varepsilon\left\{
\pi[s,X(s),u(s)]+g[s,X(s)]+
\frac{\partial}{\partial s}g[s,X(s)]\right.\right.\notag \\ & &\left.\left.+
\mu[s,X(s),u(s)]\frac{\partial}{\partial X}g[s,X(s)]+
\mbox{$\frac{1}{2}$}\sigma^2[s,X(s),u(s)]\frac{\partial^2}{\partial X^2}g[s,X(s)]\right\}\right]\notag\\ & & \times\Psi_s[X(s)]dX(s)+o(\varepsilon^{1/2}),
\end{eqnarray}
as $\varepsilon\downarrow 0$. Taking a first order Taylor series expansion on the left hand side of Equation (\ref{action9}) yields
\begin{align}\label{action10}
\Psi_s^\tau(X)+\varepsilon  \frac{\partial \Psi_s^\tau(X)}{\partial s}+o(\varepsilon)&=\frac{1}{L_\varepsilon}\int_{\mathbb{R}} \exp\biggr\{-\varepsilon \big[\pi[s,X(s),u(s)]\notag\\&+g[s,X(s)]+\frac{\partial}{\partial s}g[s,X(s)]+\frac{\partial}{\partial X}g[s,X(s)]\mu[s,X(s),u(s)]\notag\\&+\mbox{$\frac{1}{2}$}\sigma^2[s,X(s),u(s)]\frac{\partial^2}{\partial X^2}g[s,X(s)]\big]\biggr\}  \Psi_s[X(s)] dX(s)+o(\varepsilon^{1/2}).
\end{align}
For fixed $s$ and $\tau$ set $X(s)=X(\tau)+\xi$.  Furthermore, if $\xi$ is not very close to zero we cannot do Taylor series expansion. Therefore, when $\xi$ is not around zero, for a positive number $\eta<\infty$ we assume 
\begin{align}\label{action10.1}
\left\{\left[\frac{X(s)}{\epsilon}\right]^{\frac{1}{2}}\xi\right\}\leq \eta\implies \xi^2\leq\frac{\eta\epsilon}{X(s)}\implies |\xi|\leq\sqrt{\frac{\eta\epsilon}{X(s)}}.
\end{align}
Condition (\ref{action10.1}) implies that, for $\epsilon\downarrow 0$, $\xi$ takes even smaller values such that $0<X(s)\leq\eta\varepsilon/\xi^2$ (for a detailed explanation see section $7.3$ of \cite{johnson2000}). Since, our stochastic isoperimetric non-holonomic constraint follows Theorem \ref{thmf} along with Assumptions \ref{as1}, \ref{as2} in the Appendix, and $d\xi$ is a cylindrical measure where $\xi$ contributes significantly, $\Psi_\tau[X(\xi)] $ of Equation (\ref{action10}) can be expanded using a Taylor series of $\xi$ around $0$.
Therefore,	
\begin{align}
\Psi_s^\tau(X)+\varepsilon  \frac{\partial \Psi_s^\tau(X)}{\partial s}+o(\varepsilon)&= \frac{1}{L_\varepsilon}\int_{\mathbb{R}} \left[\Psi_s^\tau(X)+\xi\frac{\partial \Psi_s^\tau(X)}{\partial X}+o(\varepsilon)\right] \exp\biggr\{-\varepsilon \big[\pi[s,X(\tau)+\xi,u(s)]\notag\\&+g[s,X(\tau)+\xi]+\frac{\partial}{\partial s}g[s,X(\tau)+\xi]+\frac{\partial}{\partial x}g[s,X(\tau)+\xi] \mu[s,X(\tau)+\xi,u(s)]\notag\\&+\mbox{$\frac{1}{2}$}\sigma^2[s,X(\tau)+\xi,u(s)] \frac{\partial^2}{\partial X^2}g[s,X(\tau)+\xi]\big]\biggr\} d\xi+o(\varepsilon^{1/2}).\notag
\end{align}

Let
\begin{eqnarray*}
	f[s,\xi,u(s)]&=&\pi[s,X(\tau)+\xi,u(s)]+g[s,X(\tau)+\xi]+\frac{\partial}{\partial s}g[s,X(\tau)+\xi]\\ & & +\frac{\partial}{\partial x}g[s,X(\tau)+\xi] \mu[s,X(\tau)+\xi,u(s)]\\ & &+\mbox{$\frac{1}{2}$}\sigma^2[s,x(\tau)+\xi,u(s)]\ \frac{\partial^2}{\partial X^2}g[s,X(\tau)+\xi]+o(1),
\end{eqnarray*}
so that
\begin{multline*}
\Psi_s^\tau(X)+\varepsilon  \frac{\partial \Psi_s^\tau(X)}{\partial s}+o(\varepsilon)=\Psi_s^\tau(X) \frac{1}{L_\epsilon} \int_{\mathbb{R}} \exp\big\{-\varepsilon f[s,\xi,u(s)]\} d\xi\\
+\frac{\partial \Psi_s^\tau(X)}{\partial X}\frac{1}{L_\varepsilon}\int_{\mathbb{R}} \xi \exp \big\{-\varepsilon f[s,\xi,u(s)]\big\} d\xi+o(\varepsilon^{1/2}).
\end{multline*}
where
\begin{equation*}
f[s,\xi,u(s)]=f[s,X(\tau),u(s)]+\frac{\partial}{\partial X}f[s,X(\tau),u(s)][\xi-X(\tau)]+\mbox{$\frac{1}{2}$}\frac{\partial^2}{\partial X^2}f[s,X(\tau),u(s)][\xi-X(\tau)]^2+o(\varepsilon),
\end{equation*}
where $\varepsilon\downarrow 0$ and $\Delta X\downarrow 0$.

Define $m=\xi-X(\tau)$ so that $ d\xi=dm$, then standard integration techniques can be used to show that
\begin{multline*}
\int_{\mathbb{R}} \exp\left\{-\varepsilon f[s,\xi,u(s)]\right\} d\xi\notag=\exp\{-\varepsilon f[s,X(\tau),u(s)]\}\\
\times\int_{\mathbb{R}} \exp\biggr\{-\varepsilon \biggr[\frac{\partial}{\partial X}f[s,X(\tau),u(s)]m+\mbox{$\frac{1}{2}$}\frac{\partial^2}{\partial X^2}f[s,X(\tau),u(s)]m^2\biggr]\biggr\} dm.
\end{multline*}
Then 
\begin{equation*}
\Psi_s^\tau(X) \frac{1}{L_\varepsilon} \int_{\mathbb{R}} \exp\big\{-\varepsilon f[s,\xi,u(s)]\} d\xi=\Psi_s^\tau(X) \frac{1}{L_\varepsilon}  \sqrt{\frac{\pi}{\varepsilon a}}\exp\left\{\varepsilon \left[\frac{b^2}{4a^2}-f[s,X(\tau),u(s)]\right]\right\}, 
\end{equation*}
where $a=\frac{1}{2} \frac{\partial^2}{\partial X^2}f[s,X(\tau),u(s)]$ and $b=\frac{\partial}{\partial X}f[s,X(\tau),u(s)]$.

\bigskip

Similarly, it can be shown that
\begin{equation*}
\frac{\partial\Psi_s^\tau(X)}{\partial X} \frac{1}{L_\varepsilon}\int_{\mathbb{R}} \xi \exp\left[-\varepsilon f[s,\xi,u(s)]\right] d\xi=\frac{\partial\Psi_s^\tau(x)}{\partial X} \frac{1}{L_\varepsilon} \exp\left\{\varepsilon \left[\frac{b^2}{4a^2}-f[s,X(\tau),u(s)]\right]\right\} \biggr[X(\tau)-\frac{b}{2a}\biggr]\sqrt{\frac{\pi}{\varepsilon a}}.
\end{equation*}
Therefore
\begin{multline*}
\Psi_s^\tau(X)+\varepsilon  \frac{\partial \Psi_s^\tau(X)}{\partial s}+o(\varepsilon)=\Psi_s^\tau(X) \frac{1}{L_\varepsilon} \sqrt{\frac{\pi}{\varepsilon a}}\exp\left\{\varepsilon \left[\frac{b^2}{4a^2}-f[s,X(\tau),u(s)]\right]\right\}\\
+\frac{\partial\Psi_s^\tau(X)}{\partial X} \frac{1}{L_\varepsilon} \sqrt{\frac{\pi}{\varepsilon a}} \exp\left\{\varepsilon \left[\frac{b^2}{4a^2}-f[s,X(\tau),u(s)]\right]\right\} \biggr[X(\tau)-\frac{b}{2a}\biggr]+o(\varepsilon^{1/2}).
\end{multline*}
Assuming $L_\varepsilon=\sqrt{\frac{\pi}{\varepsilon a}}>0$ and after expanding exponential function up to the first order we get,
\begin{equation*}
\Psi_s^\tau(X)+\varepsilon \frac{\partial \Psi_s^\tau(X)}{\partial s}+o(\varepsilon)=\left\{1+\varepsilon \left[\frac{b^2}{4a^2}-f[s,X(\tau),u(s)]\right]\right\} \biggr\{\Psi_s^\tau(X)+\left[X(\tau)-\frac{b}{2a}\right]\frac{\partial\Psi_s^\tau(X)}{\partial X}+o(\varepsilon^{1/2})\biggr\}.
\end{equation*}

The term $b/(2a)$ is the ratio of the first derivative to the second derivative with respect to $X$ of $f$. As $f$ is in a Schwartz space, the derivatives of $f$ are rapidly falling and they satisfy Assumptions \ref{as1} and \ref{as2}, and therefore it is reasonable to assume, $0<|b|\leq\eta\varepsilon$ and $0<|a|\leq\mbox{$\frac{1}{2}$}(1-\xi^{-2})^{-1}$. Hence, using $X(s)-X(\tau)=\xi$, and Condition \ref{action10.1} we get,
\begin{align}\label{action25.2}
X(\tau)-\frac{b}{2a}&=X(s)-\xi-\frac{b}{2a}\notag\\
&\leq\left|\frac{\eta\epsilon}{\xi^2}-\xi-\eta\epsilon\left(1-\frac{1}{\xi^2}\right)\right|=\left|\frac{2\eta\epsilon}{\xi^2}-(\xi+\eta\epsilon)\right|\leq \frac{2\eta\epsilon}{\xi^2},
\end{align}
for sufficiently large $\eta$. Hence,
\begin{align*}
\bigg|X(\tau)-\frac{b}{2a}\bigg|\leq\frac{2\eta\epsilon}{\xi^2}.
\end{align*}

Therefore, letting $\varepsilon\ra 0$, the Wick rotated Schr\"odinger type equation is,
\begin{align}\label{action25.4}
\frac{\partial \Psi_s^\tau(X)}{\partial s}&=\left[\frac{b^2}{4a^2}-f[s,X(\tau),u(s)]\right]\Psi_s^\tau(X).
\end{align}
If we differentiate Equation (\ref{action25.4}) with respect to $u$, then the solution of the new equation will be a Walrasian optimal strategy in the stochastic case. That is,
\begin{equation}\label{w18}
\left[\frac{2\frac{\partial}{\partial X}f(s,X,u)}{\frac{\partial^2}{\partial X^2}f(s,X,u)}\left(\frac{\frac{\partial^2}{\partial X^2}f(s,X,u)\frac{\partial}{\partial X\partial u}f(s,X,u)-\frac{\partial}{\partial X}f(s,X,u)\frac{\partial^3}{\partial u\partial X^2}f(s,X,u)}{\left[\frac{\partial^2}{\partial X^2}f(s,X,u)\right]^2}\right)\right.\\
\left.-\frac{\partial}{\partial u}f(s,X,u)\right]\Psi_s^\tau(X)=0.
\end{equation}
Since the function $f$ is defined on $g$, it is a twice differentiable  function. Hence, $\frac{\partial^3}{\partial u\partial X^2}f(s,X,u)=0$. Furthermore, as the wave function $\Psi_s^\tau(X)$ cannot be zero, to satisfy the Equation (\ref{w18}) the parenthesis term must be zero. Therefore, an optimal Walrasian strategy is found by setting Equation (\ref{w18}) equal to zero obtains,
\begin{align*}
\frac{\partial}{\partial u}f(s,X,u) \left[\frac{\partial^2}{\partial X^2}f(s,X,u)\right]^2=2\frac{\partial}{\partial X}f(s,X,u) \frac{\partial^2}{\partial X\partial u}f(s,X,u).
\end{align*}
A unique solution to Equation (\ref{action25.4}) can be found using a Fourier transformation, as $\Psi_s(X)=I(X) \exp[sv(X,u)]$, which can be verified by direct differentiation \citep{pramanik2020}. $\square$

\subsection{Proof of Proposition \ref{p2}}

Euclidean action function is,
\begin{multline*}
\mathcal{A}_{0,t}(K,V)= \int_0^t \E_s\bigg\{\pi\bigg[s,H[s,K(s),V(s)],u(s)\bigg]\ ds+\lambda_1(s) \big[K(s+ds)-K(s)\\-\mu_1[s,u(s)] K(s) ds-\sigma_1[s,u(s)] K(s) dB_1(s)\big] \\
+\lambda_2(s)\big[V(s+ds)-V(s)-\mu_2[s,u(s)]\ V(s)\ ds-\sigma_2[s,u(s)] V(s) dB_2(s)\big]\bigg\}.
\end{multline*}
Following arguments similar to those used to prove Proposition \ref{p1}, define $\Delta s=\varepsilon>0$, and for $L_\varepsilon>0$ Lemma \ref{l1} in the Appendix implies, 
\begin{align}\label{m3}
\Psi_{s,s+\varepsilon}(K,V)&= \frac{1}{L_\varepsilon} \int_{\mathbb{R}} \exp\biggr\{-\varepsilon  \mathcal{A}_{s,s+\varepsilon}(K,V)\biggr\} \Psi_s(K,V) dK(s)\times dV(s),
\end{align}
as $\varepsilon\downarrow 0$ where $\Psi_s(K,V)$ is the wave function at time $s$ and states $K(s)$ and $V(s)$ respectively with initial condition $\Psi_0(K,V)=\Psi_0$.

The action function in $[s,\tau]$ where $\tau=s+\varepsilon$ with the Lagrangian is,
\begin{multline*}
\mathcal{A}_{s,\tau}(K,V)= \int_s^\tau\ \E_s\bigg\{\pi\bigg[\nu,H[\nu,K(\nu),V(\nu)],V(\nu),u(\nu)\bigg] d\nu\\
+\lambda_1(\nu) \big[K(\nu+d\nu)d\nu-K(\nu)d\nu-\mu_1[\nu,u(\nu)] K(\nu) d\nu-\sigma_1[\nu,u(\nu)]\ K(\nu) dB_1(\nu)\big] \notag\\
+\lambda_2(\nu) \big[V(\nu+d\nu)d\nu-V(\nu)d\nu-\mu_2[\nu,u(\nu)] V(\nu) d\nu-\sigma_2[\nu,u(\nu)] V(\nu) dB_2(\nu)\big]\bigg\},
\end{multline*}
with initial conditions $K(0)=K_0$ and $V(0)=V_0$, where $\lambda_1$ and $\lambda_2$ are two Lagrangian multipliers corresponding to the two constraints. The conditional expectation is valid when the strategy $u(\nu)$ is determined at time $\nu$, and hence only depends on the initial time point of this time interval.
Let $\Delta K(\nu)=K(\nu+d\nu)-K(\nu)$ and, $\Delta V(\nu)=V(\nu+d\nu)-V(\nu)$, then Fubini's Theorem implies,
\begin{align}\label{m5}
\mathcal{A}_{s,\tau}(K,V)&=\E_s\bigg\{\int_s^\tau\ \pi\bigg[\nu,H[\nu,K(\nu),V(\nu)],V(\nu),u(\nu)\bigg] d\nu\notag \\
&+\lambda_1(\nu) \big[\Delta K(\nu)d\nu-\mu_1[\nu,u(\nu)] K(\nu) d\nu-\sigma_1[\nu,u(\nu)] K(\nu)\ dB_1(\nu)\big]\notag\\
& +\lambda_2(\nu)\big[\Delta V(\nu)d\nu-\mu_2[\nu,u(\nu)]\ V(\nu)\ d\nu-\sigma_2[\nu,u(\nu)]\ V(\nu)\ dB_2(\nu)\big]\bigg\}.
\end{align}
Because $K(\nu)$ and $V(\nu)$ are It\^o processes, Theorem 4.1.2 of \cite{oksendal2003} implies that there exists a function \\$g[\nu,K(\nu),V(\nu)]\in C^2([0,\infty)\times\mathcal K\times \mathcal V)$ that satisfies Theorem \ref{thmf} in the Appendix, Assumptions \ref{as0}-\ref{as2}, such that $Y(\nu)=g[\nu,K(\nu),V(\nu)]$ where $Y(\nu)$ is an It\^o process. If we assume 
\begin{multline*}
g[\nu+\Delta \nu,K(\nu)+\Delta K(\nu),V(\nu)+\Delta V(\nu)]=\lambda_1(\nu) \big[\Delta K(\nu)d\nu-\mu_1[\nu,u(\nu)] K(\nu) d\nu-\sigma_1[\nu,u(\nu)] K(\nu) dB_1(\nu)\\
+\lambda_2 (\nu) \big[\Delta V(\nu)d\nu-\mu_2[\nu,u(\nu)]\ V(\nu) d\nu-\sigma_2[\nu,u(\nu)]\ V(\nu)\ dB_2(\nu)+o(1),
\end{multline*} 
Equation (\ref{m5}) becomes,
\begin{equation}\label{m6}
\mathcal{A}_{s,\tau}(K,V)=\E_s \bigg\{ \int_{s}^\tau \pi\bigg[\nu,H[\nu,K(\nu),V(\nu)],V(\nu),u(\nu)\bigg]\ d\nu+g[\nu+\Delta \nu,K(\nu)+\Delta K(\nu),V(\nu)+\Delta V(\nu)]\bigg\}.
\end{equation}
It\^o's Lemma and Equation (\ref{m6}) of \cite{baaquie1997} imply
\begin{align}
\mathcal{A}_{s,\tau}(K,V) &=\pi\bigg[s,H[s,K(s),V(s)],V(s),u(s)\bigg]+g[s,K(s),V(s)]\notag\\ &+\frac{\partial}{\partial s}g[s,K(s),V(s)]+\frac{\partial}{\partial S}g[s,K(s),V(s)] \mu_1[s,u(s)] K(s)\notag\\&+\frac{\partial}{\partial V}g[s,K(s),V(s)] \mu_2[s,u(s)] V(s) \notag\\&+\mbox{$\frac{1}{2}$}\bigg[\sigma_1^2[s,u(s)] K^2(s) \frac{\partial^2}{\partial K^2}g[s,K(s),V(s)]\notag\\&+2\gamma\sigma_1^3[s,u(s)] K(s) \frac{\partial^2}{\partial K\partial V}g[s,K(s),V(s)]\notag\\&+\sigma_2^2[s,u(s)] V^2(s) \frac{\partial^2}{\partial V^2}g[s,K(s),V(s)]\bigg]+o(1),\notag
\end{align}
where we have used the fact that $[\Delta K(s)]^2=\Delta V(s)]^2=\varepsilon$, and $\E_s[\Delta B_1(s)]=\E_s[\Delta B_2(s)]$, as $\varepsilon\downarrow 0$ with initial conditions $K_0$ and $V_0$. Using Equation (\ref{m3}), the transition wave function in $[s,\tau]$ becomes,
\begin{align}
&\Psi_{s,\tau}(K,V)\notag\\&=\frac{1}{L_\varepsilon}\int_{\mathbb{R}^2} \exp\bigg\{-\varepsilon \bigg[\pi\bigg[s,H[s,K(s),V(s)],V(s),u(s)\bigg]+g[s,K(s),V(s)]\notag\\&+\frac{\partial}{\partial s}g[s,K(s),V(s)]+\frac{\partial}{\partial K}g[s,K(s),V(s)] \mu_1[s,u(s)] K(s)\notag\\&+\frac{\partial}{\partial V}g[s,S(s),V(s)] \mu_2[s,u(s)] V(s) +\mbox{$\frac{1}{2}$}\bigg[\sigma_1^2[s,u(s)] K^2(s) \frac{\partial^2}{\partial K^2}g[s,K(s),V(s)]\notag\\&+2\gamma\sigma_1^3[s,u(s)] K(s) \frac{\partial^2}{\partial K\partial V}g[s,K(s),V(s)]\notag\\&+\sigma_2^2[s,u(s)] V^2(s) \frac{\partial^2}{\partial V^2}g[s,K(s),V(s)]\bigg]\bigg] \bigg\} \Psi_s(K,V) dK(s) dV(s)+o(\varepsilon^{1/2}),\notag
\end{align}
as $\varepsilon\downarrow 0$. 

Therefore,
\begin{align}
&\Psi_s^\tau(K,V)+\varepsilon \frac{\partial \Psi_s^\tau(K,V)}{\partial\notag s}+o(\varepsilon)\notag\\&=\frac{1}{L_\varepsilon}\int_{\mathbb{R}^2} \exp\bigg\{-\varepsilon \bigg[\pi\bigg[s,H[s,K(s),V(s)],V(s),u(s)\bigg]+g[s,K(s),V(s)]\notag\\&+\frac{\partial}{\partial s}g[s,K(s),V(s)]+\frac{\partial}{\partial K}g[s,K(s),V(s)] \mu_1[s,u(s)]\ K(s)\notag\\&+\frac{\partial}{\partial V}g[s,K(s),V(s)] \mu_2[s,u(s)] V(s) +\mbox{$\frac{1}{2}$}\bigg[\sigma_1^2[s,u(s)] K^2(s)\notag\\ &\times \frac{\partial^2}{\partial K^2}g[s,K(s),V(s)]+2\gamma\sigma_1^3[s,u(s)] K(s) \frac{\partial^2}{\partial K\partial V}g[s,K(s),V(s)]\notag\\&+\sigma_2^2[s,u(s)] V^2(s) \frac{\partial^2}{\partial V^2}g[s,K(s),V(s)]\bigg]\bigg] \bigg\}\Psi_s(K,V) dK(s) dV(s)+o(\varepsilon^{1/2}),\notag
\end{align}
as $\varepsilon\downarrow 0$.

\bigskip

For fixed $s$ and $\tau$ suppose that $K(s)=K(\tau)+\xi_1$, and $V(s)=V(\tau)+\xi_2$. For positive numbers $\eta_1<\infty$ and $\eta_2<\infty$  assume that $|\xi_1|\leq\sqrt{\frac{\eta_1\epsilon}{K(s)}}$ and $|\xi_2|\leq\sqrt{\frac{\eta_2\epsilon}{V(s)}}$. Here, security and volatility are $K(s)\leq\eta_1\epsilon/\xi_1^2$ and $V(s)\leq\eta_2\epsilon/\xi_2^2$, respectively. Furthermore, Theorem \ref{thmf} in the Appendix and and Assumptions \ref{as1}-\ref{as2} imply
\begin{align}
&\Psi_s^\tau(K,V)+\varepsilon \frac{\partial \Psi_s^\tau(K,V)}{\partial\notag s}+o(\varepsilon)\notag\\&=\frac{1}{L_\varepsilon}\int_{\mathbb{R}^2} \left[\Psi_s^\tau(K,V)+\xi_1\frac{\partial \Psi_s^\tau(K,V)}{\partial K}+\xi_2\frac{\partial \Psi_s^\tau(K,V)}{\partial V}+o(\varepsilon)\right]\notag\\& \exp\bigg\{-\varepsilon \bigg[\pi\bigg[s,H[s,K(\tau)+\xi_1,V(\tau)+\xi_2],V(\tau)+\xi_2,u(s)\bigg]\notag\\&+g[s,K(\tau)+\xi_1,V(\tau)+\xi_2]+\frac{\partial}{\partial s}g[s,K(\tau)+\xi_1,V(\tau)+\xi_2]\notag\\&+g_K[s,K(\tau)+\xi_1,V(\tau)+\xi_2] \mu_1[s,u(s)] (K(\tau)+\xi_1)\notag\\&+\frac{\partial}{\partial V}g[s,K(\tau)+\xi_1,V(\tau)+\xi_2] \mu_2[s,u(s)] (V(\tau)+\xi_2) \notag\\&+\mbox{$\frac{1}{2}$}\bigg[\sigma_1^2[s,u(s)](K(\tau)+\xi_1)^2 \frac{\partial^2}{\partial K^2}g[s,K(\tau)+\xi_1,V(\tau)+\xi_2]\notag\\&+2\gamma\sigma_1^3[s,u(s)] (K(\tau)+\xi_1) \frac{\partial^2}{\partial K\partial V}g[s,K(\tau)+\xi_1,V(\tau)+\xi_2]\notag\\&+\sigma_2^2[s,u(s)] (V(\tau)+\xi_2)^2 \frac{\partial^2}{\partial V^2}g[s,K(\tau)+\xi_1,V(\tau)+\xi_2]\bigg]\bigg] \bigg\}\notag\\& \Psi_\tau[K(\xi_1),V(\xi_2)]\ d\xi_1 d\xi_2+o(\varepsilon^{1/2}),\notag
\end{align}
as $\varepsilon\downarrow 0$.

Define $f[s,\xi_1,\xi_2,u(s)]$ as in Equation (\ref{m13.0}), then 
\begin{align}
&\Psi_s^\tau(K,V)+\varepsilon \frac{\partial \Psi_s^\tau(K,V)}{\partial\notag s}+o(\varepsilon)\notag\\&=\frac{1}{L_\varepsilon} \Psi_s^\tau(K,V) \int_{\mathbb{R}^2} \exp \big\{-\varepsilon f[s,\xi_1,\xi_2,u(s)]\big\}d\xi_1 d\xi_2\notag\\&+\frac{1}{L_\varepsilon} \frac{\partial \Psi_s^\tau(K,V)}{\partial K} \int_{\mathbb{R}^2} \xi_1 \exp \big\{-\varepsilon f[s,\xi_1,\xi_2,u(s)]\big\} d\xi_1 d\xi_2\notag\\&+\frac{1}{L_\varepsilon} \frac{\partial \Psi_s^\tau(K,V)}{\partial V} \int_{\mathbb{R}^2} \xi_2 \exp \big\{-\varepsilon f[s,\xi_1,\xi_2,u(s)]\big\} d\xi_1 d\xi_2+o(\varepsilon^{1/2}).\notag
\end{align}
Assume that $f$ is a $C^2$ function, then
\begin{align}
& f[s,\xi_1,\xi_2,u(s)]\notag\\&=f[s,K(\tau),V(\tau),u(s)]+[\xi_1-K(\tau)] \frac{\partial}{\partial K}f[s,K(\tau),V(\tau),u(s)]\notag\\&+[\xi_2-V(\tau)] \frac{\partial}{\partial V}f[s,K(\tau),V(\tau),u(s)]\notag\\&+\mbox{$\frac{1}{2}$}\bigg[[\xi_1-K(\tau)]^2\frac{\partial^2}{\partial K^2}f[s,K(\tau),V(\tau),u(s)]\notag\\&+2 [\xi_1-K(\tau)] [\xi_2-V(\tau)] \frac{\partial^2}{\partial K\partial V}g[s,K(\tau),V(\tau),u(s)]\notag\\&+[\xi_2-V(\tau)]^2 \frac{\partial^2}{\partial V^2}g[s,K(\tau),V(\tau),u(s)]\bigg]+o(\varepsilon),\notag
\end{align}
as $\varepsilon\downarrow 0$ and $\Delta u\downarrow 0$.
Define $m_1=\xi_1-K(\tau)$ and $m_2=\xi_2-V(\tau)$ so that $d\xi_1=dm_1$ and $d\xi_2=dm$ respectively so that
\begin{align}\label{m16}
\int_{\mathbb{R}^2} \exp \big\{-\varepsilon f[s,\xi_1,\xi_2,u(s)]\big\} d\xi_1 d\xi_2&=\int_{\mathbb{R}^2} \exp \biggr\{-\varepsilon \bigg[f[s,K(\tau),V(\tau),u(s)]+m_1 \frac{\partial}{\partial K}f[s,K(\tau),V(\tau),u(s)]\notag\\&\hspace{.5cm}+m_2\frac{\partial}{\partial V}f[s,K(\tau),V(\tau),u(s)]+\mbox{$\frac{1}{2}$} m_1^2 \frac{\partial^2}{\partial K^2}g[s,K(\tau),V(\tau),u(s)]\notag\\&\hspace{1cm}+m_1 m_2 \frac{\partial^2}{\partial K\partial V}f[s,K(\tau),V(\tau),u(s)]\notag\\&\hspace{2cm}+\mbox{$\frac{1}{2}$} m_2^2 \frac{\partial^2}{\partial V^2}f[s,K(\tau),V(\tau),u(s)]\bigg]\biggr\} dm_1 dm_2.
\end{align}
Let
\[
\Theta=\begin{bmatrix}
\mbox{$\frac{1}{2}$} \frac{\partial^2}{\partial K^2}f[s,K(\tau),V(\tau),u(s)] & \mbox{$\frac{1}{2}$} \frac{\partial^2}{\partial K\partial V}g[s,K(\tau),V(\tau),u(s)] \\ \mbox{$\frac{1}{2}$} \frac{\partial^2}{\partial K\partial V}f[s,K(\tau),V(\tau),u(s)]& \mbox{$\frac{1}{2}$} \frac{\partial^2}{\partial V^2}g[s,K(\tau),V(\tau),u(s)]
\end{bmatrix},
\]
and
\[
m=\begin{bmatrix}
m_1\\ m_2
\end{bmatrix},
\] 
and 
\[ 
-v_1=\begin{bmatrix}
\frac{\partial}{\partial K}f[s,K(\tau),V(\tau),u(s)]\\ \frac{\partial}{\partial V}f[s,K(\tau),V(\tau),u(s)]
\end{bmatrix},
\]
where we assume that $\Theta$ is positive definite, then the integrand in Equation (\ref{m16}) becomes a shifted Gaussian integral, 
\begin{align}
\int_{\mathbb{R}^2} \exp \bigg\{-\varepsilon \left(f- v_1^T\ m+m^T \Theta m\right)\bigg\}\ dm &=\exp\left(-\varepsilon f\right) \int_{\mathbb{R}^2} \exp \bigg\{(\varepsilon v_1^T) m-m^T (\varepsilon \Theta) m\bigg\} dm\notag\\&=\frac{\pi}{\sqrt{\varepsilon |\Theta|}} \exp\left[\frac{\varepsilon}{4}v_1^T \Theta^{-1} v_1-\varepsilon f\right],\notag
\end{align}
where $v_1^T$ and $m^T$ are the transposes of vectors $v_1$ and $m$ respectively. Therefore,
\begin{align}\label{m19}
\frac{1}{L_\varepsilon} \Psi_s^\tau(K,V) \int_{\mathbb{R}^2} \exp \big\{-\epsilon f[s,\xi_1,\xi_2,u(s)]\big\} d\xi_1 d\xi_2&=\frac{1}{L_\varepsilon} \Psi_s^\tau(K,V) \frac{\pi}{\sqrt{\varepsilon |\Theta|}}\exp\left[\frac{\varepsilon}{4} v_1^T \Theta^{-1} v_1-\varepsilon f\right],
\end{align}
such that inverse matrix $\Theta^{-1}>0$ exists. Similarly,
\begin{align}\label{m21}
& \frac{1}{L_\varepsilon} \frac{\partial \Psi_s^\tau(K,V)}{\partial K} \int_{\mathbb{R}^2} \xi_1 \exp \big\{-\varepsilon f[s,\xi_1,\xi_2,u(s)]\big\} d\xi_1 d\xi_2\notag\\&=\frac{1}{L_\varepsilon} \frac{\partial \Psi_s^\tau(K,V)}{\partial K} \frac{\pi}{\sqrt{\varepsilon |\Theta|}} \left(\mbox{$\frac{1}{2}$} \Theta^{-1}+K\right) \exp\left[\frac{\varepsilon}{4}\ v_1^T\ \Theta^{-1}\ v_1-\varepsilon f\right],
\end{align}
and 
\begin{align}\label{m22}
& \frac{1}{L_\varepsilon} \frac{\partial \Psi_s^\tau(K,V)}{\partial V} \int_{\mathbb{R}^2} \xi_2 \exp \big\{-\varepsilon f[s,\xi_1,\xi_2,u(s)]\big\} d\xi_1\times d\xi_2\notag\\&=\frac{1}{L_\varepsilon}\frac{\partial \Psi_s^\tau(K,V)}{\partial V} \frac{\pi}{\sqrt{\varepsilon |\Theta|}} \left(\mbox{$\frac{1}{2}$} \Theta^{-1}+V\right) \exp\left[\frac{\varepsilon}{4} v_1^T \Theta^{-1}\ v_1-\varepsilon f\right].
\end{align}
Equations (\ref{m19}), (\ref{m21}) and (\ref{m22}) imply that the Wick rotated Schr\"odinger type equation is,
\begin{multline*}
\Psi_s^\tau(K,V)+\varepsilon \frac{\partial \Psi_s^\tau(K,V)}{\partial\notag s}+o(\varepsilon)\\=\frac{1}{L_\varepsilon} \frac{\pi}{\sqrt{\varepsilon |\Theta|}} \exp\left[\frac{\varepsilon}{4} v_1^T \Theta^{-1} v_1-\varepsilon f\right] \bigg[\Psi_s^\tau(K,V)+\left(\mbox{$\frac{1}{2}$} \Theta^{-1}+K\right) \frac{\partial \Psi_s^\tau(K,V)}{\partial K}\\+\left(\mbox{$\frac{1}{2}$} \Theta^{-1}+V\right) \frac{\partial \Psi_s^\tau(K,V)}{\partial V} \bigg]+o(\varepsilon^{1/2}),
\end{multline*}
as $\varepsilon\downarrow 0$.

Assuming $L_\varepsilon=\pi/\sqrt{\varepsilon\ |\Theta|}>0$, 
\begin{multline*}
\Psi_s^\tau(K,V)+\varepsilon \frac{\partial \Psi_s^\tau(K,V)}{\partial\notag s}+o(\varepsilon)=\left[1+\varepsilon\left(\frac{1}{4} v_1^T \Theta^{-1} v_1- f\right)\right] \bigg[\Psi_s^\tau(K,V)+\left(\mbox{$\frac{1}{2}$} \Theta^{-1}+K\right) \frac{\partial \Psi_s^\tau(K,V)}{\partial K}\\+\left(\mbox{$\frac{1}{2}$} \Theta^{-1}+V\right) \frac{\partial \Psi_s^\tau(K,V)}{\partial V} \bigg]+o(\varepsilon^{1/2}).
\end{multline*}

As $K(s)\leq \eta_1\varepsilon/\xi_1^2$, assume $|\Theta^{-1}|\leq 2\eta_1\varepsilon(1-\xi_1^{-2})$ such that $|(2\Theta)^{-1}+K|\leq\eta_1\varepsilon$. For $V(s)\leq \eta_2\varepsilon/\xi_2^2$ we assume $|\Theta^{-1}|\leq 2\eta_2\varepsilon(1-\xi_2^{-2})$ such that $|(2\Theta)^{-1}+V|\leq\eta_2\varepsilon$. Therefore, $|\Theta^{-1}|\leq 2\varepsilon\min\left\{\eta_1(1-\xi_1^{-2}),\eta_2(1-\xi_2^{-2})\right\}$ such that, $|(2\Theta)^{-1}+K|\ra 0$ and $|(2\Theta)^{-1}+V|\ra 0$. Hence
\begin{align*}
\Psi_s^\tau(K,V)+\varepsilon \frac{\partial \Psi_s^\tau(K,V)}{\partial s}+o(\varepsilon)&=(1-\varepsilon f) \Psi_s^\tau(K,V)+o(\varepsilon^{1/2}). 
\end{align*}
Therefore, the Wick-rotated Schr\"odinger type Equation is,
\begin{equation*}
\frac{\partial \Psi_s^\tau(K,V)}{\partial s}=-f[s,\xi_1,\xi_2,u(s)]\ \Psi_s^\tau(K,V).
\end{equation*}
Therefore, the solution of
\begin{equation}\label{m27}
-\frac{\partial f[s,\xi_1,\xi_2,u(s)]}{\partial u}\ \Psi_s^\tau(K,V)=0,
\end{equation}
is a Walrasian optimal strategy, by \cite{pramanik2020} it has the form \[\Psi_s(K,V)=\exp\left\{-s f[s,\xi_1,\xi_2,u(s)]\right\} I(K,V).\] As the transition function $\Psi_s^\tau(K,V)$ is the solution to Equation (\ref{m27}), the result follows. $\square$

\subsection{Proof of Proposition \ref{p3}}

The Euclidean action function for firm $\rho$ under Pareto optimality in continuous time interval $[0,t]$ is, 
\begin{align*}
\mathcal{A}_{0,t}(X)&=  \int_0^t \E_s\bigg\{\sum_{\rho=1}^k \a_\rho\pi_\rho[s,X(s),u(s)]\ ds+\lambda(s) \big[X(s+ds)ds-X(s)ds-\mu[s,X(s),u(s)] ds-\sigma[s,X(s),u(s)] dB(s)\big] \bigg\}.
\end{align*} 

Similar argument of Proposition \ref{p1} implies
\begin{align}
\mathcal{A}_{s,\tau}(X)&= \E_s \bigg\{ \int_{s}^{\tau} \sum_{\rho=1}^k \a_\rho \pi_\rho[\nu,X(\nu),u(\nu)] d\nu+\lambda(\nu) \big[\Delta X(\nu)d\nu-\mu[\nu,X(\nu),u(\nu)] d\nu-\sigma[\nu,X(\nu),u(\nu)] dB(\nu)\big] \bigg\},\notag
\end{align}
where $\tau=s+\varepsilon$.

Since $X(\nu)$ is an It\^o process, by Theorem 4.1.2 of \cite{oksendal2003} there exists a $p$-dimensional vector valued function $g[\nu,X(\nu)]\in C^2([0,t]\times\mathcal X)$ which satisfies Theorem \ref{thmf} in the Appendix, Assumptions \ref{as0}-\ref{as2}, and $Y(\nu)=g[\nu,X(\nu)]$ where $Y(\nu)$ is an It\^o process. Assume 
\begin{equation*}
g[\nu+\Delta \nu,X(\nu)+\Delta x(\nu)]=\lambda(\nu) \big[\Delta X(\nu)d\nu-\mu[\nu,X(\nu),u(\nu)]\ d\nu-\sigma[\nu,X(\nu),u(\nu)] dB(\nu)+o(1),
\end{equation*} 
as $\varepsilon\downarrow 0$, then the generalized It\^o's Lemma implies,
\begin{multline*}
\mathcal{A}_{s,\tau}(X) \varepsilon= \E_s \bigg\{\sum_{\rho=1}^k \a_\rho \pi_\rho[s,X(s),u(s)] \varepsilon+g[s,X(s)] \varepsilon+\frac{\partial}{\partial s}g[s,X(s)] \varepsilon\\ 
+\sum_{\rho=1}^{k}\frac{\partial}{\partial X_\rho}g[s,X(s)] \mu[s,X(s),u(s)] \varepsilon+\sum_{\rho=1}^{k}\frac{\partial}{\partial X_\rho}g[s,X(s)] \sigma[s,X(s),u(s)] \varepsilon \Delta B(s)\\
+\mbox{$\frac{1}{2}$} \sum_{\rho=1}^{k}\sum_{j=1}^{m} \sigma^{\rho j}[s,X(s),u(s)]\frac{\partial^2}{\partial X_\rho X_j}g[s,X(s)] \varepsilon+o(\varepsilon)\bigg\},
\end{multline*}
where $\sigma^{\rho j}[s,X(s),u(s)]$ represents $\{\rho,j\}^{th}$ component of the variance-covariance matrix, and we used the conditions $\Delta B_\rho \Delta B_j=\delta^{\rho j} \varepsilon$, $\Delta B_\rho \varepsilon=\varepsilon \Delta B_\rho=0$, and $\Delta X_\rho(s) \Delta X_j(s)=\varepsilon$, where $\delta^{\rho j}$ is the Kronecker delta function.  Hence,
\begin{multline*}
\mathcal{A}_{s,\tau}(X)= \bigg[\sum_{\rho=1}^k \a_\rho\pi_\rho[s,X(s),u(s)]+g[s,X(s)]+\frac{\partial}{\partial s}g[s,X(s)]+\sum_{\rho=1}^{k}\frac{\partial}{\partial X_\rho}g[s,X(s)] \mu[s,X(s),u(s)]\\
+\mbox{$\frac{1}{2}$} \sum_{\rho=1}^{k}\sum_{j=1}^{m} \sigma^{\rho j}[s,X(s),u(s)]\frac{\partial^2}{\partial X_\rho X_j}g[s,X(s)] +o(1)\bigg],
\end{multline*}
where $\E_s[\Delta B(s)]=0$ and $\E_s[o(\varepsilon)]/\varepsilon\ra 0$ as $\varepsilon\downarrow 0$ with the vector of initial conditions $x_{0_{nk\times 1}}$. Expanding $\Psi_{s,\tau}(X)$ yields,
\begin{multline*}
\Psi_s^\tau(X)+\varepsilon  \frac{\partial \Psi_s^\tau(X)}{\partial s}+o(\varepsilon)=\frac{1}{L_\varepsilon} \int_{\mathbb{R}^{k\times m}} \exp\biggr\{-\varepsilon  \bigg[\sum_{\rho=1}^k \a_\rho\pi_\rho[s,X(s),u(s)]+g[s,X(s)]\\
+\frac{\partial}{\partial s}g[s,X(s)]+\sum_{\rho=1}^{k}\frac{\partial}{\partial X_\rho}g[s,X(s)] \mu[s,X(s),u(s)]\\
+\mbox{$\frac{1}{2}$} \sum_{\rho=1}^{k}\sum_{j=1}^{m} \sigma^{\rho j}[s,X(s),u(s)] \frac{\partial^2}{\partial X_\rho X_j}g[s,X(s)]\bigg]\biggr\} \Psi_s(X) dX(s)+o(\varepsilon^{1/2}).
\end{multline*}
Let $X(s)_{km\times 1}=X(\tau)_{km\times 1}+\xi_{km\times 1}$ and assume $||\xi||\leq\eta\epsilon [X'(s)]^{-1}$ for some $\eta>0$. Following previous arguments imply
\begin{align}
&\Psi_s^\tau(X)+\varepsilon  \frac{\partial \Psi_s^\tau(X)}{\partial s}+o(\varepsilon)\notag\\
&= \frac{1}{L_\varepsilon}\int_{\mathbb{R}^{k\times m}} \left[\Psi_s^\tau(X)+\xi\frac{\partial \Psi_s^\tau(X)}{\partial X}+o(\varepsilon)\right] \exp\biggr\{-\varepsilon\bigg[\sum_{\rho=1}^k \a_\rho\pi_\rho[s,X(\tau)+\xi,u(s)]\notag\\
&+g[s,X(\tau)+\xi]+\frac{\partial}{\partial s}g[s,X(\tau)+\xi]+\sum_{\rho=1}^{k}\frac{\partial}{\partial X_\rho}g[s,X(\tau)+\xi]\ \mu[s,X(\tau)+\xi,u(s)]\notag\\
&+\mbox{$\frac{1}{2}$}\sum_{\rho=1}^{k}\sum_{j=1}^{m} \sigma^{\rho j}[s,X(\tau)+\xi,u(s)] \frac{\partial^2}{\partial X_\rho \partial X_j}g[s,X(\tau)+\xi]\bigg]\biggr\} d\xi+o(\varepsilon^{1/2}).\notag
\end{align}
Let 
\begin{multline*}
f[s,\xi,u(s)]=\sum_{\rho=1}^k \a_\rho\pi_\rho[s,X(\tau)+\xi,u(s)]+g[s,X(\tau)+\xi]+\frac{\partial}{\partial s}g[s,X(\tau)+\xi]+\sum_{\rho=1}^{k}\frac{\partial}{\partial X_\rho}g[s,X(\tau)+\xi]\ \mu[s,X(\tau)+\xi,u(s)]\\
+\mbox{$\frac{1}{2}$}\sum_{\rho=1}^{k}\sum_{j=1}^{m}\ \sigma^{\rho j}\ [s,X(\tau)+\xi,u(s)]\frac{\partial^2}{\partial X_\rho\partial X_j}g[s,X(\tau)+\xi],
\end{multline*}
then
\begin{multline}\label{p17}
\Psi_s^\tau(X)+\varepsilon  \frac{\partial \Psi_s^\tau(X)}{\partial s}+o(\varepsilon)= \frac{1}{L_\varepsilon} \Psi_s^\tau(X) \int_{\mathbb{R}^{k\times m}} \exp\bigg\{-\varepsilon f[s,\xi,u(s)]\bigg\}  d\xi\\+\frac{1}{L_\varepsilon} \frac{\partial \Psi_s^\tau(X)}{\partial X} \int_{\mathbb{R}^{k\times m}} \xi\exp\bigg\{-\varepsilon f[s,\xi,u(s)]\bigg\} d\xi+o(\varepsilon^{1/2}).
\end{multline}
Expanding $f[s,\xi,u(s)]$ and defining $m_{km\times 1}=\xi_{km\times 1}-X(\tau)_{km\times 1}$ so that $ d\xi=dm$, first integral on the right hand side of Equation (\ref{p17}) becomes, 
\begin{multline*}
\int_{\mathbb{R}^{k\times m}} \exp\bigg\{-\varepsilon f[s,\xi,u(s)]\bigg\} d\xi=\exp\bigg\{-\varepsilon f[s,X(\tau),u(s)]\bigg\} \int_{\mathbb{R}^{k\times m}} \exp\biggr\{-\varepsilon \biggr[\sum_{\rho=1}^{k} \frac{\partial}{\partial X_\rho}f[s,X(\tau),u(s)]\ m_\rho\\
+\mbox{$\frac{1}{2}$} \sum_{\rho=1}^{k}\sum_{j=1}^{m} \frac{\partial^2}{\partial X_\rho \partial X_j}f[s,X(\tau),u(s)]\ m_\rho m_j\biggr]\biggr\} dm+ o(\varepsilon).
\end{multline*}
Assume for $k=m$ there exists a symmetric, positive definite and non-singular Hessian matrix $\theta_{k\times m}$ and a vector ${w}_{k\times 1}$ such that,
\begin{equation*}
\int_{\mathbb{R}^{k\times m}}\ \exp\bigg\{-\varepsilon f[s,\xi,u(s)]\bigg\} d\xi=\sqrt{\frac{(2\pi)^{km}}{\varepsilon |\theta|}} \exp\bigg\{-\varepsilon f[s,X(\tau),u(s)]+\frac{\varepsilon}{2} w'\theta^{-1}w\bigg\},
\end{equation*}
where, 
\[
\theta=\begin{bmatrix}
\frac{\partial^2}{\partial X_1\partial X_1}f & \frac{\partial^2}{\partial X_1\partial X_2}f & \dots & \frac{\partial^2}{\partial X_1\partial X_{m}}f\\ \frac{\partial^2}{\partial X_2\partial X_1}f & \frac{\partial^2}{\partial X_2\partial X_2}f & \dots & \frac{\partial^2}{\partial X_2\partial X_{m}}f\\ \vdots & \vdots & \ddots &\vdots\\
\frac{\partial^2}{\partial X_{k}\partial X_1}f & \frac{\partial^2}{\partial X_{k}\partial X_2}f & \dots & \frac{\partial^2}{\partial X_{k}\partial X_{m}}f
\end{bmatrix},
\]
and 
\[
w[s,X(\tau),u(s)]=\begin{bmatrix}
-\frac{\partial}{\partial X_1}f[s,X(\tau),u(s)]\\-\frac{\partial}{\partial X_2}f[s,X(\tau),u(s)]\\\vdots\\-\frac{\partial}{\partial X_{k}}f[s,X(\tau),u(s)]
\end{bmatrix}.
\]
The second integral on the right hand side of Equation (\ref{p17}) becomes,
\begin{equation*}
\int_{\mathbb{R}^{k\times m}} \xi \exp\bigg\{-\varepsilon f[s,\xi,u(s)]\bigg\} d\xi=\sqrt{\frac{(2\pi)^{km}}{\varepsilon |\theta|}} \exp\bigg\{-\varepsilon f[s,X(\tau),u(s)]+\frac{\varepsilon}{2} w'\theta^{-1}w\bigg\} \bigg[X(\tau)+\mbox{$\frac{1}{2}$}\ (\theta^{-1}\ w)\bigg].
\end{equation*}
So that
\begin{multline*}
\Psi_s^\tau(X)+\varepsilon  \frac{\partial \Psi_s^\tau(X)}{\partial s}+o(\varepsilon)=\frac{1}{L_\varepsilon} \sqrt{\frac{(2\pi)^{km}}{\varepsilon |\theta|}}\exp\bigg\{-\varepsilon f[s,X(\tau),u(s)]+\mbox{$\frac{1}{2}$}\varepsilon w'\theta^{-1}w\bigg\}\\
\times\bigg[\Psi_s^\tau(X)+\left[X(\tau)+\mbox{$\frac{1}{2}$} (\theta^{-1}\ w)\right]\frac{\partial \Psi_s^\tau(X)}{\partial x}\bigg]+o(\varepsilon^{1/2}).
\end{multline*}
Assume $L_\varepsilon=\sqrt{(2\pi)^{km}/(\varepsilon |\theta|)}>0$, then
\begin{equation*}
\Psi_s^\tau(X)+\varepsilon  \frac{\partial \Psi_s^\tau(X)}{\partial s}+o(\varepsilon)=\left\{1-\varepsilon f[s,X(\tau),u(s)]+\mbox{$\frac{1}{2}$}\varepsilon w'\theta^{-1}w\right\} \bigg[\Psi_s^\tau(X)+\left[X(\tau)+\mbox{$\frac{1}{2}$} (\theta^{-1} w)\right]\frac{\partial \Psi_s^\tau(X)}{\partial x}\bigg]+o(\varepsilon^{1/2}).
\end{equation*}
For any finite positive number $\eta$ we know $X(\tau)\leq\eta\varepsilon|\xi'|^{-1}$, and there exists $|\theta^{-1}w|\leq 2 \eta\varepsilon|1-\xi'|^{-1}$ such that for $\varepsilon\downarrow 0$ we have, $\big|X(\tau)+\mbox{$\frac{1}{2}$} (\theta^{-1} w)\big|\leq\eta\varepsilon$, and
\begin{align*}
\frac{\partial \Psi_s^\tau(x)}{\partial s}&=\left\{- f[s,X(\tau),u(s)]+\mbox{$\frac{1}{2}$} w'\theta^{-1}w\right\} \Psi_s^\tau(X). 
\end{align*}
Taking $\varepsilon\downarrow 0$, the Wick-rotated Schr\"odinger type equation is 
\begin{align*}
\frac{\partial \Psi_s^\tau(X)}{\partial s}&=- f[s,X(\tau),u(s)]\ \Psi_s^\tau(X),
\end{align*}
with the Wheeler-Di Witt type equation,
\begin{align*}
-\frac{\partial f[s,X(\tau),u(s)]}{\partial u_\rho}\ \Psi_s^\tau(X)=0,
\end{align*}
whose solution with respect to $u_\rho$ gives $\rho^{th}$ firm's cooperative Pareto Optimal strategy $\phi_p^{\rho*}[s,X^*(s)]$. $\square$	

\subsection{Proof of Proposition \ref{fixed}}
We have divided the proof into two cases.

$\mathbf{Case\ I}$: There are total $k$ firms such that $\rho=1,2,...,k$. We assume that $m\subset \mathbb N$, a set $\beth$ with condition $|\beth|=m+1$, and affinely independent state variable and strategy $\{Z_\rho(s)\}_{\rho\in \beth}\subset \mathbb R^3$ such that $\widetilde \Xi$ coincides with the simplex convex set of $\{ Z_\rho(s)\}_{\rho\in \beth}$. For each $Z(s)\subset \Xi$, there is a unique way in which the vector $Z(s)$ can be written as a convex combination of the extreme valued state variable and strategy , namely, $Z(s)=\sum_{\rho\in\beth}\alpha_\rho(s,Z)Z_\rho(s)$ such that $\sum_{\rho\in\beth}\alpha_\rho(s,Z)=1$ and $\alpha_\rho(s,Z)\geq 0,\ \forall \rho\in\beth$ and $s\in[0,t]$. For each firm $\rho$, define a set 
\[
\widetilde\Xi_\rho:=\left\{Z\in\widetilde\Xi:\a_\rho[\mathcal L_\rho(s,Z)]\leq\a_\rho(s,Z)\right\}.
\]
By the continuity of the Lagrangian of $\rho^{th}$ firm $\{\mathcal L_\rho\}_{\rho\in\beth}$, $\widetilde\Xi_\rho$ is closed. Now we claim that, for every $\tilde \beth \subset \beth$, the convex set consists of $\{Z_\rho\}_{\rho\in\tilde\beth}$ is proper subset of $\bigcup_{\rho\in\tilde\beth}\widetilde\Xi_\rho$. Suppose $\tilde\beth\subset\beth$ and $Z(s)$ is also in the non-empty, convex set consists of the state variables and the strategy $\{Z_\rho(s)\}_{\rho\in\tilde\beth}$. Therefore, there exists $\rho\in\tilde\beth$ such that $\a_\rho(s,Z)\geq\a_\rho\left[\mathcal L_\rho(s,Z)\right]$ which implies $Z(s)\in\tilde\Xi\subset \bigcup_{l\in\tilde\beth}\tilde\Xi_l$. By \emph{Knaster-Kuratowski-Mazurkiewicz Theorem}, there is $\bar Z_\rho^*\in \bigcap_{\rho\in\beth}\tilde\Xi_\rho$, in other words, the condition $\a_\rho\left[\mathcal L_\rho(s,\bar Z_\rho^*)\right]\leq\a_\rho(s,\bar Z_\rho^*)$ for all $\rho\in\beth$ and for each $s\in[0,t]$ \citep{gonzalez2010}. Hence, $\mathcal L_\rho(s,\bar Z_\rho^*)=\bar Z_\rho^*$ or $\mathcal L_\rho$ has a fixed-point. 

$\mathbf{Case\ II}$: Again consider $\widetilde\Xi\subset \mathbb R^3$ is a non-empty, convex and compact set. Then for $m\subset \mathbb N$, a set $\beth$ with condition $|\beth|=m+1$, and affinely independent state variable and strategy $\{Z_\rho(s)\}_{\rho\in \beth}\subset \mathbb R^3$ such that $\widetilde\Xi$ is a proper subset of the convex set based on $\{Z_\rho(s)\}_{\rho\in\beth}$ for all $s\in[0,t]$. Among all the simplices, suppose $\hat\aleph$ is the set with smallest $m$. Let $\tilde Z(s)$ be a dynamic point in the $m$-dimensional interior of $\hat\aleph$. Define ${\hat{\mathcal L}}_\rho$, an extension of $\mathcal L_\rho$ to the whole simplex $\hat\aleph$, as follows. For every $Z(s)\in\hat\aleph$, let 
\[
\bar\zeta(s,Z):\max\left\{\bar\zeta\in[0,1]:(1-\bar\zeta)\tilde Z(s)+\bar\zeta Z(s)\in\widetilde\Xi\right\},\ \forall s\in[0,t],
\]
and,
\[
{\hat{\mathcal L}}_\rho(s,Z):\mathcal L_\rho\left\{\left[1-\bar\zeta(s,Z)\right]\tilde Z(s)+\bar\zeta(s,Z) Z(s)\right\}.
\]
Therefore, $\bar\zeta$ is continuous which implies ${\hat{\mathcal L}}_\rho(s,Z)$ is continuous. Since the codomain of ${\hat{\mathcal L}}_\rho(s,Z)$ is in $\tilde\Xi$, every fixed-point of ${\hat{\mathcal L}}_\rho(s,Z)$ is also a fixed-point of $\mathcal L_\rho$. Now by \emph{$\mathbf{Case\  I}$}, ${\hat{\mathcal L}}_\rho(s,Z)$ has a fixed-point and therefore, $\mathcal L_\rho$ also does. $\square$

\subsection{Proof of Proposition \ref{pr1}}

Following \cite{chow1996} the Euclidean action function of firm $\rho$ in $[0,t]$ is, 
\begin{align}
\mathcal{A}_{0,t}^\rho(X)&=  \int_0^t \E_s\bigg\{ \pi_\rho[s,X(s),u_\rho(s),\hat{\mathbf{u}}^*_{-\rho}(s)] ds+\lambda_\rho(s) \big[X(s+ds)ds-X(s)ds\notag\\
&-\mu[s,X(s),u_\rho(s),\hat{\mathbf{u}}^*_{-\rho}(s)] ds-\sigma[s,X(s),u_\rho(s),\hat{\mathbf{u}}^*_{-\rho}(s)] dB(s)\big] \bigg\}.\notag
\end{align}
Let $\Delta s=\varepsilon>0$, and for $L_\varepsilon>0$ from Lemma \ref{l1} in the Appendix, the transition wave function of firm $\rho$ is 
\begin{align}\label{na4}
\Psi_{s,s+\varepsilon}^\rho(X)&= \frac{1}{L_\varepsilon} \int_{\mathbb{R}^{km}} \exp\biggr\{-\varepsilon  \mathcal{A}_{s,s+\varepsilon}^\rho(X)\biggr\} \Psi_s^\rho(X) dX(s),
\end{align}
for interval $[s,s+\varepsilon]$, where $\varepsilon\downarrow 0$, and $\Psi_s^\rho(X)$ is the value of firm $\rho$'s transition function at time $s$ and states $X(s)$ with initial conditions $\Psi_0^\rho(X)=\Psi_0^\rho$. In Equation (\ref{na4}), $\mathbb{R}^{km}$ represents $k\times m$-dimensional strategy space of firm $\rho$. Let $\Delta X(\nu)=X(\nu+d\nu)-X(\nu)$, then the Euclidean action function of firm $\rho$ is,
\begin{align}\label{na6}
\mathcal{A}_{s,\tau}^\rho(X)&= \E_s\ \bigg\{ \int_{s}^{\tau} \pi_\rho[\nu,X(\nu),u_\rho(\nu),\hat{\mathbf{u}}^*_{-\rho}(\nu)] d\nu+\lambda_\rho(\nu) \big[\Delta x(\nu)d\nu\notag\\
&\hspace{2cm}-\mu[\nu,X(\nu),u(\nu)] d\nu-\sigma[\nu,X(\nu),u_\rho(\nu),\hat{\mathbf{u}}^*_{-\rho}(\nu)] dB(\nu)\big] \bigg\}.
\end{align}
By Theorem 4.1.2 of \cite{oksendal2003},  there exists a $k$-dimensional vector valued function $$g^\rho[\nu,X(\nu)]\in C^2([0,\infty)\times\mathcal X)$$ that satisfies Theorem \ref{thmf} in the Appendix, Assumptions \ref{as0}-\ref{as2}, and $Y^\rho(\nu)=g^\rho[\nu,X(\nu)]$ where $Y^\rho(\nu)$ is firm $\rho$'s It\^o process. Assume 
\begin{multline*}
g^\rho[\nu+\Delta \nu,X(\nu)+\Delta X(\nu)]\approx \lambda_\rho(\nu) \big[\Delta X(\nu)d\nu-\mu[\nu,X(\nu),u_\rho(\nu),\hat{\mathbf{u}}^*_{-\rho}(\nu)] d\nu-\sigma[\nu,X(\nu),u_\rho(\nu),\hat{\mathbf{u}}^*_{-\rho}(\nu)] dB(\nu)+o(1),
\end{multline*} 
Equation (\ref{na6}) becomes,
\begin{align}
\mathcal{A}_{s,\tau}^\rho(X)&=\E_s\ \bigg\{ \int_{s}^{\tau}\ \pi_\rho[\nu,X(\nu),u_\rho(\nu),\hat{\mathbf{u}}^*_{-\rho}(\nu)]\ d\nu+g^\rho[\nu+\Delta \nu,X(\nu)+\Delta X(\nu)]\bigg\}.\notag
\end{align}
Generalized It\^o's Lemma implies
\begin{multline*}
\mathcal{A}_{s,\tau}^\rho(X)= \bigg[\pi_\rho[s,X(s),{u}_\rho(s),\hat{\mathbf{u}}^*_{-\rho}(s)]+g^\rho[s,X(s)]+\frac{\partial}{\partial s}g^\rho[s,X(s)]+\sum_{\rho=1}^k\frac{\partial}{\partial X_\rho}g^\rho[s,X(s)] \mu[s,X(s),u_\rho(s),\hat{\mathbf{u}}^*_{-\rho}(s)]\\
+\mbox{$\frac{1}{2}$} \sum_{\rho=1}^k\sum_{j=1}^m \sigma^{\rho j}\left[s,X(s),{u}_\rho(s),\hat{\mathbf{u}}^*_{-\rho}(s)\right] \frac{\partial^2}{\partial X_\rho\partial X_j}g^\rho[s,X(s)] +o(1)\bigg],
\end{multline*}
where $\E_s[\Delta B(s)]=0$ and $\E_s[o(\varepsilon)]/\varepsilon\ra 0$ as $\varepsilon\downarrow 0$ with the vector of initial conditions $x_0^\rho$, where $\sigma^{\rho j}[s,X(s),u_\rho(s),\hat{\mathbf{u}}^*_{-\rho}(s)]$ represents $\{\rho,j\}^{th}$ component of the variance-covariance matrix, and $\Delta B_\rho\Delta B_j=\delta^{\rho j} \varepsilon$, $\Delta B_\rho \varepsilon=\varepsilon \Delta B_\rho=0$, and $\Delta X_\rho(s) \Delta X_j(s)=\varepsilon$. A Taylor series expansion of the vector valued transition function $\Psi_{s,\tau}^\rho$ yields
\begin{multline*}
\Psi_s^{\tau, \rho}(X)+\varepsilon  \frac{\partial \Psi_s^{\tau,\rho}(X)}{\partial s}+o(\varepsilon)=\frac{1}{L_\varepsilon} \int_{\mathbb{R}^{km}} \exp\biggr\{-\varepsilon  \bigg[\pi_\rho[s,X(s),u_\rho(s),\hat{\mathbf{u}}^*_{-\rho}(s)]+g^\rho[s,X(s)] \mu[s,X(s),u_\rho(s),\hat{\mathbf{u}}^*_{-\rho}(s)]\\
+\mbox{$\frac{1}{2}$} \sum_{\rho=1}^k\sum_{j=1}^{m} \sigma^{\rho j}[s,X(s),u_\rho(s),\hat{\mathbf{u}}^*_{-\rho}(s)] \frac{\partial^2}{\partial X_\rho\partial X_j}g^\rho[s,X(s)]\bigg]\biggr\}\Psi_s^\rho(X) dX(s)+o(\varepsilon^{1/2}),
\end{multline*}
as $\varepsilon\downarrow 0$. Let $X(s)_{n\times 1}=X(\tau)_{k\times 1}+\xi_{k\times 1}$. There exists a positive number $\eta<\infty$ such that, $||\xi||\leq\eta\varepsilon [X'(s)]^{-1}$, and $[X'(s)]^{-1}$ exists and not equal to zero. Following our previous arguments
\begin{align}\label{na13}
\Psi_s^{\tau,\rho}(X)+\varepsilon \frac{\partial \Psi_s^{\tau,\rho}(X)}{\partial s}+o(\varepsilon)&= \frac{1}{L_\varepsilon}\int_{\mathbb{R}^{km}} \left[\Psi_s^{\tau,\rho}(X)+\xi\frac{\partial \Psi_s^{\tau,\rho}(X)}{\partial X}+o(\varepsilon)\right] \notag\\ 
&\times\exp\biggr\{-\varepsilon \bigg[\pi_\rho[s,X(\tau)+\xi,u_\rho(s),\hat{\mathbf{u}}^*_{-\rho}(s)]+g^\rho[s,X(\tau)+\xi]\notag\\
&+\frac{\partial}{\partial s}g^\rho[s,X(\tau)+\xi]+\sum_{rho=1}^k \frac{\partial}{\partial X_\rho}g^\rho[s,X(\tau)+\xi]\ \mu[s,X(\tau)+\xi,u_\rho(s),\hat{\mathbf{u}}^*_{-\rho}(s)]\notag\\
&\hspace{1cm}+\mbox{$\frac{1}{2}$}\sum_{\rho=1}^k\sum_{j=1}^{m} \sigma^{\rho j} [s,X(\tau)+\xi,u_\rho(s),\hat{\mathbf{u}}^*_{-\rho}(s)]\notag\\ &\hspace{2cm}\times \frac{\partial^2}{\partial X_\rho\partial X_j}g^\rho[s,X(\tau)+\xi]\bigg]\biggr\} d\xi+o(\epsilon^{1/2}).
\end{align}
Let
\begin{multline*}
f^\rho[s,\xi,u_\rho(s),\mathbf{u}_{-\rho}^*(s)]=\pi_\rho[s,X(\tau)+\xi,u_\rho(s),\hat{\mathbf{u}}^*_{-\rho}(s)]+g^\rho[s,X(\tau)+\xi]\\
+\frac{\partial}{\partial s}g^\rho[s,X(\tau)+\xi]+\sum_{\rho=1}^k\frac{\partial}{\partial X_\rho}g^\rho[s,X(\tau)+\xi]\ \mu[s,X(\tau)+\xi,u_\rho(s),\hat{\mathbf{u}}^*_{-\rho}(s)]\\
+\mbox{$\frac{1}{2}$}\sum_{\rho=1}^k\sum_{j=1}^{m}\ \sigma^{\rho j} [s,X(\tau)+\xi,u_\rho(s),\hat{\mathbf{u}}^*_{-\rho}(s)] \frac{\partial^2}{\partial X_\rho\partial X_j}g^\rho[s,X(\tau)+\xi]. 
\end{multline*}
Equation (\ref{na13}) implies
\begin{align}
\Psi_s^{\tau,\rho}(X)+\varepsilon  \frac{\partial \Psi_s^{\tau,\rho}(X)}{\partial s}+o(\varepsilon)&= \frac{1}{L_\varepsilon} \Psi_s^{\tau,\rho}(X) \int_{\mathbb{R}^{km}} \exp\bigg\{-\varepsilon f^\rho[s,\xi,u_\rho(s),\hat{\mathbf{u}}^*_{-\rho}(s)]\bigg\}  d\xi\notag\\
&+\frac{1}{L_\varepsilon} \frac{\partial \Psi_s^{\tau,\rho}(X)}{\partial X} \int_{\mathbb{R}^{km}} \xi\exp\bigg\{-\varepsilon f^\rho[s,\xi,u_\rho(s),\hat{\mathbf{u}}^*_{-\rho}(s)]\bigg\} d\xi+o(\varepsilon^{1/2}),\notag
\end{align}
where
\begin{multline*}
f^\rho[s,\xi,u_\rho(s),\hat{\mathbf u}_{-\rho}^*(s)]=f^\rho[s,X(\tau),u_\rho(s),\hat{\mathbf{u}}^*_\rho(s)]+\sum_{\rho=1}^k \frac{\partial}{\partial X_\rho}f^\rho[s,X(\tau),u_\rho(s),\hat{\mathbf{u}}^*_{-\rho}(s)] [\xi_\rho-X_\rho(\tau)]\\
+\mbox{$\frac{1}{2}$} \sum_{\rho=1}^k\sum_{j=1}^{m} \frac{\partial^2}{\partial X_\rho\partial X_j}f^\rho[s,X(\tau),u_\rho(s),\hat{\mathbf{u}}^*_{-\rho}(s)][\xi_\rho-X_\rho(\tau)] [\xi_j-X_j(\tau)] +o(\varepsilon),
\end{multline*}
as $\varepsilon\downarrow 0$. Define $m_{k\times 1}:=\xi_{k\times 1}-X(\tau)_{k\times 1}$ so that $ d\xi=dm$, then
\begin{align}
\int_{\mathbb{R}^{km}} \exp\bigg\{-\varepsilon f^\rho[s,\xi,u_\rho(s),\hat{\mathbf{u}}^*_{-\rho}(s)]\bigg\} d\xi&=\sqrt{\frac{(2\pi)^{km}}{\varepsilon |\tilde{\theta}|}} \exp\bigg\{-\varepsilon f^\rho[s,X(\tau),u_\rho(s),\hat{\mathbf{u}}^*_{-\rho}(s)]+\mbox{$\frac{1}{2}$}\varepsilon \widehat{w}'\tilde{\theta}^{-1}\widehat{w}\bigg\},\notag
\end{align} 
where for $k=m$. Let $\tilde{\theta}_{k\times k}$ be a symmetric, positive definite and non-singular Hessian matrix
\[
\tilde{\theta}=\begin{bmatrix}
\frac{\partial^2}{\partial X_1\partial X_1}f^\rho &  \frac{\partial^2}{\partial X_1\partial X_2}f^\rho& \dots & \frac{\partial^2}{\partial X_1\partial X_k}f^\rho\\ \frac{\partial^2}{\partial X_2\partial X_1}f^\rho & \frac{\partial^2}{\partial X_2\partial X_2}f^\rho& \dots & \frac{\partial^2}{\partial X_2\partial X_k}f^\rho\\ \vdots &\vdots&\ddots & \vdots\\\frac{\partial^2}{\partial x_n\partial x_1}f^\rho & \frac{\partial^2}{\partial X_k\partial X_2}f^\rho&\dots& \frac{\partial^2}{\partial X_k\partial X_k}f^\rho
\end{bmatrix},
\]
and 
\[
\widehat{w}[s,X(\tau),u_\rho(s),\hat{\mathbf{u}}^*_{-\rho}(s)]=\begin{bmatrix}
-\frac{\partial}{\partial X_1}f^\rho[s,X(\tau),u_\rho(s),\hat{\mathbf{u}}^*_{-\rho}(s)]\\-\frac{\partial}{\partial X_2}f^\rho[s,X(\tau),u_\rho(s),\hat{\mathbf{u}}^*_{-\rho}(s)]\\\vdots\\-\frac{\partial}{\partial X_k}f^\rho[s,X(\tau),u_\rho(s),\hat{\mathbf{u}}^*_{-\rho}(s)]
\end{bmatrix}.
\]
Similarly,
\begin{multline*}
\int_{\mathbb{R}^{km}} \xi \exp\bigg\{-\varepsilon f^\rho[s,\xi,u_\rho(s),\hat{\mathbf{u}}^*_{-\rho}(s)]\bigg\} d\xi\\
=\sqrt{\frac{(2\pi)^{km}}{\varepsilon |\tilde{\theta}|}} \exp\bigg\{-\varepsilon f^\rho[s,X(\tau),u_\rho(s),\hat{\mathbf{u}}^*_{-\rho}(s)]+\mbox{$\frac{1}{2}$}\varepsilon \widehat{w}'\tilde{\theta}^{-1}\widehat{w}\bigg\} \bigg[X(\tau)+\mbox{$\frac{1}{2}$} (\theta^{-1} \widehat{w})\bigg].
\end{multline*}
Therefore
\begin{multline*}
\Psi_s^{\tau,\rho}(X)+\varepsilon \frac{\partial \Psi_s^{\tau,\rho}(X)}{\partial s}+o(\varepsilon)=\frac{1}{L_\varepsilon} \sqrt{\frac{(2\pi)^{km}}{\varepsilon |\tilde{\theta}|}} \exp\bigg\{-\varepsilon f^\rho[s,X(\tau),u_\rho(s),\hat{\mathbf{u}}^*_{-\rho}(s)]+\mbox{$\frac{1}{2}$}\varepsilon \widehat{w}'\tilde{\theta}^{-1}\widehat{w}\bigg\}\\
\times\bigg[\Psi_s^{\tau,\rho}(X)+\left[X(\tau)+\mbox{$\frac{1}{2}$} (\tilde{\theta}^{-1} \widehat{w})\right]\frac{\partial \Psi_s^{\tau,\rho}(X)}{\partial X}\bigg]+o(\varepsilon^{1/2}).
\end{multline*}
Assuming $L_\varepsilon=\sqrt{(2\pi)^n/(\varepsilon |\tilde{\theta}|)}>0$, the Wick-rotated Schr\"odinger type equation is,
\begin{multline*}
\Psi_s^{\tau,\rho}(X)+\varepsilon  \frac{\partial \Psi_s^{\tau,\rho}(X)}{\partial s}+o(\varepsilon)=\left\{1-\varepsilon f^\rho[s,X(\tau),u_\rho(s),\hat{\mathbf{u}}^*_{-\rho}(s)]+\mbox{$\frac{1}{2}$}\varepsilon \widehat{w}'\tilde{\theta}^{-1}\widehat{w}\right\}\\ \times\bigg[\Psi_s^{\tau,\rho}(X)+\left[X(\tau)+\mbox{$\frac{1}{2}$} (\tilde{\theta}^{-1} \widehat{w})\right]\frac{\partial \Psi_s^{\tau,\rho}(X)}{\partial X}\bigg]+o(\varepsilon^{1/2}).
\end{multline*}
For any finite positive number $\eta$ we know $X(\tau)\leq\eta\varepsilon|\xi'|^{-1}$, and there exists a $|\theta^{-1}w|\leq 2 \eta\varepsilon|1-\xi'|^{-1}$ such that for $\varepsilon\downarrow 0$ we have $\big|X(\tau)+\mbox{$\frac{1}{2}$}(\theta^{-1} w)\big|\leq\eta\varepsilon$. Hence, 
\begin{align*}
\frac{\partial \Psi_s^{\tau,\rho}(X)}{\partial s}&=- f^\rho[s,X(\tau),u_\rho(s),\hat{\mathbf{u}}^*_{-\rho}(s)] \Psi_s^{\tau,\rho}(X), 
\end{align*}
with the Wick-rotated Schr\"odinger-type equation as,
\begin{align}
-\frac{\partial f^\rho[s,X(\tau),u_\rho(s),\hat{\mathbf{u}}^*_{-\rho}(s)]}{\partial u_\rho}\ \Psi_s^{\tau,\rho}(X)=0,\notag
\end{align}
whose solution with respect to $u_\rho$ gives $\rho^{th}$ firm's non-cooperative feedback Nash equilibrium strategy\\ $\phi_N^{\rho*}[s,X^*(s),u_\rho^*(s),\hat{\mathbf{u}}^*_{-\rho}(s)]$.
$\square$

\section{Concluding remarks.}

The optimal control theory has been developed by \citep{boltyanskiy1962,pontryagin1966}, and hence the introduction of \emph{Pontryagin maximum principle}. The type of problem considered in this paper is \emph{Lagrange problem} in optimal control theory. The Maximum principle is generally stated either for the somewhat simpler so-called \emph{Meyer} problem or the more general \emph{Bolza} problem \citep{intriligator1971}. Although, all of these three problems are equivalent, and when the problem is formulated in vector form, one can easily go back and forth between these different problems \citep{acemoglu2008}. In control theory HJB approach is popular as it deals with a partial differential equation with a value function. The problem of this approach is that, for very high dimensional state variable, a computer simulation of the HJB equation is very difficult \citep{yang2014path}. An alternative approach is path integral control which utilizes \emph{Feynman-Kac} lemma \citep{kappen2005,theodorou2010}. There are two major problems of this approach; (i) if the Cauchy problem does not have a \emph{sufficiently integrable} solution, then this approach does not make sense, and (ii) the solution is not unique \citep{lindstrom2018}. Apart from that, the \emph{Feynman-Kac} needs a value function to work with. Our \emph{Feynaman}-type approach does not need the value function. Instead, we utilize the fact of existance of an integrating factor to emply it\^o lemma. We replace the functional form of the SDE by a $C^{1,2}$ It\^o process $g$. Then we subdivide $[0,t]$ into $n$ number of equal subintervals $[s,\tau]$ so that $\tau=s+\epsilon$ and $\epsilon=t/n$. If there are $k$ firms in the economy, then $k<n$. We construct a \emph{stochastic Lagrangian} and we fit a Feynman action function for each $[s,\tau]$. Further, we implement a few first order Taylor series expansions for $g$ and $f$. Once, the Wick rotated Schr\"odinger-type equation is found, the first order condition with respect to the strategy $u(s)$ yields the stochastic control. Again like path integral control approach, our approach does not give a unique solution which we have discussed through different examples. We also compared our results with Pontryagin maximum priciple generated by HJB examples in \cite{yeung2006}.

Recent advances in reinforcement learning (RL) and adaptive dynamic programming have significantly influenced the development of modern optimal control frameworks, particularly in stochastic and multi-agent environments. For example, \cite{zhang2025reinforcement} provides a comprehensive overview of the evolution of reinforcement learning as an interdisciplinary field combining artificial intelligence and control science, emphasizing its growing role as a central paradigm for solving optimal control problems. Their study analyzes the conceptual relationship between single-agent reinforcement learning (SARL) and multi-agent reinforcement learning (MARL), beginning with the historical development of RL and clarifying how it differs from other machine learning paradigms. Building upon the theoretical foundations of SARL, the authors extend the discussion to multi-agent system (MAS) cooperative control and show how algorithmic structures in SARL correspond to those in MARL \citep{pramanik2016tail}. In particular, SARL methods are reorganized into dynamic programming, value-function decomposition, and policy gradient approaches, while MARL algorithms are categorized into behavior analysis, centralized learning, communication learning, and collaborative learning paradigms. This mapping between single-agent and multi-agent formulations provides a unified framework for understanding cooperative decision-making in large-scale systems and highlights ongoing challenges in scaling RL-based control methods. 

Complementary developments can be found in the control-theoretic reinforcement learning literature. For instance, \cite{qin2025reinforcement} introduces a secure tracking control strategy for nonlinear interconnected systems under safety constraints and mismatched disturbances. In their approach, the tracking problem is transformed into a stabilization problem through system augmentation, and the control task is subsequently formulated as a zero-sum game inspired by philosophical principles of balanced competition. A control barrier function (CBF) is embedded in the cost function to ensure that system trajectories remain within a prescribed safe set, while an event-triggered mechanism leads to the derivation of a safety Hamilton–Jacobi–Isaacs equation \citep{pramanik2022stochastic,khan2023myb}. An adaptive single evaluation neural network combined with experience replay is used to approximate the optimal solution, and Lyapunov analysis establishes the uniform ultimate boundedness of both tracking and learning errors. Similarly, \cite{qin2023barrier} studies nonzero-sum differential games for uncertain nonlinear systems subject to state constraints and stochastic disturbances \citep{pramanik2022lock}. Their framework integrates control barrier functions with adaptive robust control and neural network learning to transform a robust multiplayer regulation problem into an optimal control problem with safety guarantees. Unlike classical actor–critic architectures, the method employs only a critic neural network for each player to approximate the cost function, thereby eliminating the need for an initially stabilizing control policy. Using Lyapunov stability theory, the authors prove that both the system states and the neural network weight estimation errors remain uniformly ultimately bounded despite the presence of uncertainty and random disturbances \citep{pramanik2021optimal}.

In addition, safety-oriented reinforcement learning methods have also been developed for systems with actuator faults and input saturation. In this direction, \cite{qin2025observer} proposes an event-triggered safety fault-tolerant control framework based on adaptive dynamic programming for nonlinear systems with constrained states. Their method introduces a smooth mapping to transform the constrained system into an unconstrained representation, designs a fault observer to estimate unknown actuator faults, and compensates for such faults in real time \citep{hertweck2023clinicopathological,vikramdeo2023profiling}. The optimal safety value function is approximated through a single critic neural network, and an event-triggered mechanism allows the optimal control law to be computed without explicitly constructing an event-triggered HJB equation. Furthermore, by adjusting a triggering parameter, the framework balances resource utilization with control performance, while theoretical analysis guarantees boundedness of all system signals \citep{dasgupta2023frequent,kakkat2023cardiovascular,pramanik2023path}. These developments demonstrate the growing interaction between reinforcement learning, game theory, and stochastic control. In the context of the present paper, which develops a path-integral-based stochastic control framework for cooperative economic systems, these studies highlight how modern RL approaches can complement classical optimal control methodologies such as dynamic programming and \emph{Pontryagin's maximum principle} \citep{pramanik2021consensus,pramanik2023consensus}. In particular, the algorithmic ideas of sampling-based optimization, robustness to uncertainty, and multi-agent coordination discussed in the RL literature provide an important conceptual bridge to the stochastic path-integral and game-theoretic control formulations investigated in this work.

Moreover, in this paper we use our method to find optimal strategies for dynamic profit functions quadratic in time with a stochastic differential market dynamics for infinite dimensional vector spaces (i.e. Walrasian equilibrium) and finite dimensional vector spaces (i.e. Pareto optimality and Nash equilibrium). In Proposition \ref{p2} we show in the generalized non-linear case like the Merton-Garman-Hamiltonian \citep{belal2007,merton1973} Equation we are able to obtain an optimal strategy where employing HJB approach is very difficult. Furthermore, in Example \ref{ex2}  where both the profit function and the drift of market dynamics are linear to strategy, we are still able to find an optimal strategy of a firm. Again in this case, we cannot use Pontryagin's maximum principle because after doing differentiation with respect to control, the strategy term vanishes and optimal strategy cannot be found. According to the Generalized Weierstrass Theorem we know, solution exists when both the objective function and market dynamics are linear in terms of control \citep{intriligator1971,pramanik2023path}. Under Proposition \ref{pr1} we determine a non-cooperative feedback Nash equilibrium and in the future we plan to extend this result to cooperative Nash equilibria.

\appendix

\section{Appendix.}

This appendix outlines the complete assumptions required to develop the results. Throughout this paper we are considering Euclidean quantum field theory which requires further assumptions on Equation (\ref{mkt}). A quantum field is an operated valued distribution $\mathfrak{F}[s, X(s)]$ to the unbounded operators on a Hilbert space following the Garding-Wightman axioms \citep{simon1979}. Consider a measure $d\xi\equiv dX(s)$ on an Euclidean free field $\mathfrak{L}(\mathbb{R}^2)$ (The dimension is two for space-time under a Walrasian system) \citep{khan2023myb}, whose moments are the candidates of Schwinger functions. For a real valued tempered distribution $T$, let
\begin{align}
\left(\int\ \mathfrak{F}(y)\ f(y)\ d^2y \right)(T)&=T(f),\notag
\end{align}
be a random variable, where $f$ is a real valued test function of $y\in\mathbb{R}$ takes Schwinger function
\begin{align}
S_n(y_1,...,y_n)&=\int\ \mathfrak{F}(y_1)...\ \mathfrak{F}(y_n)\ dX\notag
\end{align}
which implies
\begin{align}
\int S_n(y_1,...,y_n)\ f_1(y_1)...\ f_n(y_n)\ d^{2n}y&=\int T(f_1)...\ T(f_n)\ dX(T).\notag
\end{align}

\begin{theorem}\label{thmf}
	(Fr\"ohlich's Reconstruction Theorem \citep{simon1979}) Let $d\xi$ be a cylindrical measure on Euclidean free field $\mathfrak{L}(\mathbb{R}^2)$ obeying the following properties:\\
	(i) The measure $d\xi$ is invariant with respect to proper Euclidean motions of the form $T(x)\mapsto T(Ex+h)$, where $ h\in\mathbb{R}^2$ and $ E\in \boldsymbol{SO}(2)$, where $\boldsymbol{SO}(2)$ represents Lie Special orthogonal group of dimension $2$.\\
	(ii) Osterwalder-Schrader positivity: For a given real valued test function $f$ in Garding-Wightman field $\mathfrak{L}'(\mathbb{R})$ or $f\in \mathfrak{L}'(\mathbb{R})$ with $\text{support}\ f\subset\{(s,x),\ s>0\}$, let $(\theta f)(s,x)=f(-s,x)$, where $\theta$ is a parameter. Then for real valued $f_1, f_2,...,\ f_n$ with the above support and for the set of complex numbers $z_1, z_2,..., z_n\in \mathbb{C}$ we have
	\begin{align}
	\sum_{j=1}^n\ \sum_{k=1}^n\ \overline{z}_kz_j\ \int\ \exp\left\{\mathfrak{i}[\mathfrak{F}(f_k)-\mathfrak{F}(\theta f_j)]\right\} d\xi\geq 0.\notag 
	\end{align} \\
	(iii) For any real valued test function in Euclidean free field $f\in\mathfrak{L}(\mathbb{R}^2)$, $$\int\ \exp[\mathfrak{F}(f)]\ d\xi<\infty,$$ and the action of the translations $[s,x(s)]\rightarrow[s+\varepsilon,x(s+\varepsilon)]$ is ergodic.
\end{theorem}

Assume Equation (\ref{mkt}) is in Euclidean free field and it satisfies three conditions in above Theorem \ref{thmf}. Hence, the measure $d\xi$ is cylindrical and the feasible set of Equation (\ref{mkt}) satisfies
\begin{align}\label{w3}
dX(s)\geq \mu[s,X(s),u(s)] ds+\sigma[s,X(s),u(s)] dB(s).
\end{align}
As $G[s,X(s),u(s)]=dX(s)- \mu[s,X(s),u(s)]\ ds-\sigma[s,X(s),u(s)]\ dB(s)$, Equation (\ref{w3}) implies $G[s,X(s),u(s)]\geq 0$. The
dynamic Walrasian system then satisfies
\begin{align}
&\max_{u \in \mathcal U}\ \Pi(u,t)=\max_{u \in \mathcal U}\ \E_0 \left\{\int_{0}^{t}\ \pi[s,X(s),u(s)]\biggr|\mathcal{F}_0^X\right\} ds,\notag
\end{align} 
with constraint $dX(s)=\mu[s,X(s),u(s)]\ ds+\sigma[s,X(s),u(s)]\ dB(s)$, and initial condition $x(0)=x_0$. Following  \cite{chow1996} at time $s'\in[0,t']$, the stochastic Lagrangian function is 
\begin{align}\label{w4}
&\int_0^{t'} \E_{s'}\biggr\{{\pi}[s',X(s'),u(s')]-\lambda(s') \tilde{G}[s',X(s'),u(s')]\biggr\} ds',
\end{align}
where $\lambda(s')$ is the non-negative Lagrangian multiplier, $$\tilde{G}[s',X(s'),u(s')]\ ds'={G}[s',X(s'),u(s')],$$ and $\E_{s'}$ is the conditional expectation on time $s'$, $\E_{s'}(.)=\E[.|X(s')]$. Since we are interested in a forward looking process, at time $s'$ only the information up to $s'$ is available, and based on this we forecast the state of the system at time $s'+ds'$. Furthermore, in the path integral approach \citep{pramanik2024bayes}, Equation (\ref{w4}) corresponds to the Lagrangian function of the Feynman action functional in Minkowski space-time with imaginary time $s'$. In order to get a Euclidean path integral we need to perform the Wick rotation on Equation (\ref{w4}). Suppose, there exists dynamic non-negative measurable profit function $\hat{\pi}$ such that ${\pi}[s',X(s'),u(s')]=d^2 \hat{\pi}[s',X(s'),u(s')]/(ds')^2$ \citep{pramanik2024estimation}. For imaginary time $s'$, the Feynman action functional becomes
\begin{align}\label{w5}
\mathcal{A}_{0,t'}^F(X)&= \int_0^{t'} \E_{s'}\biggr\{{\pi}[s',X(s'),u(s')]-\lambda(s') \tilde{G}[s',x(s'),u(s')]\biggr\} ds'. 
\end{align}
Multiplying both sides of Equation (\ref{w5}) by $\mathfrak{i}$ and substituting $s'=-\mathfrak{i} s$ (so that $ds'=-\mathfrak{i}\ ds$) yields,
\begin{align}\label{w6}
\mathfrak{i} \mathcal{A}_{0,t'}^F(X)&=\mathfrak{i}  \int_0^{t} \E_{s}\biggr\{\left(\frac{d}{-\mathfrak{i} ds}\right)^2\hat{\pi}[s,X(s),u(s)]-\lambda(s) \tilde{G}[s,X(s),u(s)]\biggr\} (-\mathfrak{i} ds)
\end{align}
so that
\begin{align}
\mathcal{A}_{0,t}(X)&=  \int_0^t \E_{s}\biggr\{{\pi}[s,X(s),u(s)]+\lambda(s) \tilde{G}[s,X(s),u(s)]\biggr\} ds.\notag
\end{align}
In Equation (\ref{w6}), $\mathcal{A}_{0,t}(X)$ is defined as Euclidean action functional after the Wick rotation. Theorem \ref{thmf} and Condition (\ref{w3}) imply that if $G[s,X(s),u(s)]\geq 0$ then $\tilde{G}[s,X(s),u(s)]\geq 0$ and the parenthesis term of conditional expectation at real time $s$ is always non-negative. Now we assume some further conditions on $G[s,X(s),u(s)]$.

\begin{as}\label{as1}
	Suppose $G[s,X(s),u(s)]$ is a non-negative real valued continuous function of $(s,X,u)\in[0,t]\times \mathcal X\times\mathbb{R}$ and infinitely differentiable with respect to $X$ and $u$ if $s\in[0,t]$ is fixed and $\a^{th}$ order derivatives $\partial_X^\a G[s,X,u]$ and $\partial_u^\a G[s,X,u]$, respectively are continuous functions of $(s,X,u)$ for any $\a$. Moreover, for any integer $m\geq 2$ there exist positive constants $v_{m}^1$ and $v_{m}^2$ such that, $
	\big|\partial_X^\a G[s,X,u]\big|\leq v_{ m}^1$,  $\big|\partial_u^\a G[s,X,u]\big|\leq v_{m}^2,$
	if $\a$ is an integer with $2\leq\a\leq m$ and $(s,X,u)\in[0,t]\times\mathcal X\times\mathbb{R}$.
\end{as}

From Assumption \ref{as1} there exists positive constants $ v_{\mathfrak 0}^1$,  $v_{\mathfrak 1}^1$ and $v_{\mathfrak 1}^2$ such that  for all $(s,X,u)\in[0,t]\times \mathcal X\times\mathbb{R}$, $\big| G(s,X,u)\big|\leq v_{\mathfrak 0}^1(1+|X|)^2$, $ \big|\partial_X G(s,X,u)\big|\leq v_{\mathfrak 1}^1(1+|X|)$ and, $ \big|\partial_u G(s,X,u)\big|\leq v_{\mathfrak 1}^2(1+|X|)$.

\begin{definition}\label{d1}
	For all $X\in[0,1]$ define a Wick-rotated wave integral $I(\Psi)$ with Euclidean action function $\mathcal{A}(X)$ such that 
	\begin{align}
	I(\Psi)=\int_{\mathbb{R}}\ \exp\{-\mathcal{A}(X)\}\ \Psi(X)\ dX,\notag 
	\end{align}
	where $\Psi(X)$ is a real valued wave function of $X$. 
\end{definition}
The integration defined in Definition \ref{d1} may not converge absolutely, and we need following definition \citep{fujiwara2017}.

\begin{definition}\label{d2}
	For $\varepsilon>0$ consider a family of $C^{\infty}$, $\omega_{\varepsilon}(X)$ which follows the properties given in Definition 3.1 of \cite{fujiwara2017}. The Wick rotated wave integral is
	\begin{align}
	I(\Psi)=\lim_{\varepsilon\ra 0}\ \int_{\mathbb{R}} \omega_\varepsilon \exp\{-\mathcal{A}(X)\}\ \Psi(X) dX, \notag
	\end{align}	
	as long as\\
	(i) For any family of $\omega_{\varepsilon}(X)$ the integral $I(\omega_\varepsilon)$ converges absolutely and,\\
	(ii) The right hand side limit of Equation (\ref{w}) exists and independent of choice of $\{\omega_\varepsilon\}$. 
\end{definition} 
After using Proposition $3.1$ in \cite{fujiwara2017} and Definition \ref{d2} we conclude that, integral $I(\Psi)$ in Definition \ref{d1} is absolutely convergent.

\begin{as}\label{as2}
	Suppose, $x\in \mathcal X$ such that;\\
	(i) The Euclidean action $\mathcal{A}(X)$ is a $C^\infty$ function. If $|\a|\geq 1$, then there exists a positive constant $\mathcal{C}_\a$ such that,
	\begin{align}
	\big|\partial_X^\a \mathcal{A}(X)\big|\leq\mathcal{C}_\a.\notag
	\end{align}
	(ii) The wave function $\Psi(X)$ which depends on $X$ is infinitely differentiable with respect to $X$. There exists a constant $\rho\geq 0$ such that for any $\a$
	\begin{align}
	\sup_{X\in \mathcal X}\ (1+|X|)^{-\rho}\ \big|\partial_X^\a \Psi(X)\big|<\infty .\notag
	\end{align}	
\end{as}

\begin{lem}\label{l1}
	[Convergence of Euclidean path integral \citep{fujiwara2017}] Consider small real time interval $[s,s+\Delta s]\subset[0,t]$ such that for some positive number $\delta>0$ we have $|\Delta s|\leq\delta$ and let $\Delta:s=s_0<s_1<...<s_J<s_{J+1}=s+\Delta s$ be an arbitrary division of interval $[s,s+\Delta s]$. Suppose $\tau_j=s_j-s_{j-1}$, $|\Delta|=\max_{1\leq j\leq j+1}\ \tau_j$ and for $X\in\mathcal X$ define transition function
	\begin{align}\label{w7}
	\Psi_{0,t}(X)&=\int_{A} \exp\big[-\mathcal{A}_{0,t}(X)\big] \mathfrak{D}_X,
	\end{align}
	where $A$ is the space of all paths that connect $x(0)$ to $x(t)$ and $\mathfrak{D}_X$ is a uniform measure on the space $A$. Let us define a local transition function in the interval $[s,s+\Delta]$ such that
	\begin{align}\label{w8}
	\Psi_{s,s+\Delta s}(X):=\frac{1}{L_\varepsilon} \int_{\mathbb{R}} \exp\biggr\{-\mathcal{A}_{s,s+\Delta s}(X)\biggr\} \Psi_s(X)dX
	\end{align}
	which satisfies Definitions \ref{d1} and \ref{d2} with
	\begin{align}\label{w9}
	I(\Delta,X,s,s+\Delta s):=\ \frac{1}{(L_\varepsilon)^k} \int_{\mathbb{R}^n}\ \exp\biggr\{ \sum_{j=1}^n - \mathcal{A}\big[x(s_{j-1},s_j)\big] \biggr\}\ \Psi_s(x) \prod_{j=1}^n dX(s_j),
	\end{align}
	where $\mathcal{A}_{s_{j-1},s_j}(x)$ is the Euclidean action function in $[s,s+\Delta s]$ and it is the Euclidean action function of $\tau_j$. If Equations (\ref{w7})-(\ref{w9}) satisfy Assumptions \ref{as1} and \ref{as2} then the following limit exists
	\begin{align*}
	\Psi_{0,t}(X)&=\lim_{|\Delta|\ra 0}\ I(\Delta,X,s,s+\Delta s).
	\end{align*}
\end{lem}

\subsection*{Proof of Lemma \ref{l0}.}
For all $\rho=1,..,k$, $\mathbf{x}^*\in\mathcal{X}\in\mathbb{R}^k$ and for the filtration $\mathcal{F}_0^{X}$ define a map $A:\mathcal{U}\ra2^{\mathcal{U}}$ by\\
\[(i)\ A_{\mathbf{u}^1}=\left\{\mathbf{u}^2\in\mathcal{U}\big|u_\rho\in M_\rho(\hat{\mathbf{u}}_\rho),\ \forall \rho=1,...,k\right\}=\prod_{\rho=1}^k M_\rho(\hat{\mathbf{u}}_\rho).\]
For all $\mathbf{u}^2\in\mathcal{U}$
\[(ii)\ A_{\mathbf{u}^2}^{-1}=\left\{\mathbf{u}^1\in\mathcal{U}\big|u_\rho\in M_\rho(\hat{\mathbf{u}}_\rho),\ \forall \rho=1,...,k\right\}=\bigcap_{\rho=1}^k\mathcal{U}_\rho\times M_\rho(\hat{\mathbf{u}}_\rho),\]
where $\mathbf{u}^1$ and $\mathbf{u}^2$ are two arbitrary strategy vectors in $\mathcal{U}$. From $(i)$ and $(ii)$ we know that the values of $A$ are convex and non-empty and, the values of $A^{-1}$ are open. Therefore, $A$ is a Fan map, and by Fan-Browder Theorem $A$ has a set of fixed points $\{u_\rho^*,\hat{\mathbf{u}}_{-\rho}^*\}$ \citep{granas2004}. Hence, $\{u_\rho^*,\hat{\mathbf{u}}_{-\rho}^*\}\in \bigcap_{\rho=1}^k\mathcal{G}_\rho$. $\square$

\subsection*{Proof of Lemma \ref{l0.1}.}

For  filtration $\mathcal{F}_0^X$ define a map $M_\rho:\mathcal{U}^\rho\ra 2^{\mathcal{U}_\rho}$ by 
\[
M_\rho(\hat{\mathbf{u}}_{-\rho})=\left\{u_\rho\in\mathcal{U}_\rho\big|l_\rho(s,\mathbf{x}^*,u_\rho,\hat{\mathbf{u}}_{-\rho})>0\right\},\ \forall\rho=1,...,k.
\]
Since each $M_\rho(\hat{\mathbf{u}}_{-\rho})$ is convex set by the quasi-concavity of  ${u}_\rho\mapsto l_\rho\left(s,\mathbf{x}^*,u_\rho,\hat{\mathbf{u}}_{-\rho}\right)$ and is non-empty because of $(c)$, this $M_\rho$ is a Fan map. Furthermore, since 
\[
M_\rho^{-1}(u_\rho)=\left\{\hat{\mathbf{u}}_{-\rho}\in\mathcal{U}^\rho\big|l_\rho(s,\mathbf{x}^*,u_\rho,\hat{\mathbf{u}}_{-\rho})>0\right\},
\]
each value from this inverse mapping (i.e. fiber of $M_\rho$) is open by $(a)$. Therefore, $M_1,...,M_k$ satisfy Lemma \ref{l0}. Hence, there exists a set of optimal strategies $\left\{u_\rho^*,\hat{\mathbf{u}}_{-\rho}^*\right\}\in\mathcal{U}$ such that, $\hat{\mathbf{u}}_\rho\in M_\rho(\hat{\mathbf{u}}_\rho)$, which implies
\[
l_\rho\left(s,\mathbf{x}^*,u_\rho^*,\hat{\mathbf{u}}_{-\rho}^*\right)=l_\rho\left(s,\mathbf{x}^*,u_\rho,\hat{\mathbf{u}}_{-\rho}\right)>0,\ \forall \rho=1,...,k.\ \square
\]
	
	\bibliographystyle{apalike}
	\bibliography{mybibfile}
	\end{document}